\documentclass[a4paper,11pt]{article}
\pdfoutput=1 % if your are submitting a pdflatex (i.e. if you have
\usepackage{jheppub} % for details on the use of the package, please see the JHEP-author-manual
\makeatletter
\RenewDocumentCommand{\author}{O{}m}{
    \IfBlankTF{#1}
    {\auth@toks=\expandafter{\the\auth@toks#2\ }}
    {\auth@toks=\expandafter{\the\auth@toks#2\unboldmath$^{#1}$\ }}
}
\makeatother

\NewExpandableDocumentCommand{\paperTitle}{}{%
    Finite-coupling spectrum of \texorpdfstring{\(\OO(N)\)}{O(N)} model in \texorpdfstring{\AdS{}}{AdS}%
}
\NewExpandableDocumentCommand{\paperKeywords}{}{%
    \texorpdfstring{\(\OO(N)\)}{O(N)} Model, %
    \texorpdfstring{\(1/N\)}{1/N} Expansion, %
    \texorpdfstring{\QFT{} in \AdS{}}{QFT in AdS}, %
    Conformal Field Theory
}

\hypersetup{
    bookmarksnumbered,   % include chapter/section numbers in PDF outline
    pdfauthor   = {Jonáš Dujava, Petr Vaško},
    pdftitle    = {\paperTitle},
    pdfsubject  = {Study of the the boundary CFT corresponding to the O(N) model in AdS.},
    pdfkeywords = {\paperKeywords},
    pdfdisplaydoctitle, % https://tex.stackexchange.com/a/435434/120853
}

\usepackage[T1]{fontenc} % proper encoding (of accented characters)
\usepackage{lmodern} % Latin Modern font (improved version of default Computer Modern)

\DeclareEmphSequence{\slshape,\itshape}

\usepackage[table]{xcolor} % color support
\usepackage{booktabs}   % improved horizontal lines in tables
\usepackage{multirow}   % multi-row cells in tables
\usepackage{array}      % custom column types in tables
\usepackage{tabularray} % enhanced `tblr` tables
\usepackage{codehigh}   % verbatim in tables (with `\fakeverb` macro)
\UseTblrLibrary{amsmath,booktabs}
\usepackage{enumerate}
\usepackage{adjustbox}

\usepackage{microtype}  % improve typography
\usepackage[capitalise,nameinlink]{cleveref}   % enhanced cross-referencing
\usepackage{csquotes}   % context-sensitive quotation facilities

\usepackage{caption}    % customizing captions
\usepackage{subcaption}
\graphicspath{      % default paths to figures
    {./figures/}
}

\usepackage{fetamont}   % "logo" font
\NewDocumentCommand{\githubrepository}{O{[jdujava/ONinAdS]}}{%
    \href{https://github.com/jdujava/ONinAdS}{\ttfamily #1}%
}
\NewDocumentCommand{\wolframlogo}{}{\raisebox{-0.4ex}{\hspace{0.1em}\includegraphics[height=2.2ex]{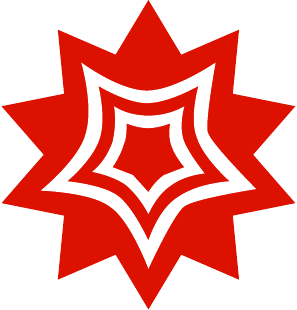}\,}}
\NewDocumentCommand{\wmathematica}{}{\wolframlogo\textffm{Wolfram Mathematica}}

\NewDocumentCommand{\notebook}{}{\githubrepository[\wolframlogo\textcolor{black}{\textffm{Notebook}}]}

\NewDocumentCommand{\todo}{O{} >{\TrimSpaces} m}{%
    \textcolor{red}{%
        \textsf{TODO\IfBlankF{#1}{(#1)}}%
        \IfBlankTF{#2}{.}{: #2}%
    }%
}
\NewDocumentCommand{\Note}{O{} >{\TrimSpaces} m}{%
    \textcolor{green!50!black}{%
        \textsf{Note\IfBlankF{#1}{(#1)}}: #2%
    }%
}

\NewDocumentCommand{\grayenclose}{s O{(} m O{)}}{%
    \ifmmode
        \let\colorcmd\mathcolor% TODO: maybe do this everywhere for math mode
        \DeclareCommandCopy{\Left}{\left}%
        \DeclareCommandCopy{\Right}{\right}%
    \else%
        \let\colorcmd\textcolor%
        \DeclareCommandCopy{\Left}{\big}%
        \DeclareCommandCopy{\Right}{\big}%
    \fi%       (just use one command everywhere)
    \IfBooleanTF{#1}{%
        \colorcmd{gray}{\Left#2}#3\colorcmd{gray}{\Right#4}%
    }{%
        \colorcmd{gray}{#2}#3\colorcmd{gray}{#4}%
    }%
}

\makeatletter
\input{preamble/tikz-diagrams.tex} % TikZ diagrams
\usepackage{mathtools}
\usepackage{amsmath,amsfonts,amssymb,bbold,bm}
\usepackage{physics}
\usepackage{scalerel}   % scaling of math symbols (\ThisStyle, \SavedStyle, ...)

\usepackage{mathfixs}   % fix some odd behaviour in math mode, add math macros
\ProvideMathFix{autobold}
\ProvideMathFix{greekcaps=it}
\ProvideMathFix{frac,rfrac,vfrac,vfracskippre=4mu}
\NewCommandCopy{\dotaccent}{\.}
\RenewDocumentCommand{\.}{}{\TextOrMath{\dotaccent}{\mspace{1mu}}}
\NewCommandCopy{\acuteaccent}{\'}
\renewcommand*{\'}{\TextOrMath{\acuteaccent}{\mspace{-1mu}}}

\usepackage{siunitx}
\NewDocumentCommand{\eqend}{}{\,.}
\NewDocumentCommand{\eqcomma}{}{\,,}

\NewDocumentCommand{\eqdim}{O{} m}{
    \IfBlankTF{#1}
    {\mathrel{\overset{\mathcolor{black!70}{d\.=\.#2}}{\scalebox{2.8}[1]{\(=\)}}}}
    {\mathrel{\overset{\mathcolor{black!70}{#2}}{\scalebox{#1}[1]{\(=\)}}}}
}
\NewDocumentCommand{\relphantom}{m}{\mathrel{\phantom{#1}}}
\NewDocumentCommand{\graytimes}{}{\mathcolor{gray}{\times}}

\NewCommandCopy{\olddet}{\det}
\RenewDocumentCommand{\det}{}{\olddet\nolimits}
\DeclareMathOperator{\Det}{Det}

\NewCommandCopy{\transpose}{\intercal}
\DeclareMathOperator{\vol}{vol}

\NewDocumentCommand{\Id}{}{\bm{1}}
\NewDocumentCommand{\const}{}{\text{const.}}

\NewDocumentCommand{\bdry}{}{\partial}

\NewDocumentCommand{\exchange}{O{\quad}}{\xleftrightarrow{#1}}

\NewDocumentCommand{\I}{}{\mathring{\imath}}
\NewDocumentCommand{\E}{s}{\IfBooleanTF{#1}{\mathrm{e}}{\mathinner{\mathrm{e}}}}

\NewDocumentCommand{\bigO} {l m}{\fbraces#1{\lparen}{\rparen}                   {O} {#2}}
\NewDocumentCommand{\biggO}{l m}{\fbraces#1{\lparen}{\rparen}{\raisemath{-0.1ex}{O}}{#2}}

\NewDocumentCommand{\R}{}{\mathbb{R}}

\NewDocumentCommand{\N}{}{\mathbb{N}}

\NewCommandCopy{\leftOrig}{\left}
\NewCommandCopy{\rightOrig}{\right}
\NewCommandCopy{\middleOrig}{\middle}
\RenewDocumentCommand{\left}{}{\mathopen{}\mathclose\bgroup\leftOrig}
\RenewDocumentCommand{\right}{}{\aftergroup\egroup\rightOrig}
\RenewDocumentCommand{\middle}{sm}{\IfBooleanTF{#1}{\.\middleOrig#2\.}{\mathrel{}\middleOrig#2\mathrel{}}}
\NewDocumentCommand{\innermiddle}{m}{\mathinner{}\middleOrig#1\mathinner{}}

\NewDocumentCommand{\Ext}{e{^}}{
    \scalerel*{\bigwedge}{\xmathstrut[-0.1]{0.1}} % scale down the \bigwedge a bit
    \IfValueT{#1}{^{\mspace{-2mu}#1}} % shift possible superscript little bit to the left
}
\NewDocumentCommand{\Odot}{e{^}}{
    \scalerel*{\bigodot}{\xmathstrut[-0.1]{0.1}} % scale down the \bigodot a bit
    \IfValueT{#1}{^{\mspace{-0mu}#1}} % (do not) shift possible superscript little bit to the left
}

\DeclareDocumentCommand{\raisemath}{m}{\mathpalette{\raisemathAux{#1}}}% \raisemath{<len>}{...}
\DeclareDocumentCommand{\raisemathAux}{mmm}{\raisebox{#1}{\(#2#3\)}}
\DeclareDocumentCommand{\scalemath}{m}{\mathpalette{\scalemathAux{#1}}}% \scalemath{<factor>}{...}
\DeclareDocumentCommand{\scalemathAux}{mmm}{\scalebox{#1}{\(#2#3\)}}

\NewDocumentCommand{\sbullet}{O{0.5}}{%
    \mathbin{\ThisStyle{\vcenter{\hbox{\scalebox{#1}{\(\SavedStyle\bullet\)}}}}}
}

\NewDocumentCommand{\idot}{s}{\mathcolor{darkgray}{\IfBooleanTF{#1}{\.\sbullet[0.4]\.}{\sbullet[0.4]}}}

\NewDocumentCommand{\dind}{}{\mathcolor{black!60}{\sbullet[1.2]}}
\NewDocumentCommand{\ind}{}{\mathcolor{lightgray}{\bullet}}

\NewDocumentCommand{\argument}{s O{} O{1}}{%
    \def\squaresize{1.0}%                   % default size
    \IfBooleanT{#1}{\def\squaresize{0.7}}%  % if starred, set size corresponding to scriptstyle
    \IfBlankF{#2}{\def\squaresize{#2}}%     % if optional argument is passed, set size to custom value
    \scalebox{\squaresize}{%
        \begin{tikzpicture}[baseline=-#3*0.6ex]
            \node(char)[draw, shape=rectangle, dash=on 1.2pt off 1.05pt phase 0.5pt, dash expand off,
                inner ysep=2pt, inner xsep=2pt, minimum size=0.6em, rounded corners=2pt] {};
        \end{tikzpicture}%
    }%
}

\tikzset{
    leadsto/.style={-{Stealth[length=0.6em,open,round]},decorate,decoration={snake,amplitude=0.20ex,segment length=0.5em,pre length=0.2em,post length=0.6em}},
    toleads/.style={{Stealth[length=0.6em,open,round]}-,decorate,decoration={snake,amplitude=0.20ex,segment length=0.5em,pre length=0.6em,post length=0.2em}},
    correspondsto/.style={{Stealth[length=0.6em,open,round]}-{Stealth[length=0.6em,open,round]},decorate,decoration={snake,amplitude=0.20ex,segment length=0.5em,pre length=0.7em,post length=0.7em}},
}
\NewDocumentCommand{\longleadsto}{s O{} O{}}{%
    \ensuremath{\mathrel{
            \tikz[baseline=-0.5ex, inner sep=0ex, font=\scriptsize]{
                \node[minimum width=2.15em, inner xsep=0.6em, align=center] (a){\hphantom{#2}\\[0ex]\hphantom{#3}};
                \IfBlankF{#2}{\node[inner sep=0.3ex, above=0.3ex, xshift=-0.1em] at (a){#2};}
                \IfBlankF{#3}{\node[inner sep=0.3ex, below=0.3ex, xshift=-0.1em] at (a){#3};}
                \IfBooleanTF{#1}
                {\draw[line width=0.6pt, toleads] (a.west) -- (a.east);}
                {\draw[line width=0.6pt, leadsto] (a.west) -- (a.east);}
            }
        }}%
}
\NewDocumentCommand{\correspondsto}{O{}O{}}{%
    \ensuremath{\mathrel{
            \tikz[baseline=-0.5ex, inner sep=0ex, font=\scriptsize]{
                \node[minimum width=3.48em, inner xsep=0.8em, align=center] (a){\hphantom{#1}\\[0ex]\hphantom{#2}};
                \IfBlankF{#1}{\node[inner sep=0.3ex, above=0.3ex] at (a){#1};}
                \IfBlankF{#2}{\node[inner sep=0.3ex, below=0.3ex] at (a){#2};}
                \draw[line width=0.6pt, correspondsto] (a.west) -- (a.east);
            }
        }}%
}

\NewDocumentCommand{\bigslant}{O{0.2}O{1.7}mm}{
    \left.\mspace{-1mu}\raisemath{#1em}{#3}
    \mspace{-#2mu} \middleOrig/ \mspace{-\fpeval{#2+1}mu}
    \raisemath{-#1em}{#4} \mspace{-\fpeval{5*#1}mu} \right.
}

\DeclareDocumentCommand{\Id}{}{\mathbb{1}}

\NewDocumentCommand{\dAlembertian}{s}{\mathord{\raisemath{-0.05ex}{\square}}}

\DeclareMathOperator{\OO}{\mathsf{O}} % orthogonal group
\DeclareMathOperator{\SO}{\mathsf{SO}}

\NewDocumentCommand{\Sphere}{}{\ensuremath{\mathbb{S}}}

\NewDocumentCommand{\manifold}{}{\mathcal{M}}

\NewDocumentCommand{\dS}  {}  {\ensuremath{\mathsf{dS}}}
\NewDocumentCommand{\AdS} {}  {\ensuremath{\mathsf{AdS}}}
\NewDocumentCommand{\EAdS}{}  {\ensuremath{\mathsf{EAdS}}}

\NewDocumentCommand{\QFT}{O{}}{\ensuremath{\mathsf{#1QFT}}}
\NewDocumentCommand{\CFT}{O{}}{\ensuremath{\mathsf{#1CFT}}}
\NewDocumentCommand{\AdSCFT}{}{\textsf{AdS/CFT}}

\NewDocumentCommand{\Action}{}{\ensuremath{\mathcal{S}}}
\NewDocumentCommand{\qAction}{}{\ensuremath{\Gamma}}

\NewDocumentCommand{\Lagr}{}{\ensuremath{\mathcal{L}}}

\DeclareDocumentCommand{\correlator}{ l m o }{
    \braces#1{\langle}{\rangle\IfValueT{#3}{_{\mspace{-2mu}#3}}}{#2}
}

\NewDocumentCommand{\OperSymbol}{e{_^}}{ % tweaks default spacing of \Oper subscript 
    \mathcal{O\xmathstrut[-0.1]{-1}}\IfValueT{#1}{_{\mspace{-2mu}#1}}\IfValueT{#2}{^{#2}}
}
\NewDocumentCommand{\Oper}{s O{} e{_^}}{ % generic local quantum operator
    \IfBooleanTF{#1}{\mspace{2mu}\widehat{\mspace{-2mu}\OperSymbol}}{\OperSymbol}%
    \IfBlankF{#2}{^{#2}}\IfValueT{#3}{_{\mspace{-2mu}#3}}\IfValueT{#4}{^{#4}}
}
\NewExpandableDocumentCommand{\Opr}{}{\Oper^{\prime}} % generic local quantum operator (primed)
\NewDocumentCommand{\Ostar}{O{}}{\OperSymbol_{\star}\IfBlankF{#1}{^{(#1)}}} % generic local quantum operator (star subscript)
\NewDocumentCommand{\Oprstar}{O{}}{\OperSymbol^{\prime}_{\star}\IfBlankF{#1}{^{(#1)}}} % generic local quantum operator (prime and star subscript)
\NewDocumentCommand{\Obullet}{O{}}{\OperSymbol_{\bullet}\IfBlankF{#1}{^{(#1)}}} % generic local quantum operator (bullet subscript)
\NewDocumentCommand{\Ophi}{O{}}{\OperSymbol_{\mspace{-1.5mu}\phi^{#1}}} % single-trace operator
\NewDocumentCommand{\Osigma}{O{}}{\widehat{\sigma}}

\NewDocumentCommand{\SpecNoArgs}{O{}}{\operatorname{\mathsf{Spec}}\IfBlankF{#1}{_{#1}}} % spectral function
\NewDocumentCommand{\Spec}{O{s} O{\D} O{\mathcolor{gray}{0}} m}{ % spectral function
    \SpecNoArgs[\IfBlankF{#1}{#1\!}]
    \left\lbrack
    \IfBlankF{#2}{
        \begin{+matrix}[cells={c},rowsep=-0.5pt,colsep=2pt]
        #2 \\ #3
        \end{+matrix}
        \mspace{-2mu} \mathcolor{black!60}{\innermiddle|}
    }
    \smallerWDtrue #4\smallerWDfalse
    \right\rbrack
}
\NewDocumentCommand{\ope}  {l m}{\fbraces#1{\lbrack}{\rbrack}{\operatorname{\mathsf{ope}}}{#2}} % operator product expansion coefficient
\NewDocumentCommand{\opesq}{l m}{\fbraces#1{\lbrack}{\rbrack}{\operatorname{\mathsf{ope^{2}}}}{#2}} % operator product expansion coefficient squared

\NewDocumentCommand{\Bubble}{s}{\opbraces{\widetilde{B}\IfBooleanT{#1}{^{\raisemath{0.4ex}{\.\prime\!}}}}}

\NewDocumentCommand{\G}{e{^} O{}}{%
    \IfBlankTF{#2}%
    {\opbraces{\mathop{\Gamma\xmathstrut{-0.1}}\IfValueT{#1}{^{#1}}}}%
    {\mathop{\Gamma\xmathstrut{-0.1}}\nolimits\IfValueT{#1}{^{#1}}_{\!#2}}%
}
\RenewDocumentCommand{\digamma}{e{^} O{}}{%
    \IfBlankTF{#2}%
    {\opbraces{\psi\IfValueT{#1}{^{#1}}}}%
    {\psi\IfValueT{#1}{^{#1}}_{\mspace{-2mu}#2}}%
}

\NewDocumentCommand{\widetildesmashAux}{O{0ex}m}{ % ignore extra height of \widetilde and shift it down
    \mathrlap{\smash{\raisemath{#1}{\widetilde{\phantom{#2}}}}}#2
}
\NewDocumentCommand{\widetildesmash}{O{}m}{
    \IfBlankTF{#1}{%
        \mathchoice%
        {\widetildesmashAux[0ex]{#2}}%
        {\widetildesmashAux[-0.04ex]{#2}}%
        {\widetildesmashAux[-0.07ex]{#2}}%
        {\widetildesmashAux[-0.15ex]{#2}}%
    }{%
        \widetildesmashAux[#1]{#2}%
    }%
}
\NewDocumentCommand{\D}   {s O{}}{\IfBooleanTF{#1}{\widetildesmash{\Delta}}{\Delta}\IfBlankF{#2}{^{\!#2}}}
\NewDocumentCommand{\Dpr} {s}{\IfBooleanTF{#1}{\widetildesmash{\Delta}}{\Delta}^{\mspace{-2mu}\prime}}
\NewDocumentCommand{\Dstar}{s O{}}{\IfBooleanTF{#1}{\widetildesmash{\Delta}}{\Delta}_{\star}\IfBlankF{#2}{^{\!#2}}}
\NewDocumentCommand{\Dbullet}{s O{}}{\IfBooleanTF{#1}{\widetildesmash{\Delta}}{\Delta}_{\bullet}\IfBlankF{#2}{^{\!#2}}}
\NewDocumentCommand{\Dprstar}{s}{\IfBooleanTF{#1}{\widetildesmash{\Delta}}{\Delta}^{\mspace{-2mu}\prime}_{\star}}
\NewDocumentCommand{\Dphi}{s}{\IfBooleanTF{#1}{\widetildesmash{\Delta}}{\Delta}_{\phi}}

\NewDocumentCommand{\twistmin}{}{\tau_{\text{min}}}
\NewDocumentCommand{\Dmin}{}{\D_{\text{min}}}
\NewDocumentCommand{\Jmin}{}{J_{\text{min}}}

\NewDocumentCommand{\MFT}{s}{\ensuremath{\IfBooleanTF{#1}{\text{MFT}}{(\text{MFT})}}}
\NewDocumentCommand{\OPEcoef}{O{}}{f\IfBlankF{#1}{^{(#1)}}}
\NewDocumentCommand{\OPEfunc}{O{}}{C\IfBlankF{#1}{^{(#1)}}}
\NewDocumentCommand{\OPEsq}{O{}}{C\IfBlankF{#1}{^{(#1)}}}
\NewDocumentCommand{\anomdim}{O{1}}{\gamma\IfBlankF{#1}{^{(#1)}}}

\NewDocumentCommand{\Cnorm}{O{\smash{\Dphi}}}{\mathfrak{C}_{\mspace{-1mu}#1}}
\NewDocumentCommand{\Gbulkbulk}{O{\Dphi}}{G^{bb}_{\mspace{-2mu}#1}}
\NewDocumentCommand{\Gbulkbdry}{O{\Dphi}}{G^{b\partial}_{\mspace{-2mu}#1}}
\NewDocumentCommand{\Gbdrybdry}{O{\Dphi}}{G^{\partial\partial}_{\mspace{-2mu}#1}}

\NewDocumentCommand{\HypGeoNoArgs}{s O{2} O{1}}{
    \mathop{{}_{#2}\bm{\IfBooleanTF{#1}{\widetilde{\mathsf{F}}}{\mathsf{F}}}_{#3}}
}
\NewDocumentCommand{\HypGeo}{s O{2} O{1} m m O{1}}{
    \IfBooleanTF{#1}{\HypGeoNoArgs*[#2][#3]}{\HypGeoNoArgs[#2][#3]}
    \left\lbrack
    \begin{+matrix}[cells={c},rowsep=0.5pt,colsep=2pt]
    #4 \\ #5 \\
    \end{+matrix}
    \mathcolor{black!60}{\innermiddle|} #6
    \mspace{2mu}\right\rbrack
}
\NewDocumentCommand{\anomdimFfunction}{O{d} m m}{
    \operatorname{\mathsf{F}}^{\smash{\raisemath{0.4ex}{\mathcolor{darkgray}{(#1)}}}}_{\!#2,\.#3}
}

\NewDocumentCommand{\Repre}    {s}{\ensuremath{\mathtt{\IfBooleanTF{#1}{R }{(R) }}}}
\NewDocumentCommand{\Singlet}  {s}{\ensuremath{\mathtt{\IfBooleanTF{#1}{S }{(S) }}}}
\NewDocumentCommand{\AntiSym}  {s}{\ensuremath{\mathtt{\IfBooleanTF{#1}{AS}{(AS)}}}}
\NewDocumentCommand{\SymTrless}{s}{\ensuremath{\mathtt{\IfBooleanTF{#1}{ST}{(ST)}}}}
\NewDocumentCommand{\NonSinglet}{s}{\SymTrless/\AntiSym}
\NewDocumentCommand{\ARepre}{O{\Repre}}{\mathcal{A}_{#1}}
\NewDocumentCommand{\ASinglet}  {}{\ARepre[\Singlet]}
\NewDocumentCommand{\AAntiSym}  {}{\ARepre[\AntiSym]}
\NewDocumentCommand{\ASymTrless}{}{\ARepre[\SymTrless]}

\NewDocumentCommand{\FourPtNonSinglet}{O{1}}{\FourPt[\(\substack{\SymTrless*\\\AntiSym*}\)][#1]}

\NewDocumentCommand{\Proj}{O{} o}{%
    \IfValueTF{#2}%
    {\fbraces{\lparen}{\rparen}{\mathcal{P}_{#1}}{#2}}%
    {\mathcal{P}_{#1}}%
}
\NewDocumentCommand{\PRepre}     {}{\Proj[\Repre]}
\NewDocumentCommand{\PSinglet}   {}{\Proj[\Singlet]}
\NewDocumentCommand{\PAntiSym}   {}{\Proj[\AntiSym]}
\NewDocumentCommand{\PSymTrless} {}{\Proj[\SymTrless]}

\NewDocumentCommand{\Knorm}{O{\D*,J}}{\operatorname{\mathnormal{K}}_{\xmathstrut[-1]{0.33}\mspace{-4mu}#1}}
\NewDocumentCommand{\Snorm}{O{\D*,J}}{\operatorname{\mathnormal{S}}_{\xmathstrut[-1]{0.33}\mspace{-2mu}#1}}
\NewDocumentCommand{\ConfBlock}{s O{\D,J} O{s}}{
    \operatorname{\mathnormal{G}}
    \IfBooleanTF{#1}
    {_{#2}^{(#3)}}
    {_{\xmathstrut[-1]{0.33}\mspace{-2mu}#2}^{\raisemath{0.5ex}{(#3)}}}
}
\NewDocumentCommand{\CPWnorm}{O{\D,J}}{n_{#1}}
\NewDocumentCommand{\sixjsymbol}{s}{%
    \IfBooleanTF{#1}
    {\ensuremath{\mathsf{6j}\text{--Symbol}}}
    {\ensuremath{\mathsf{6j}\text{--symbol}}}%
}
\NewDocumentCommand{\CrK}{O{\D,J}O{\Dpr,J'}O{20mu}}{\operatorname{\mathsf{CrK}}\IfBlankTF{#1}{^{s\leftarrow t}}{^{\mspace{#3}s\,\longleftarrow\,t}_{\bra{#1}\ket{#2}}}}
\makeatother
\title{\paperTitle}
\author[a]{Jonáš Dujava}\emailAdd{jonas.dujava@gmail.com}
\author[b]{and Petr Vaško}\emailAdd{petr.vasko@matfyz.cuni.cz}

\affiliation[a]{Institute of Theoretical Physics,          Charles University,\\V Holešovičkách 2, 180\,00 Prague, Czech Republic}
\affiliation[b]{Institute of Particle and Nuclear Physics, Charles University,\\V Holešovičkách 2, 180\,00 Prague, Czech Republic}

\abstract{%
    We determine the scaling dimensions in the boundary \(\CFT_{d}\) corresponding to the \(\OO(N)\) model in \(\EAdS_{d+1}\).
    The \CFT{} data accessible to the 4-point boundary correlator of fundamental fields are extracted in \(d=2\) and \(d=4\), at a finite coupling, and to the leading nontrivial order in the \(1/N\) expansion.
    We focus on the non-singlet sectors, namely the anti-symmetric and symmetric traceless irreducible representations of the \(\OO(N)\) group, extending the previous results that considered only the singlet sector.
    Studying the non-singlet sector requires an understanding of the crossed-channel diagram contributions to the \(s\)-channel conformal block decomposition.
    Building upon an existing computation, we present general formulas in \(d=2\) and \(d=4\) for the contribution of a \(t\)-channel conformal block to the anomalous dimensions of \(s\)-channel double-twist operators, derived for external scalar operators with equal scaling dimensions.
    Up to some technical details, this eventually leads to the complete picture of \(1/N\) corrections to the \CFT{} data in the interacting theory.
}


\arxivnumber{2503.16345}

\begin{document}
\pdfbookmark[2]{Title Page}{titlepage} % add PDF outline entry
\maketitle
\flushbottom

\section{Introduction}

Recently, various \QFT{} models have been studied in \AdS{} (or \dS{}) spacetime at finite coupling.
Rather than the usual weak coupling expansion, the large \(N\) expansion together with \AdSCFT{} intuition was employed as an alternative handle to perform explicit calculations.
This movement started with \(\OO(N)\) and Gross--Neveu Model \cite{Carmi:2018qzm}, and was continued also for scalar QED \cite{Ankur:2023lum}.
% \todo[Maybe Mention]{While formulas appearing in \AdS{} are generally (much) more complicated than in flat space, there are also certain advantages --- possibility to use ideas of holography/\CFT{} techniques, propagators/correlators have simpler pole structure (usually no branch cuts), ...}

While these studies have addressed key questions such as the phase structure, exact propagators, and the boundary correlators (4-point functions) in the large \(N\) limit, they have primarily focused on the \emph{singlet spectrum} of the \CFT{} on the boundary.

Largely unexplored is the \emph{non-singlet spectrum} corresponding to operators transforming in nontrivial representations of the global internal (large \(N\)) symmetry group.
We will thus primarily focus on extending the analysis of the \(\OO(N)\) model in \AdS{} at finite coupling also to the non-singlet sector --- namely the \emph{rank-2 anti-symmetric} and \emph{symmetric traceless} representations of \(\OO(N)\) group.

Building on the already established results for the singlet sector, we were able to extract anomalous dimensions of operators in the non-singlet sector by utilizing the \sixjsymbol{}s (also known as the \emph{crossing kernel}) \cite{Liu:2018jhs}.
It appears that these methods (in particular the Lorentzian inversion formula) missed some contributions to anomalous dimensions of scalar (spin zero) operators.
Further discussion of this subtlety can be found in \Cref{sub:Dependence of anomalous dimensions on the coupling}.

More broadly, there are essentially three main pieces of motivation in this line of research.
They were nicely summarized in the Introduction sections of~\cite{Carmi:2018qzm,Ankur:2023lum}, which the reader is invited to check out for more details.
Even after receiving some attention, numerous issues require deeper investigation:
\begin{itemize}
    \item Free theory in \AdS{} bulk corresponds to GFF/MFT at the asymptotic boundary, that is a \CFT{} with all correlation functions simply given just by products of 2-point functions.
          While a great deal has been said about small deformations in a bulk coupling, less explored are large deformations of the \CFT{} data away from MFT.
          Large \(N\) techniques are helpful in this respect, since they retain more of the nonlinear structure of the exact theory compared to the ordinary lowest-order perturbation theory.
          As such, they can shed some light on interesting phenomena like appearance of new operators in the spectrum (corresponding to the existence of bound states), \(\AdS{}\) analogs of resonances, or level crossing.
    \item For a \QFT{} in \(\AdS_{d+1}\) with a nontrivial phase structure, it is interesting to determine how \(\CFT_{d}\) data of the dual theory at the boundary change, in particular as borderlines separating different phases of the bulk theory are crossed (for instance by deforming the critical bulk theory at a given fixed point of the renormalization group by a relevant operator, \(S_{\mathrm{CFT}}\mapsto S_{\mathrm{CFT}}+g\int\mspace{-2mu} \Oper^{(\mathrm{bulk)}}_{\Delta}\), triggering an RG-flow to the new desired phase).
          Bulk couplings \(g\) are generically dimensionful, as exactly marginal couplings corresponding to families of bulk \CFT{}s are very rare.
          Since \AdS{} provides a natural length scale, its radius \(\ell\), it is convenient to form dimensionless combinations \(\widehat{g}=g\.\ell^{\Delta-(d+1)}\).
          Thus the bulk RG-flow of \(g\) can be thought of as a flow in the \AdS{} radius \(\ell\).

          The framework for studying it, with a particular emphasis on the flat space limit \(\ell\to\infty\), was laid out in~\cite{Paulos:2016fap}.
          More detailed studies about constraining the bulk RG-flow by boundary \CFT{} axioms followed.
          By now there is already a significant amount of literature, for an invitation to this topic with further references included, see~\cite{Hogervorst:2021spa,Antunes:2021abs,Antunes:2024hrt}.
          The role of certain 3-point correlation functions (\AdS{} form factors) of two boundary operators with one bulk field in the change of the \AdS{} length scale \(\ell\) was investigated in~\cite{Levine:2023ywq}.
          The important special case of the bulk operator being the stress tensor was done in~\cite{Meineri:2023mps}.

          A distinct instance of the above discussion occurs when the flow is initiated in a free conformal bulk theory (for example a conformally coupled scalar) corresponding to a MFT at the boundary, which ubiquitously possesses higher spin currents (all the double-twist families saturate unitarity bound and form thus short conformal multiplets).
          Then one can naturally wonder about their fate once interactions are turned on.
          General theorems constraining their existence in arbitrary interacting \CFT{}s were given in~\cite{Maldacena:2011jn,Maldacena:2012sf}.
          On the other hand, one can ask a similar question, just not at the boundary but instead in the bulk.
          A complete breaking of bulk higher spin currents by interactions in \(\AdS{}_{2}\) was analyzed in~\cite{Antunes:2025iaw}.

    \item In the special case when the bulk theory in \(\AdS_{d+1}\) is critical (becoming a \(\CFT_{d+1}\)), one can perform (in the Euclidean signature) a Weyl symmetry transformation from \EAdS{} to flat half-space \(\R^{d} \times \R_{\geq}\).
          This is easily done using the Poincaré coordinates, which cover \EAdS{} globally.
          The \CFT{} with a boundary --- \(\CFT[B]_{d+1}\) --- on this flat half-space can thus be studied using \AdS{} methods, whose recent advances have made such approach rather efficient.

          Moreover, one can generalize it from boundaries (\CFT[B]) to defects (\CFT[D]).
          A Weyl equivalence between an \((n-1)\)-dimensional defect in \(\R^{n+m}\) and \(\AdS_{n}\times \Sphere^{m}\) allowed~\cite{Cuomo:2021kfm} to use results of~\cite{Carmi:2018qzm} for the singlet spectrum to extract interesting \CFT[D] data.
          More concretely, they studied the critical \(\OO(N)\) model in the presence of a localized magnetic external field understood as an \(1\)-dimensional defect, thus \(\AdS_{2}\) was relevant.
          Such data is important to understand phase transitions of real-world systems.
    \item Finally, there are attempts to replace the LSZ axioms for the flat-space \(S\)-matrix by the flat-space limit of asymptotic boundary observables in \AdS{}~\cite{vanRees:2022zmr,Polchinski:1999ry,Gary:2009mi,Hijano:2019qmi}.
          Asymptotic boundary correlators for massive \QFT{}s in \(\AdS_{d+1}\) obey \(\CFT_{d}\) axioms --- except for the existence of a stress tensor --- that are mathematically more rigorous than the current ones for the \(S\)-matrix in flat space.
          These asymptotic correlators are by definition holographic as is the flat-space \(S\)-matrix, and reduce to it in the flat-space limit (in a sense that still requires a more rigorous definition).

          Since the curvature of \AdS{} also acts as an IR regulator, it could potentially cure the possible IR divergences, which make the flat \(S\)-matrix an ill-defined object \cite{Duary:2022pyv,Duary:2022afn}.
          In practice, one often has to recourse to IR safe inclusive observables, which however require taking the experimental setup into account, and are thus not clean theoretical observables.
          More discussion about the IR issues of the flat \(S\)-matrix and another attempt how to solve it can be found in~\cite{Strominger:2017zoo}.
\end{itemize}

Our work focuses on a specific topic in the above vast landscape.
We examined large deformations of the free MFT spectrum for the \(\OO(N)\) model in the unbroken phase.
While primary outcome is the computation and analysis of the non-singlet spectrum, along the way we also obtained some additional results.

First, we completed the singlet \CFT{} data by extracting the OPE coefficients.
Furthermore, we discussed extension of the spectrum (even the singlet one) to \(d=4\), where the necessary regularization complicates the picture.
To partly deal with the associated renormalization scheme ambiguity --- finite part of a counterterm --- we resorted to the critical bulk theory describing an ordinary phase transition.

Formally extending the critical point beyond its upper critical dimension --- also considered in~\cite[Section 3.1 and Figure 6]{Giombi:2020rmc} --- we discovered an intriguing pattern in the singlet spectrum.
It is related with the appearance of emergent operators at a strong enough coupling, which are not connected with the usual MFT-type spectrum found at weak coupling.
Based on this analysis, a formula for the critical singlet spectral function in all even (boundary) dimensions reproducing the desired critical spectrum was proposed.

The next natural step would be to deal with the broken phase and especially with the critical theory in the bulk, whose analysis could result in extraction of non-singlet \CFT[D] data along the lines of~\cite{Cuomo:2021kfm}.
Hopefully, our results can add a concrete small piece to the above complex mosaic.
% \Note[Broken Phase]{This just boils down to reading poles of slightly different function (due to modified exact propagator of \(\sigma\)), right?}
% \Note[Conformal Phase]{Just substitute \(\Dphi=1, \lambda=\infty\) --- taking residues we can also extract OPE coefficients, as was done in \cite{Carmi:2018qzm}.}

\paragraph{Outline of the paper.}

The subsequent sections are organized as follows.
% Projection of the \(4\)-point boundary correlator to non-singlet sectors is given by appropriate combinations of \(t\)-channel and \(u\)-channel exchange Witten diagrams.
% Therefore, the main task of extracting the non-singlet anomalous dimensions then consists of performing the \(s\)-channel \emph{conformal block decomposition} of the \(t\)- and \(u\)-channel exchange diagrams, which is achieved using the \sixjsymbol{}s from the known \(t\)-channel decomposition of the \(t\)-channel exchange diagram (which is given by resolving the singlet spectrum).

In \Cref{sec:review-ONmodelinAdS} we define the \(\OO(N)\) model, mainly focusing on its formulation suitable for a systematic large \(N\) expansion.
Then we introduce the main observable --- the \(4\)-point boundary correlator of fundamental fields \(\phi^{\ind}\) in \AdS{}.
The rest of the section is devoted to fixing our conventions, particularly those for OPE channels, and some comments about renormalization scheme are made as well.

In \Cref{sec:CFTgeneralities} we will review relevant generalities of \CFT{}s, focusing on the extraction of the \CFT{} data from the \(4\)-point correlator.
In particular, we will present general formulas (in \(d=2\) and \(d=4\)) for the contribution of a single \(t\)-channel conformal block to anomalous dimensions of \(s\)-channel double-twist operators, applicable for arbitrary twists and spins.

These formulas will be utilized in \Cref{sub:non-singlet_spectrum} to calculate the leading \(1/N\) contributions to the non-singlet scaling dimensions of the \(\OO(N)\) model in \AdS{} at finite coupling.
This computation requires the singlet spectrum as an input, whose properties are summarized in the previous parts of~\Cref{sec:spectrum-ONinAdS}, together with the decomposition of the \(4\)-point correlator into \(\OO(N)\) irreducible representations.
Criticality in the bulk is also briefly discussed in \Cref{sub:Criticality in the bulk}, which suggests a possible extension beyond its upper critical dimension.

In \Cref{sec:analysis_of_the_non-singlet_sector} we present the results for the non-singlet spectrum, mostly in the form of various plots.
The main one is the twist--spin plot showing a characteristic organization of the spectrum into Regge trajectories.
Various limits are studied, specifically a cross-check regarding the large spin asymptotics.
In \Cref{sub:Dependence of anomalous dimensions on the coupling}, an issue with spin zero symmetric traceless operators is discussed, regarding missed contributions to their anomalous dimensions due to limitations of computational methods used.

An executive summary and synthesis of results is given in~\Cref{sec:summary}, together with possible future directions.

Various detailed computations and the implementation of main formulas can be found in the accompanying \notebook{} (links to a \textsf{GitHub} repository \githubrepository{}).

\section{Review of the \texorpdfstring{\(\OO(N)\)}{O(N)} model in \texorpdfstring{\AdS{}}{AdS}}%
\label{sec:review-ONmodelinAdS}

We start by introducing the \(\OO(N)\) model in \Cref{sub:Generalities of ON model}.
Some general features are reviewed, in particular how its large \(N\) expansion enables calculations at finite coupling.
In \Cref{sub:CFT on the boundary of AdS} we define our main observable --- the boundary \(4\)-point correlator in \(\AdS\) --- which can be viewed as an observable in the \CFT{} living on the boundary of \AdS{}.
% It can be expressed in a form particularly suitable for the study of the \CFT{} spectrum by utilizing the spectral representation and related techniques, which we briefly describe in \Cref{sub:Utilizing Spectral representation}.
In the following \Cref{sub:Utilizing Spectral representation} we briefly describe how the spectral representation can be utilized to express the boundary correlator in a form particularly suitable for the study of the \CFT{} spectrum.

While most of the material in this section is well-known for experts in the field, with some more specific details thoroughly discussed already in~\cite{Carmi:2018qzm}, we believe that an alternative account (stressing some additional points) can be beneficial.
In the meanwhile, we also fix the notation and conventions that will be used throughout the paper.

\subsection{Generalities of the \texorpdfstring{\(\OO(N)\)}{O(N)} model}%
\label{sub:Generalities of ON model}

The Euclidean action of the \(\OO(N)\) model on a \(\grayenclose{d+1}\)--dimensional Riemannian manifold \(\manifold\) takes the form \textcolor{darkgray}{(we implicitly contract both spacetime and internal indices)}
\begin{align}
    \label{eq:Action_ON}
    \Action[\phi^{\ind}] = \int_{\manifold} \dd[d+1]{x} \sqrt{g}\, \left[
        {\frac{1}{2}} (\partial \phi^{\ind})^{2}
        + {\frac{1}{2}}\. m^{2} (\phi^{\ind})^{2}
        + {\frac{\lambda}{2N}}\.\left((\phi^{\ind})^{2}\right)^{\!2\.}
    \right]
    \eqcomma
\end{align}
where \(\phi^{\ind}=(\phi^{1},\ldots,\phi^{N})\) is an \(N\)-tuple of real scalar fields transforming in the vector representation of the global internal \(\OO(N)\) symmetry group.
In the case of maximally symmetric spacetimes --- in our case \EAdS{} --- the possible coupling to the curvature can be absorbed into the mass term \(m^{2}\).
Results for Lorentzian \AdS{} can be obtained by analytical continuation, so we will use \EAdS{}/\AdS{} interchangeably.

The interaction term was introduced in such a way that the model admits a large \(N\) expansion with the \(\lambda\) coupling fixed and finite.
We will review details relevant to us.
It was also outlined in~\cite[Section 2]{Carmi:2018qzm}, and more thorough accounts in flat space can be found in a specialized review~\cite{Moshe:2003xn} or in the original paper~\cite{Coleman:1974jh}.

Feynman rules for \eqref{eq:Action_ON} assign a propagator to each internal line, and the interaction vertex can be diagrammatically represented as \textcolor{darkgray}{(up to numerical factors)}
\begin{equation}
    \label{eq:ON_vertex}
    \vertexPhi \sim \frac{\lambda}{N} \left( \vertexPhiChannelS + \vertexPhiChannelT + \vertexPhiChannelU \right)
    \eqcomma
\end{equation}
where the connected lines on the right-hand side indicate Kronecker deltas in the indices of the corresponding \(\phi\) fields.
As usual, all internal vertices throughout the paper are to be integrated over \(\manifold\) with the geometric measure including the \(\sqrt{g}\) density.

Order of \(1/N\) in a given diagram is not simply given just by the number of interaction vertices, since after their expansion \eqref{eq:ON_vertex} a number of closed index loops can form.
Each such index loop carries an additional factor of \(N\), since all \(\{\phi^{i}\}_{i=1}^{N}\) circulating in the loop contribute equally.

\paragraph{Hubbard--Stratonovich transformation.}%
\label{par:Hubbard--Stratonovich transformation.}

For the purposes of large \(N\) expansion, it turns out more efficient to proceed via a reparametrization of the interaction term under the path integral by introducing an auxiliary Hubbard--Stratonovich field \(\sigma\) --- equivalent on-shell (up to normalization) to the composite operator \((\phi^{\ind})^{2}\) --- thus obtaining the action
\begin{equation}
    \label{eq:ON_HS_Action}
    \Action_{\text{HS}}[\phi^{\ind}\!,\sigma] = \int_{\manifold} \dd[d+1]{x} \sqrt{g}\, \left[
        {\frac{1}{2}}(\partial\phi^{\ind})^{2}
        + {\frac{1}{2}}\. m^{2} (\phi^{\ind})^{2}
        - \frac{1}{2\lambda}\sigma^{2}
        + \frac{1}{\sqrt{N}}\sigma(\phi^{\ind})^{2}
    \right]
    \eqend
\end{equation}
The interaction vertex \eqref{eq:ON_vertex} is now substituted by the following rules
\begin{equation}
    \label{eq:ON_HS_rules}
    \SigmaFreePropag \equiv -\lambda \Id \eqcomma \qquad
    \vertexSigma* \equiv \frac{2}{\sqrt{N}}\delta^{ij} \eqcomma
\end{equation}
first representing the free \(\sigma\)-propagator, and second the \(\sigma\phi^{2}\) vertex.
We suppressed the position dependence of \(\SigmaFreePropag\equiv \correlator{\sigma\sigma}[\text{free}]\) by writing it --- up to factor \((-\lambda)\) --- as an operator \(\Id\) acting on functions through the convolution with the kernel \(\Id(x,y)\equiv \delta(x,y)\).
Here we use the \(\delta\)-distribution normalized with respect to the geometric measure, so the \(\Id\) operator is really acting on functions as an identity.

\paragraph{Effective action.}%
\label{par:Effective action}

Now, we want to \enquote{integrate out the loops} of the \(\phi\) fields, which contribute at the leading order in the large \(N\) expansion.
Since our new Lagrangian \(\Lagr_{\text{HS}}\) is only quadratic in \(\phi\), the \(\phi\)-loop subdiagrams have the form of a loop with arbitrary number of \(\sigma\) lines attached to it, thus inducing non-local contributions to the 1PI effective interaction vertices between \(\sigma\) fields of the form
\begin{equation} \label{eq:phiLoop}
    \begin{aligned}
        \qAction^{\text{1-loop}}[\sigma]
         & = \sum_{n=0}^{\infty} \phiLoop
        = \const - N \sum_{n=1}^{\infty} \frac{(-1)^{n}}{2n} \Tr \left[ \left( \frac{1}{(-\dAlembertian + m^{2})\.\Id}\circ\frac{2}{\sqrt{N}}\sigma \right)^{\!n\,} \right] \\[-0.2ex]
         & = \frac{N}{2} \Tr \ln \left( (-\dAlembertian + m^{2})\.\Id + \frac{2}{\sqrt{N}}\sigma \right)
        = - \ln \Det^{-\frac{N}{2}}\left( (-\dAlembertian + m^{2})\.\Id + \frac{2}{\sqrt{N}}\sigma \right)
        \eqend \mspace{-10mu}
    \end{aligned}
\end{equation}
The prefactor of \(N\) comes from the already performed trace over the indices in the closed loop.
Also other combinatorial/symmetry factors are properly accounted for.
This can be alternatively seen as a functional determinant coming from Gaussian integration over \(\phi\) rewritten as contribution to the Euclidean effective action.

Note, that in \eqref{eq:phiLoop} we understand the argument of \(\Tr \ln (\argument)\) as an operator acting on functions, which we can represent as a convolution with an associated kernel.
In particular, the kernel corresponding to \((-\dAlembertian+m^{2})\Id\) is \([(-\dAlembertian+m^{2})\Id](x,y)\equiv (-\dAlembertian_{x}+m^{2})\delta(x,y)\), and the action of pointwise multiplication with \(\sigma\) has the kernel \(\sigma(x,y)\equiv \sigma(x)\delta(x,y)\).

We can now introduce a \grayenclose{Euclidean} effective action
\begin{equation}
    \label{eq:effectiveAction}
    \begin{aligned}
        \qAction[\phi^{\ind}\!,\sigma]
         & \equiv \qAction^{\text{0-loop}}[\phi^{\ind}\!,\sigma] + \qAction^{\text{1-loop}}[\sigma]
        \equiv \Action_{\text{HS}}[\phi^{\ind}\!,\sigma] + \eqref{eq:phiLoop} \\
         & = \int_{\manifold} \dd[d+1]{x} \sqrt{g}\, \left[
            {\frac{1}{2}}(\partial\phi^{\ind})^{2}
            + {\frac{1}{2}}\. m^{2} (\phi^{\ind})^{2}
            - \frac{1}{2\lambda}\sigma^{2}
            + \frac{1}{\sqrt{N}}\sigma(\phi^{\ind})^{2}
        \right] \\
         & \mspace{270mu} + \frac{N}{2} \Tr \ln \left( (-\dAlembertian + m^{2})\.\Id + \frac{2}{\sqrt{N}}\sigma \right)
        \eqcomma
    \end{aligned}
\end{equation}
which is indeed the effective 1PI action including all of the leading \(1/N\) contributions.
Proceeding further, we should first determine the ground state of the theory by extremizing (preferably minimizing) the effective potential, given by the effective action \(\qAction\) evaluated for constant classical fields and divided by the volume of the spacetime which factorizes.
Afterward, one expands the effective action in terms of the shifted fields \(\delta\phi\) and \(\delta\sigma\) which vanish at the found extremum.

The coefficients of this expansion are the exact propagators (quadratic terms) and 1PI vertices (higher order terms) in the leading \(1/N\) approximation.
The linear \enquote{tadpole} terms in \(\delta\phi\) and \(\delta\sigma\) are absent by virtue of expanding around the extremum of the effective potential.
Observables can be then calculated using these leading forms of exact propagators and 1PI vertices, but all diagrams which contain \(\phi\)-loops should be omitted, since those are already included in the effective \(\sigma\) self-interactions.

Alternatively, this can be viewed as a large \(N\) saddle point analysis, since all terms in the effective action are of the same order of magnitude in the neighborhood of the effective potential extremum --- they are all of order \(\bigO{N}\).

\paragraph{Phase structure.}
\label{par:phase_structure}

Phases of the \(\OO(N)\) model have been thoroughly investigated on flat space of dimension \(D\equiv d+1\).
The dynamics of the theory significantly differs for the ranges \(2<D<4\) and \(D\geq 4\).
Since we will do computations for \(d=2 \Leftrightarrow D=3\) (\(\AdS{}_{3}\)) and \(d=4 \Leftrightarrow D=5\) (\(\AdS{}_{5}\)), we need to briefly summarize both cases, in order to clearly specify which phase we treat in this work.

In the first case \(2<d+1<4\), the flat-space theory is asymptotically free.
The free unstable UV fixed point \(\CFT_{\text{UV}}\) has two relevant operators (the mass term and the interaction term) that trigger an RG flow.
When appropriately tuned, it ends in a semi-stable interacting (the strength increases as \(D\) decreases from \(4\)) IR fixed point --- the Wilson--Fisher \(\CFT_{\text{IR}}\)~\cite{Wilson:1971dc} --- with one relevant operator corresponding to the mass term.
The RG flow triggered by it, depending on the sign of the deformation, leads either to a trivially gapped phase with unbroken global \(\OO(N)\) symmetry or to spontaneous symmetry breaking \(\OO(N)\to\OO(N-1)\), whose low energy dynamics is governed by a non-linear sigma model of \(\grayenclose{N-1}\) Goldstone bosons (with a free fixed point \(\CFT_{\text{IR}}\) in the deep IR).

In the second case \(d+1\geq 4\), the flat-space theory is IR free.
For large enough \(N\), a perturbatively unitary UV fixed point was found in~\cite{Fei:2014yja}.
The authors proposed a UV completion by a cubic theory in \(D=6\) with an IR fixed point, whose equivalence to the UV fixed point of the \(\OO(N)\) model was checked using \(\epsilon\)-expansions.
Later, it was shown~\cite{Giombi:2019upv} that this fixed point is non-unitary beyond perturbation theory.
In particular, scaling dimensions receive exponentially suppressed (in large \(N\)) imaginary parts caused by instantons existing in either formulation of the fixed point.
This phenomenon is known as complex \CFT{}~\cite{Gorbenko:2018ncu}.
The gapped phase as well as the spontaneously broken phase exist as in the previous case.

The phase structure of \(\OO(N)\) model in \(\AdS_{d+1}\) for dimensions ranging in \(2<d+1<4\) was analyzed in~\cite[Section 3]{Carmi:2018qzm}, and differences from flat space were summarized there.
For example, in \(d=2\) (\(\AdS_{3}\)), there exists a region of the parameter space where both the unbroken and broken phase seem to coexist.
% \Note{This seems to be the case for \(\lambda\) large in \(d=2\), and broken phase seem to be energetically favored, but due to \(\AdS_{3}\) background both may play a role (unlike in the flat space).}
In this work, we treat almost exclusively the simplest of the phases --- the unbroken phase with the \(\OO(N)\) symmetry preserved.
We leave the broken phase and deeper analysis of the critical point for future work.

Moreover, the final results will be mainly presented for \(d=2\) (\(\AdS_{3}\)), where formulas simplify enough to actually perform the calculations.
Nevertheless, at a certain point a numerical evaluation is necessary.
In addition \(d=4\) (\(\AdS_{5}\)) will be also included, which however still requires an independent phase structure analysis.

Choice of the unbroken \(\OO(N)\)--symmetric phase specializes to the vicinity of the saddle point \(\left(\phi^{\ind}(x)=0,\,\sigma(x)=\sqrt{N}\sigma_{\star}\right)\), where \(\sigma_{\star}\) is a constant parametrizing the vacuum expectation value (VEV) of the \(\sigma\) field as \(\correlator{\sigma}\equiv \sqrt{N}\sigma_{\star}\).
The expansion of the effective action \(\qAction\) \eqref{eq:effectiveAction} around this saddle point just gets rid of the \(\sigma\)-tadpole term, and the \(\phi\)-field mass-squared is shifted to \(m_{\phi}^{2} \equiv m^{2}+2\sigma_{\star}\).
Since everything else stays the same, from now on we take all \enquote{free} \(\phi\)-field propagators with the effective mass-squared \(m_{\phi}^{2}\), and use \(\sigma\) to refer mostly just to the deviation \(\delta\sigma\equiv \sigma-\correlator{\sigma}\).

Finally, let us remark that the unbroken phase is present when the (effective) mass is above the Breitenlohner--Freedman (BF) bound \(m_{\phi}^{2}>-\frac{d^{2}}{4}\) \cite{Breitenlohner:1982jf,Klebanov:1999tb}, a point which will be also discussed in \Cref{sub:CFT on the boundary of AdS}.

\paragraph{Exact \(\sigma\)-propagator.}%
\label{par:Exact sigma-propagator}

Since the interaction terms in the expansion of \(\qAction\) are of the order \(\bigO*{1/\sqrt{N}}\) or higher, the leading \(\bigO{1}\) form of the exact \(\sigma\)-propagator can be obtained by inverting the kernel of the quadratic \(\sigma\)-part of \(\qAction\) expanded around the saddle point, or equivalently by summing up the geometric series
% \Note{at further subleading orders there are present corrections from effective \(\sigma\) self-interactions}
\begin{equation}
    \label{eq:sigma_exact_propagator}
    \begin{aligned}
        \SigmaPropag
         & = \SigmaFreePropag + \SigmaFreeBubble + \SigmaFreeBubbleTwice + \cdots \\
         & = (-\lambda \Id) + (-\lambda \Id)\circ 2B \circ(-\lambda \Id) + \dots
        = -\lambda\sum_{n=0}^{\infty} \left(- 2\lambda B \right)^{n}
        = -\left[ \frac{\Id}{\lambda} + 2B \right]^{-1}
        \eqcomma
    \end{aligned}
\end{equation}
where \(B\) is the kernel corresponding to the \textcolor{darkgray}{(half of)} \enquote{bubble} diagram given by
\begin{equation}
    \label{eq:bubblediagram}
    B(x,y) \equiv \frac{1}{2}\! \bubbleDiagram \equiv \left[ \frac{1}{(-\dAlembertian + m_{\phi}^{2})\Id}(x,y) \right]^{2}
    \eqend
\end{equation}
Note that for each bubble in \eqref{eq:sigma_exact_propagator} the \((1/\sqrt{N})^{2}\) from the vertices canceled out with the \(N\) coming from the index loop.
The diagram in \eqref{eq:bubblediagram} is really a free correlator of a scalar composite field \(\phi^{2}\equiv {:}(\phi^{i})^{2}{:}\.\) being the normal-ordered square of \(\phi^{i}\) for any fixed index \(i\), that is \(\correlator{\phi^{2}(x)\.\phi^{2}(y)}[\text{free}]=2G_{\phi}(x,y)^{2}\equiv 2B(x,y)\).
The factor of \(2\) comes from two different ways of contracting legs at both vertices.
% By \enquote{free} we mean that the propagator \(G_{\phi}\) was obtained from the leading large \(N\) form of effective action \eqref{eq:effectiveAction} --- we thus include the mass shift \(m^{2} \mapsto m_{\phi}^{2}\) from the VEV of \(\sigma\) --- without including additional contributions from the interactions with \(\sigma\)-field, which come at subleading orders in \(1/N\).
% Combinatorial factor is \(\frac{2^{2}}{2}\), where \(2^{2}\) comes from all possible naive ways how to contract legs at both vertices, but we need to divide by symmetry factor \(2\), since half of these are actually the same Wick contraction --- thus, to obtain just bubble function, we do extra division by \(2\).

From the point of view of the initial action \eqref{eq:Action_ON}, the \(\sigma\)-propagator resums an infinite class of Feynman diagrams contributing in the leading order of \(1/N\) expansion.
For example, the leading \(s\)-channel connected contribution to the \(4\)-point function of \(\phi\) fields is given by
\begin{align}
    \label{eq:bubble_resum}
    \sum_{n=0}^{\infty}\;\bubbleChain\xmathstrut[1.4]{0}=\sExchFeyn\sim \frac{1}{N} \eqend
\end{align}
To increase clarity, from now on, we will explicitly write out the \(N\)-dependence in front of the diagrams instead of including it in the diagrams themselves.

% There are two favorite large \(N\) counting schemes in the literature, differing by a normalization of the \(\sigma\)-field \todo{Not sure if we want to include this.}
% \begin{equation} \label{eq:largeN_counting}
%     \begin{alignedat}{4}
%         \vertexSigma & \sim\frac{1}{\sqrt{N}} &  & \mspace{45mu}\raisebox{-3.5ex}{or}\mspace{52mu} & \vertexSigma & \sim 1          \\[-1ex]
%         \SigmaPropag & \sim 1                 &  &                                                 & \SigmaPropag & \sim\frac{1}{N}
%         \eqend
%     \end{alignedat}
% \end{equation}
% Both these schemes lead to the same large \(N\) counting by definition.
% In particular, they imply that the result~\eqref{eq:bubble_resum} is the main object of interest for calculating the \(4\)-point function \(\correlator{\phi\phi\phi\phi}\).

\subsection{\texorpdfstring{\CFT{}}{CFT} on the boundary of \texorpdfstring{\AdS{}}{AdS}}%
\label{sub:CFT on the boundary of AdS}

Now we will specialize to the case of \(\AdS_{d+1}\) spacetime.
The isometries of \(\AdS\) act on its asymptotic/conformal boundary as conformal transformations, and by performing an appropriate boundary limit of the correlators one obtains \emph{boundary correlators} satisfying the \(\CFT_{d}\) axioms (apart from the presence of the stress tensor operator) \cite{Duetsch:2002wy}.

The boundary operator \(\Ophi^{i}\equiv \Ophi[i]\) corresponding \grayenclose{or \emph{dual}\hspace{0.05em}} to the scalar field \(\phi^{i}\) has a scaling dimension \(\Dphi\) satisfying the equation \(m_{\phi}^{2} = \Dphi(\Dphi-d)\) \cite{Witten:1998qj}.
We measure all dimensionful quantities in units of the \AdS{} radius \(\ell\), which we throughout set to \(\ell\equiv 1\).

To avoid certain subtleties, in the following we consider the positive \enquote{Dirichlet} branch
\begin{equation} \label{eq:dimension_mass-relation}
    \Dphi = \D_{+} \equiv \frac{d}{2} + \sqrt{\frac{d^{2}}{4} + m_{\phi}^{2}} \eqcomma
\end{equation}
which in Poincaré coordinates \((z,y) \in \R_{\ge}\times\R^{d}\) corresponds to the boundary condition \(\phi \sim z^{\D_{+}}\) as \(z\to 0\).
The dual operator is then defined by the boundary limit
\begin{equation}
    \label{eq:boundary_operator}
    \Ophi^{i}(P) \equiv \frac{1}{\sqrt{\Cnorm}} \lim_{s \to \infty} s^{\Dphi} \phi^{i}\Big(X\equiv s P \mathcolor{darkgray}{+ \biggO{1/s}}\Big)
    \eqcomma
\end{equation}
where \(P\) is a point on the boundary of \(\EAdS_{d+1}\) (a future directed null vector in \(\R^{1,d+1}\)) and \(X\) is a point of \(\EAdS\) in the embedding formalism \cite{Costa:2014kfa} approaching \(P\) in the limit \(s\to \infty\), where the \(\bigO{1/s}\) term enables the \(\EAdS\) condition \(X^{2}=-\ell^{2}\equiv -1\) to be satisfied while \(P^{2}=0\).
The normalization constant \(\Cnorm\) is given by
\begin{equation}
    \label{eq:boundary_operator_normalization}
    \Cnorm = \frac{\G(\Dphi)}{2\pi^{\frac{d}{2}}\G(\Dphi-\frac{d}{2}+1)}
    % \Cnorm = \frac{\G[\Dphi]}{2\pi^{d/2}\G[\Dphi-\frac{d}{2}+1]}
    \eqend
\end{equation}

Free propagator of \(\phi^{\ind}\) fields in \EAdS{} is given by the \emph{bulk-to-bulk propagator} \(\Gbulkbulk\) expressible in terms of the chordal distance \(\zeta(X,Y) \equiv (X-Y)^{2} = -2 - 2X\idot Y\) between two points in \AdS{}
% (using the embedding formalism/metric and setting \enquote{\AdS{} radius} to \(\ell\equiv 1\)) as
\begin{equation} \label{eq:bulk-to-bulk_propagator}
    \correlator{\phi^{i}(X)\phi^{j}(Y)}[\text{free}]
    \equiv \delta^{ij} \Gbulkbulk(X,Y)
    = \delta^{ij} \frac{\Cnorm}{\zeta(X,Y)^{\Dphi}}
    \HypGeo{\Dphi,\,\Dphi-\frac{d}{2}+\frac{1}{2}}{2\Dphi-d+1}[-\frac{4}{\zeta(X,Y)}]
    \eqend
\end{equation}
Taking the appropriate boundary limit of one operator --- in the sense of \eqref{eq:boundary_operator} --- one obtains the \emph{bulk-to-boundary propagator} \(\Gbulkbdry\), which has a much simpler form
\begin{equation} \label{eq:bulk-to-boundary_propagator}
    \correlator{\phi^{i}(X)\Ophi^{j}(P)}[\text{free}]
    \equiv \delta^{ij} \Gbulkbdry(X,P)
    = \frac{\sqrt{\Cnorm} \,\delta^{ij}}{(-2X\idot P)^{\Dphi}}
    \eqend
\end{equation}
Sending also the second operator to the boundary, one finds the \emph{boundary-to-boundary propagator} \(\Gbdrybdry\) with the usual form of the 2-point function in flat-space \CFT{} of scalar operator with scaling dimension \(\Dphi\), that is
\begin{equation} \label{eq:boundary-to-boundary_propagator}
    \correlator{\Ophi^{i}(P_{1})\Ophi^{j}(P_{2})}[\text{free}]
    \equiv \delta^{ij} \Gbdrybdry(P_{1},P_{2})
    = \frac{\delta^{ij}}{(-2P_{1}\idot P_{2})^{\Dphi}}
    % = \frac{\delta^{ij}}{(P_{1} - P_{2})^{2\Dphi}}
    = \frac{\delta^{ij}}{\abs{y_{1} - y_{2}}^{2\Dphi}}
    \eqcomma
\end{equation}
where \(y_{1},y_{2} \in \R^{d}\) are flat-space coordinates corresponding to the points \(P_{1},P_{2}\in \bdry{\EAdS_{d+1}}\).
We now see that the normalization in the definition of the boundary operator \eqref{eq:boundary_operator} was chosen such that the boundary \(2\)-point function is conventionally normalized.

\paragraph{Main observable --- boundary \texorpdfstring{\(4\)}{4}-point correlator.}%
\label{par:Main observable --- boundary 4-point correlator}

Let us define the main observable that we will use to probe the spectrum of the \(\CFT_{d}\) emerging at the asymptotic/conformal boundary of \(\AdS_{d+1}\).
It is the boundary limit of the \(4\)-point correlator \(\correlator{\phi\phi\phi\phi}\) of fundamental fields \(\phi^{\ind}\) in the vector representation of the global \(\OO(N)\) symmetry, that is
\begin{align}
    \label{eq:4pt_corr}
    \correlator{\Oper^{i}_1\.\Oper^{j}_2\.\Oper^{k}_3\.\Oper^{l}_4}
    \equiv  \correlator{\Ophi^{i}(P_1)\,\Ophi^{j}(P_2)\,\Ophi^{k}(P_3)\,\Ophi^{l}(P_4)}
    \equiv \FourPt* \eqcomma
\end{align}
where \(P_{\ind}\) are points (suppressed in diagrams) lying on the boundary of \(\AdS_{d+1}\), which is represented diagrammatically by a circle.
Considering it up to the order \(1/N\) in the large \(N\) expansion, but to all orders in the coupling \(\lambda\), its Witten diagram representation is given by
\begin{equation}
    \label{eq:4pt_wd}
    \begin{aligned}
        \FourPt* & = \left(\sDisc*+\tDisc*+\uDisc*\right) \\
                 & \relphantom{=} {} + \frac{1}{N}\left(\sExch*+\tExch*+\uExch*\right) + \biggO{\frac{1}{N^2}}
        \eqcomma
    \end{aligned}
\end{equation}
where we used the leading forms of the 1PI \(\sigma\phi^{2}\) vertex \eqref{eq:ON_HS_rules} and exact \(\sigma\)-propagator \eqref{eq:sigma_exact_propagator}.
The \(N\)-dependence is now explicitly written in front of the diagrams.
Lines with one or both ends on the boundary are bulk-to-boundary \eqref{eq:bulk-to-boundary_propagator} or boundary-to-boundary \eqref{eq:boundary-to-boundary_propagator} propagators, respectively.
Using these ingredients, one can easily compose explicit expressions for diagrams figuring in~\eqref{eq:4pt_wd}, where as usual we integrate over the bulk points.

\paragraph{Convention for channels.} \label{par:Convention for channels}

Our naming convention for both the OPE and also Witten diagrams of the \(4\)-point correlator is
\begin{equation} \label{eq:channel_conventions}
    \arraycolsep=1.5em
    \begin{array}{ccc}
        \text{\(s\)-channel} \mspace{10mu} {}
         & \text{\(t\)-channel} \mspace{10mu} {}
         & \text{\(u\)-channel} \mspace{10mu} {} \\
        \left(\Oper^{i}_{1}\Oper^{j}_{2}\right)\left(\Oper^{k}_{3}\Oper^{l}_{4}\right) \eqcomma
         & \left(\Oper^{i}_{1}\Oper^{k}_{3}\right)\left(\Oper^{j}_{2}\Oper^{l}_{4}\right)\eqcomma
         & \left(\Oper^{i}_{1}\Oper^{l}_{4}\right)\left(\Oper^{j}_{2}\Oper^{k}_{3}\right) \eqend
    \end{array}
\end{equation}
They correspond to the diagrams in~\eqref{eq:4pt_wd}, in respective order at each order of \(1/N\).
This is in agreement with the conventions used in~\cite{Carmi:2018qzm} and \cite[(2.53)]{Karateev:2018oml}, but \(t\)-channel and \(u\)-channel are swapped in \cite[(3.5)/(3.8)]{Liu:2018jhs}.
As we will comment at appropriate places in \Cref{sec:CFTgeneralities}, one just needs to be careful with including an additional factor of \((-1)^{J}\) for \sixjsymbol{} compared to their final formulas.

\paragraph{Renormalization scheme.}
\label{par:Renormalization scheme}
In~\eqref{eq:4pt_wd} we drew only nontrivial Witten diagrams contributing to the \(4\)-point correlator.
Those just modifying the normalization of the \(\phi\)-field and shifting its mass --- connected with the normalization of the dual boundary operator \(\Ophi\) and shifting the scaling dimension via the relation \(m_{\phi}^{2} = \Dphi(\Dphi-d)\) --- were omitted, so let us comment on them here.
Intuition about UV renormalization can be borrowed from flat space, since when distances among points are infinitesimally small, any spacetime looks approximately flat.
Explicit renormalization in \AdS{} in weak coupling perturbation theory, especially for a \(\phi^{4}\)-theory relevant for us, was done in~\cite{Bertan:2018khc,Bertan:2018afl,Banados:2022nhj,Sachs:2023eph}.

We work with the \(4\)-point correlator computed up to (and including) the order \(1/N\).
We will describe the renormalization of the classical action~\eqref{eq:ON_HS_Action}, but sometimes it is useful to reiterate what various steps mean from the perspective of the original action~\eqref{eq:Action_ON}.

First, let us once again qualify the term \enquote{free} in connection with the \(\langle\phi\phi\rangle\) propagator.
The discussion is intimately related with quantization around the nontrivial saddle point \(\sigma(x)=\sqrt{N}\sigma_{\star}\) of the effective action~\eqref{eq:effectiveAction}.
This step produces an \(\bigO{1}\) exact \(\phi\)-propagator (in the large \(N\) expansion), which is from the point of view of~\eqref{eq:Action_ON} given by a resummation of all \enquote{cactus} diagrams.
Those renormalize just the mass \(m\) of the free \(\phi\)-field and shift it to the value \(m_{\phi}^{2}=m^{2}+2\sigma_{\star}\).
Thus, the \enquote{free} \(\phi\)-propagator is associated with a free equation of motion, but with a shifted mass.

Such \(\phi\)-propagator is exact at \(\bigO{1}\) as we stated, therefore it is sufficient for all diagrams that are already of order \(\bigO{1/N}\), in particular for the exchanges of the \(\sigma\)-field in the second line of~\eqref{eq:4pt_wd}.

However, it is not sufficient for the disconnected diagrams of order \(\bigO{1}\) in the first line of~\eqref{eq:4pt_wd}.
In those diagrams, the renormalization of the \(\phi\)-propagator needs to be carried out to order \(\bigO{1/N}\).
Such a renormalization can affect both the normalization of the \(\phi\)-field and a shift in its mass.
Luckily, the type of the propagator subjected to it is the boundary-boundary one, which is completely fixed by conformal symmetry on the boundary.
We choose the two counterterms in such a way that it remains in the \enquote{free} form~\eqref{eq:boundary-to-boundary_propagator}.
This finishes renormalization of the \(\phi\)-propagator.

Now we start discussing renormalization of the \(\langle\sigma\sigma\rangle\) bulk-bulk propagator.
Its form~\eqref{eq:sigma_exact_propagator} is \(\bigO{1}\) exact and from the perspective of~\eqref{eq:ON_HS_Action} renormalizes the \enquote{mass-squared} \(\lambda^{-1}\) of the free \(\sigma\)-field.
In particular, the spectral representation --- see next~\Cref{sub:Utilizing Spectral representation} --- of the \(\sigma\)-propagator \(\left(\lambda^{-1}+2\widetilde{B}\right)^{-1}\) requires a nontrivial renormalization in \(d\geq 3\), where the bubble diagram acquires a UV divergence.
The bubble function must be then somehow regularized, and with the UV divergence absorbed by a counterterm we are left with the finite combination \(\lambda^{-1}+2\widetilde{B}\).
The poles of such renormalized spectral function determine the physical scaling dimensions of boundary operators associated with the bulk field \(\sigma\).
We will comment on fixing the subtraction scheme employed in defining the regularized bubble function \(\widetilde{B}\) in~\Cref{sub:Criticality in the bulk}.
% In the case of interest containing a UV divergence (\(d=4\)), there is one counterterm.
% However, to fix its finite part (constant shift of \(\widetilde{B}\) or equivalently of \(\lambda^{-1}\)) we would have to \enquote{measure} one of the scaling dimensions.
% We attempt to do it in the critical theory describing the ordinary transition (with the assumption that it can be continued above its upper critical dimension \(d=3\)) in~\Cref{sub:Criticality in the bulk}.
% At the critical point the boundary spectrum is determined by bulk conformal symmetry, which fixes the subtraction scheme ambiguity.

Let us also note that the renormalization of the \(\sigma\)-propagator described above, when reinterpreted from the point of view of~\eqref{eq:Action_ON}, would correspond to renormalization of the 1PI vertex \(\phi^{4}\) associated with the coupling \(\lambda\).
Classically it starts at order \(\bigO{1/N}\), and since all \enquote{bubble-chain} diagrams contribute at the same order, they need to be resummed.
This is done in~\eqref{eq:bubble_resum}, which precisely leads to the \(\bigO{1}\) exact form~\eqref{eq:sigma_exact_propagator} of the \(\sigma\)-propagator.

Finally, let us comment about the 1PI vertex \(\sigma\phi^{2}\), which is of order \(\bigO*{1/\sqrt{N}}\) classically.
It appears twice in the \(\sigma\)-exchange diagrams, thus making them of order \(\bigO{1/N}\).
Consequently, to the maximal order we consider, it does not require any renormalization, as all corrections to its classical part would be subleading in the large \(N\) expansion.

This finishes the description of the renormalization.
The renormalization scheme defined above might be called \emph{on-shell}, since it is fully determined in terms of physical scaling dimensions \(\Dphi,\,\Delta_{\Osigma_{0}}\) --- \(\Osigma_{0}\) being the lowest-dimensional boundary operator associated to the \(\sigma\)-field --- and the standard normalization of the boundary operator \(\Ophi\), leading to the canonical form of the \CFT{} \(2\)-point function~\eqref{eq:boundary-to-boundary_propagator}.

\subsection{Utilizing the spectral representation}%
\label{sub:Utilizing Spectral representation}

To proceed further with the calculation of the 4-point correlator \eqref{eq:4pt_corr}, mainly the evaluation of bulk integrals in the \(\sigma\)-exchange diagrams, it is convenient to employ the so-called \emph{spectral representation} of the \(\sigma\)-propagator.
It is an integral expansion into \emph{harmonic functions} --- eigenfunctions of the \AdS{} Laplacian --- which form a continuous basis for integrable functions depending only on the geodesic distance between two points.
This is in precise analogy to the (radial) Fourier representation of the propagator in flat space, which is also adapted to the isometries (translations and rotations) of the theory.
In the limit of large \AdS{} radius, it actually reduces to the flat-space (radial) Fourier transform.

While we will not explicitly use this technology --- the relevant results from \cite{Carmi:2018qzm} will be cited in \Cref{sec:singlet_spectrum} --- it is rather crucial in calculation and deserves a brief mention.
More details can be found in \cite[Appendix 4.C]{Penedones:2007ns} \cite[Appendix B]{Penedones:2010ue} and~\cite[Appendix B]{Carmi:2018qzm}.

Suppose we know the spectral representation \(\Bubble(\D)\) of the bubble function \(B(x,y)\) defined in \eqref{eq:bubblediagram}, that is we have a decomposition into harmonic functions \(\Omega_{\D}\)
\begin{equation} \label{eq:bubble_spectral_rep}
    B(x,y) = \int_{\R} \dd{\nu} \Bubble(\D) \Omega_{\D}(x,y)
    \equiv \int_{\frac{d}{2}+\I\R} \frac{\dd{\D}}{\I} \Bubble(\D) \Omega_{\D}(x,y) \eqcomma
\end{equation}
where we use \(\D\equiv \frac{d}{2}+\I\nu \Leftrightarrow \nu\equiv -\I(\D-\frac{d}{2})\) interchangeably.
While it is common to use the spectral parameter \(\nu\) to index the harmonic functions and also to denote the functional dependence of spectral representations, we will mostly prefer using the dimension \(\D\) directly for later notational convenience.

Ignoring the index structure, the calculation can be schematically represented as
\begin{equation}
    \label{eq:exchange_diagram_computation}
    \begin{aligned}
        \smash[b]{\sExch}
         & = 4\int_{\R} \dd{\nu} \left( \frac{-1}{\lambda^{-1}+2\Bubble(\D)} \right) \sExchHarmonic \\
         & = 4\int_{\R} \dd{\nu} \left( -\frac{1}{\lambda^{-1}+2\Bubble(\D)} \right) \sqrt{\Cnorm[\D]\Cnorm[\D*]\.} \frac{\nu^{2}}{\pi} \sExchSplit \\
         & = \int_{\frac{d}{2}+\I\R_{\ge0}} \frac{\dd{\D}}{2\pi\I} \left( -\frac{1}{\lambda^{-1}+2\Bubble(\D)} \right)
        {\color{darkgray}\left(\frac{\G[\cdots]^{2}\cdots\G[\cdots]}{{}\cdots\G[\cdots]^{2}\cdots\G[\cdots]}\right)}
        \shadowRepCPW \\[0.5ex]
         & \equiv \int_{\frac{d}{2}+\I\R_{\ge0}} \frac{\dd{\D}}{2\pi\I} \left( -\frac{1}{\lambda^{-1}+2\Bubble(\D)} \right)
        {\color{darkgray}\left(\frac{\G[\cdots]^{2}\cdots\G[\cdots]}{{}\cdots\G[\cdots]^{2}\cdots\G[\cdots]}\right)}
        \ket{\sPartWave[\D\mathcolor{gray}{,\.0}][]} \eqend
    \end{aligned}
\end{equation}
Line by line, we utilized following properties/relations (for details see \cite[Appendix C]{Carmi:2018qzm}):
\begin{enumerate}
    \item Similarly to the Fourier transform, the spectral representation transforms a convolution of functions into a product of their spectral functions.
          Thus, the \enquote{operator} inversion figuring in the \(\sigma\)-propagator \eqref{eq:sigma_exact_propagator} is represented just as a numeric inversion of the corresponding spectral function.
          The overall factor of \(4\) comes from the two \(\sigma\phi^{2}\) vertices \eqref{eq:ON_HS_rules}, we implicitly integrate over the black bulk points, and the wavy line represents the harmonic function.
    \item The harmonic function can be further rewritten by means of the \emph{split representation} \cite{Costa:2014kfa,Fitzpatrick:2011ia}, which up to a certain prefactor is given by two bulk-to-boundary propagators --- one with dimension \(\D\) and the other with the shadow dimension \(\D*\equiv d-\D\) --- with the black boundary point implicitly integrated over \(\bdry{\AdS}\).
          Remember that the rest of bulk-to-boundary propagators have the dimension \(\Dphi\).
    \item Now we can perform the bulk integrations.
          For fixed boundary points, both of the bulk integrals individually must result in a multiple of the unique \CFT{} \(3\)-point structure of the corresponding scalar operators --- these are diagrammatically represented by the blue blobs.
          We are thus left with a convolution of two such \(3\)-point structures, since we have yet to integrate over the boundary point.
    \item Such a convolution is actually the \emph{shadow representation} of the \emph{Conformal Partial Wave} (CPW) \cite{Mack:1974jjo,Dobrev:1977qv,Dolan:2011dv,Simmons-Duffin:2012juh,Simmons-Duffin:2017nub,Liu:2018jhs}, which we denote by \(\ket{\sPartWave[][][0.55]}\).
\end{enumerate}
In the end, the computation results in the \emph{Conformal Partial Wave decomposition} of the \(s\)-channel \(\sigma\)-exchange diagram.
The \(t\)-channel/\(u\)-channel diagrams are given by the same expression, just with the CPWs in the corresponding channels.

We will discuss such decompositions in more detail in the next section.
The associated coefficient/weight function (also called the \emph{spectral function}) is the main object of interest.
It encodes the \CFT{} data accessible to the \(4\)-point correlator \(\correlator{\Ophi\Ophi\Ophi\Ophi}\).

% \Note{Spectral representation is in a certain sense dual to the CPW decomposition to be discussed in the next section.
%     The spectral parameter is in the \emph{Principal Series}, the \enquote{split representation} is analogous to \emph{shadow representation} of CPWs.
%     If we attach 2 bulk-to-boundary propagators to each end-points of the harmonic function, after integration we obtain multiple of corresponding CPW, right?
%     Any nice reference where this is discussed?
% }

\section{\texorpdfstring{\CFT{}}{CFT} generalities --- \texorpdfstring{\(4\)}{4}-point correlators and anomalous dimensions}%
\label{sec:CFTgeneralities}

As the boundary limits of correlators in \AdS{} enjoy a conformal symmetry, it is only natural to study them using the language and methods of \CFT{}s.
Calculation of the exchange diagram contribution to the 4-point correlator --- directly relevant for the \(\OO(N)\) singlet \CFT{} spectrum --- naturally resulted in its \emph{conformal partial wave} decomposition, which we will review in \Cref{sub:CB and CPW Decompositions}.

Things get more involved when one wants to extract the non-singlet spectrum.
We will see in \Cref{sub:ON_irrep_decomposition} that crossed-channel diagrams --- take for example the \(t\)-channel --- turn out to be essential for this task.
Even though it is not viable to calculate directly the \(s\)-channel decomposition of the \(t\)-channel interaction diagram, its \(t\)-channel decomposition is basically known after resolving the singlet spectrum.
In \Cref{sub:CPW completeness and 6j-symbol} we will discuss how to translate the \(t\)-channel decomposition into the \(s\)-channel one via the \sixjsymbol{}s.

To fully prepare for the extraction of the leading \(1/N\) corrections to the non-singlet spectrum, in \Cref{sub:tChannelContributiontoAnomalousDimensions} we will present the general formulas (in \(d=2\) and \(d=4\)) for the contribution of a \(t\)-channel conformal block to the anomalous dimensions of \(s\)-channel double-twist operators, applicable for external scalar operators with equal scaling dimensions.

For simplicity --- and because it is the relevant case for us --- in the following we assume all external operators to be scalars \(\Ophi\) with equal scaling dimensions \(\Dphi\).
To avoid clutter, we will suppress throughout the \textcolor{darkgray}{possible} global symmetry index structure of the operators, and the dependence of correlators/conformal blocks/conformal partial waves on the scaling dimensions and positions of external operators.

It is enough to concentrate only on the nontrivial crossing between \(s\)-channel and \(t\)-channel, since the \(u\)-channel is then given by including an additional factor of \((-1)^{J}\).
Here \(J\) is the spin of the exchanged operator belonging to the symmetric traceless representation of the rotation group \(\SO(d)\).
This follows from a simple relation between \(t\)-channel and \(u\)-channel conformal blocks for equal external scaling dimensions \cite[(59)]{Poland:2018epd}, or can be deduced directly from the property of the OPE coefficients for two scalar operators \(\Oper_{1},\Oper_{2}\) and a spin \(J\) operator \(\Oper_{J}\) --- \(\ope{\Oper_{J}\Oper_{1}\Oper_{2}}=(-1)^{J}\ope{\Oper_{J}\Oper_{2}\Oper_{1}}\) --- see \cite[(25)]{Poland:2018epd}.

To summarize, the starting point is MFT, where it is known that the identity operator in the \(t\)-channel induces a double-twist family of operators \(\Oper_{n,J}\) in the \(s\)-channel decomposition, with scaling dimensions \(\D[\MFT]_{n,J} \equiv 2\Dphi+2n+J\).
The question that we want to answer in the following is: Given a \(t\)-channel contribution of an operator \(\Oper'\) with scaling dimension \(\Dpr\) and spin \(J'\), what anomalous dimensions \grayenclose{and corrections to OPE coefficients} does it induce for the double-twist family in the \(s\)-channel decomposition?
Supposing that the exchange contribution is of order \(\bigO{1/N}\), we will compute the leading form of the corrected scaling dimensions \(\D_{n,J} \equiv \D[\MFT]_{n,J} + \anomdim[]_{n,J} \equiv 2\Dphi+2n+J+\anomdim[]_{n,J}\), where the anomalous dimensions \(\anomdim[]_{n,J}\) are also of order \(\bigO{1/N}\).
Thus, as expected, this family of operators reduces to the MFT ones in the limit \(N\to \infty\).
The dimension \(\Dpr\) of the \(t\)-channel exchanged primary does not need to be parametrically close to an MFT value in the large \(N\) expansion.
In fact, it actually turns out that to contribute nontrivially at leading order, it should not be.

\subsection{Conformal Block and Conformal Partial Wave decompositions}%
\label{sub:CB and CPW Decompositions}

By grouping the operators in pairs and using the \emph{Operator Product Expansion} (OPE) twice, we can write the 4-point correlator as a discrete sum over contributions of exchanged conformal families, weighted by the \grayenclose{squares of} OPE coefficients.
Such contributions are called \emph{Conformal Blocks} (CBs), and they correspond to an exchange of a physical primary operator together with all of its descendants.
Since only symmetric traceless tensors appear in the OPE of two scalars, the exchanged families are labeled by their scaling dimension \(\Dstar\) and spin \(J_{\star}\.\), giving us the \emph{Conformal Block Decomposition}
\begin{equation} \label{eq:CB_decomposition_discrete}
    \FourPt = \sum_{\substack{\text{primary } \Ostar \\\text{with }\Dstar,J_{\star}}} \opesq{\Ophi\Ophi\Ostar} \ket{\ConfBlock[\Dstar,J_{\star}][s]}
    = \mathcolor{darkgray}{\left(\substack{\text{analogously in} \\\text{\(t\)-channel}}\right)}
    \eqcomma
\end{equation}
where \(\ope{\Ophi\Ophi\Ostar}\) is the OPE coefficient for operator \(\Ostar\) appearing in the \(\Ophi\times \Ophi\) OPE, and the corresponding \(s\)-channel conformal block is denoted by \(\ket{\ConfBlock*[\Dstar,J_{\star}][s]}\).
We suppress the dependence of CBs on conformal representations and positions of the external operators.
% \todo{Mention that we include \enquote{leg factors} in CBs.}

Alternatively, there is an expansion into Unitary Irreducible Representations (UIRs) of the (Euclidean) conformal group \(\SO(d+1,1)\), including mainly the \emph{Principal Series} \grayenclose{traceless symmetric} representations with integer spin \(J \in \N_{0}\.\) but \enquote{unphysical} complex dimensions \(\Delta\in\frac{d}{2}+\I\R_{\geq}\).
The associated eigenfunctions of conformal Casimir are called \emph{Conformal Partial Waves} (CPWs), which furthermore form a complete basis of \enquote{normalizable} functions \cite{Dobrev:1977qv}.
They come from the \emph{harmonic analysis} of the Euclidean conformal group \(\SO(1,d+1)\), and are analogous to plane waves, which are eigenfunctions of translations in the flat space.

Ignoring non-normalizable contributions such as the exchange of the identity operator (see \cite{Simmons-Duffin:2017nub} for some discussion), we have the \emph{Conformal Partial Wave Decomposition}
\begin{equation} \label{eq:CPW_decomposition}
    \FourPt
    = \sum_{J=0}^{\infty} \int_{\frac{d}{2}+\I\R_{\geq}}  \frac{\dd{\D}}{2\pi\I}   \Spec[s][\D][J]{\FourPt}    \ket{\sPartWave}
    = \mathcolor{darkgray}{\left(\substack{\text{analogously in} \\\text{\(t\)-channel}}\right)}
    \eqcomma
\end{equation}
where the \(s\)-channel conformal partial wave is denoted by \(\ket{\sPartWave[][][0.55]}\), and the spectral function \(\SpecNoArgs[s][{}\cdots]\) represents the coefficients of the corresponding CPWs.
Again, we suppress the dependence of CPWs on conformal representations and positions of the external operators.

If we were to perform decomposition in the \(t\)-channel, we would use \(t\)-channel spectral function \(\SpecNoArgs[t][{}\cdots]\) and \(t\)-channel CPWs \(\ket{\tPartWave[][][0.55]}\).
In \cite{Liu:2018jhs, Simmons-Duffin:2017nub} they use either \(\rho\) or \(I/n\) for our \(\SpecNoArgs[s]\), where the normalization \(n\) is defined in \eqref{eq:CPWnorm}.
Notice that the integration in \eqref{eq:CPW_decomposition} is over the half-line \(\D \in \frac{d}{2}+\I\R_{\geq}\), since the \emph{shadow} scaling dimensions \(\D*\equiv d-\D \in \frac{d}{2}+\I\R_{\leq}\) correspond to equivalent representations of the conformal group.

These two decompositions are closely related, as CPWs are (up to normalization) shadow-symmetric combinations of CBs
\begin{align}
    \label{eq:CPWasCB_s_or_t_channel}
    \begin{alignedat}{2}
        \ket{\sPartWave} & = \Knorm[\D*,J]    \ket{\ConfBlock[\D,J]}       &  & + \Knorm[\D,J]    \ket{\ConfBlock[\D*,J]}       \eqcomma \\
        \ket{\tPartWave} & = \Knorm[\Dpr*,J'] \ket{\ConfBlock[\Dpr,J'][t]} &  & + \Knorm[\Dpr,J'] \ket{\ConfBlock[\Dpr*,J'][t]} \eqcomma
    \end{alignedat}
\end{align}
where the normalization coefficients are given by \cite[(2.16)]{Liu:2018jhs} \cite[(A.5),(A.6)]{Simmons-Duffin:2017nub}
\begin{equation}
    \label{eq:Knorm}
    \Knorm[\D,J]
    = \frac{\pi^{\frac{d}{2}}}{(-2)^{J}}
    \frac{\G[\D-\frac{d}{2}] \G[\D+J-1]}{\G[\D-1] \G[d-\D+J]}
    \left( \frac{\G[\frac{\D*+J}{2}]}{\G[\frac{\D+J}{2}]} \right)^{\!2}
    \eqend
\end{equation}
In the \(t\)-channel we will usually use primed quantities for improved distinction.
Also, to make the expressions compact, we will often utilize the abbreviation \(\G[\argument]\equiv\G(\argument)\).

Since CPWs are shadow-symmetric, it is natural to choose the spectral function to be shadow-symmetric as well.
Substituting \eqref{eq:CPWasCB_s_or_t_channel} into \eqref{eq:CPW_decomposition}, using the aforementioned shadow-symmetry of \(\SpecNoArgs\) to extend the integration from half-line \(\frac{d}{2} + \I\R_{\geq}\) to \(\frac{d}{2} + \I\R\) together with taking only the first conformal block term in CPW, we obtain an integral decomposition in terms of conformal blocks as
\begin{equation}
    \label{eq:CB_integral_decomposition}
    \FourPt
    = \sum_{J} \int_{\frac{d}{2}+\I\R} \frac{\dd{\D}}{2\pi\I} \Spec[s][\D][J]{\FourPt} \Knorm[\D*,J] \ket{\ConfBlock[\D,J][s]}
    \eqend
\end{equation}
After enclosing the integration in the right-half plane, the residue theorem enables us to write it as a discrete sum over physical poles of \(\SpecNoArgs[s]\), thus obtaining the usual conformal block decomposition \eqref{eq:CB_decomposition_discrete}.
Certain subtleties of this procedure, in particular appearance and cancellation of additional spurious poles, are discussed in \cite[Appendix B]{Simmons-Duffin:2017nub}.

We see that (at least the accessible part of) the \CFT{} data are encoded in the spectral function \(\SpecNoArgs[s]\), namely the scaling dimensions of primary operators are given by the positions of \(\SpecNoArgs[s]\) poles, and the corresponding squared OPE coefficients in the \(s\)-channel are given by the (minus, since the contour is clockwise) residues as
\begin{equation}
    \label{eq:squaredOPEasRes}
    \opesq{\Ophi\Ophi\Ostar} = -\Res_{\D=\Dstar}\left(\Knorm[\D*,J_{\star}] \Spec[s][\D][J_{\star}]{\FourPt}\right)
    \eqend
\end{equation}

\subsection{CPW orthogonality and completeness, \texorpdfstring{\sixjsymbol{}}{6j-symbol}}%
\label{sub:CPW completeness and 6j-symbol}

As already mentioned, CPWs (along the principal series) form a complete basis of functions, which furthermore are orthogonal with respect to an appropriate conformally-invariant pairing involving integration over all four external points \cite[(1.3)]{Simmons-Duffin:2017nub}, or alternatively a closely related inner product \cite[(A.27)]{Simmons-Duffin:2017nub}.
We will use the bra-ket/inner-product notation in the following, that is
\begin{equation} \label{eq:CPW_orthogonality}
    \bra{\sPartWave} \ket{\sPartWave[\bar{\D}][\bar{J}]}
    = \CPWnorm\, 2\pi\delta(\nu-\bar{\nu})\,\delta_{J\bar{J}} \eqcomma
\end{equation}
where \(\D\equiv \frac{d}{2} + \I\nu\) and \(\bar{\D}\equiv \frac{d}{2} + \I\bar{\nu}\) with \(\nu,\bar{\nu}\ge0\) are scaling dimensions in the principal series, and \(\CPWnorm\) is the normalization \cite[(2.35)]{Liu:2018jhs}
\begin{equation} \label{eq:CPWnorm}
    \CPWnorm \equiv \frac{\Knorm[\D*,J]\Knorm[\D,J]\vol(\Sphere^{d-2})}{\mathcolor{orange!55!black}{2^{d}} \vol(\SO(d-1))} \frac{(2J+d-2) \pi \G[J+1] \G[J+d-2]}{2^{d-2} \G[J+\frac{d}{2}]^{2}}
    \eqend
\end{equation}
Note that it includes an extra \(\mathcolor{orange!55!black}{2^{-d}}\) compared to \cite[(A.14), (A.15)]{Simmons-Duffin:2017nub}.
We can thus express the completeness relation (for normalizable functions) in terms of CPWs as
\begin{align}
    \label{eq:CPW_id_decomp}
    \Id = \sum_{J} \int_{\frac{d}{2}+\I\R_{\geq}} \frac{\dd{\D}}{2\pi\I} \ket{\sPartWave} \frac{1}{\CPWnorm} \bra{\sPartWave} \eqend
\end{align}

Consider now a certain contribution to 4-point correlator, for which we are able to calculate its \(t\)-channel spectral function \(\SpecNoArgs[t]\), that is we know the decomposition
\begin{equation} \label{eq:tContrib_tChannel_CPW_decomp}
    \tExchContrib = \sum_{J'} \int_{\frac{d}{2}+\I\R_{\geq}} \frac{\dd{\Dpr}}{2\pi\I} \Spec[t][\Dpr][J']{\tExchContrib} \ket{\tPartWave}
\end{equation}
into \(t\)-channel CPWs.
To extract the associated contribution to the \CFT{} data, we first need to translate this decomposition into the \(s\)-channel.
By inserting the completeness relation~\eqref{eq:CPW_id_decomp} into the \(t\)-channel decomposition, or by directly utilizing the orthogonality~\eqref{eq:CPW_orthogonality}, we obtain the \(s\)-channel spectral function as
\begin{equation} \label{eq:tContrib_sChannel_Spec}
    \Spec[s][\D][J]{\tExchContrib}
    = \sum_{J'} \int_{\frac{d}{2}+\I\R_{\geq}} \frac{\dd{\Dpr}}{2\pi\I}
    \frac{\bra{\sPartWave}\ket{\tPartWave}}{\CPWnorm} \Spec[t][\Dpr][J']{\tExchContrib}
    \eqcomma
\end{equation}
where the \((s\leftarrow t)\)--channel translation is being performed \grayenclose{up to the normalization \(\CPWnorm\)} by the \enquote{Clebsch-Gordan coefficient} for the conformal group called \sixjsymbol{} \cite[(3.5)]{Liu:2018jhs}
\begin{align}
    \label{eq:6j}
    \begin{aligned}
        \sixjsymbol
         & \equiv \bra{\sPartWave} \ket{\tPartWave} \\
         & = \Knorm[\Dpr*,J'] \underbrace{\bra{\sPartWave} \ket{\ConfBlock[\Dpr,J'][t]}}_{\mathcal{B}^{\Dphi}_{[\D,J],[\Dpr,J']}}
        + \Knorm[\Dpr,J']  \underbrace{\bra{\sPartWave} \ket{\ConfBlock[\Dpr*,J'][t]}}_{\mathcal{B}^{\Dphi}_{[\D,J],[\Dpr*,J']}}
        \eqcomma
    \end{aligned}
\end{align}
where \(\mathcal{B}\) is the notation used in \cite[(3.35), (3.41)]{Liu:2018jhs}, and corresponds to the \(s\)-channel spectral function of a single \(t\)-channel conformal block.

% \Note{One can also write \eqref{eq:tContrib_sChannel_Spec} in the \enquote{integral} CB decomposition form.
%     Since the \enquote{crossing kernel} has no poles in the integration variable \(\Dpr\) to the right of the integration contour, we can close the contour in the right half plane (where we pick only poles from the \(t\)-channel spectral function).
%     Following shows the same, but maybe more elegantly by taking \enquote{physical} CB decomposition as a starting point.
% }

Alternatively, we could start with the \(t\)-channel conformal block decomposition of the contribution \eqref{eq:tContrib_tChannel_CPW_decomp}, given by sum over \(t\)-channel conformal blocks with some coefficients
\begin{equation} \label{eq:tContrib_tChannel_CB_decomp}
    \tExchContrib = \sum_{\substack{\text{primary } \Oprstar \\ \text{with }\Dprstar,J'_{\star}}} C_{\Oprstar} \ket{\ConfBlock[\Dprstar,J'_{\star}][t]}
    \eqend
\end{equation}
Again, utilizing the orthogonality~\eqref{eq:CPW_orthogonality}, the corresponding \(s\)-channel spectral function is given by
\begin{equation} \label{eq:tContrib_sChannel_Spec_CB}
    \begin{aligned}
        \Spec[s][\D][J]{\tExchContrib}
         & = \frac{1}{\CPWnorm} \bra{\sPartWave}\ket{\tExchContrib[][0.78]} \\
         & = \!\sum_{\substack{\text{primary } \Oprstar \\ \text{with }\Dprstar,J'_{\star}}}\!
        \OPEsq_{\Oprstar} \underbrace{\frac{1}{\CPWnorm} \bra{\sPartWave} \ket{\ConfBlock[\Dprstar,J'_{\star}][t]}}_{\displaystyle\equiv \CrK[\D,J][\Dprstar,J'_{\star}]} \eqcomma
    \end{aligned}
\end{equation}
where we introduced the \enquote{crossing kernel} \(\CrK[]\) notation for the \(s\)-channel spectral function of a single \(t\)-channel conformal block \grayenclose{including the proper normalization}, which in the following will be slightly more notationally convenient than \(\mathcal{B}\).

Strictly speaking, due to the limited validity range of the Lorentzian inversion formula utilized to calculate \(\CrK[]\) in \cite{Simmons-Duffin:2017nub}, the \(t\)-channel conformal block inversion in \eqref{eq:tContrib_sChannel_Spec_CB} seems to work only for \(J>J'\).
We will touch on this issue a little bit more in \Cref{sub:Dependence of anomalous dimensions on the coupling}, also see remarks in \cite[Section 4]{Liu:2018jhs}.
% \Note{Also interesting is \cite{Alday:2017gde}.}

From extensive studies of the Lorentzian inversion formula~\cite{Caron-Huot:2017vep,Simmons-Duffin:2017nub,Kravchuk:2018htv}, it is well known that \(\CrK[]\) has double zeros in \(\Dpr\) at the locations of MFT double-twist dimensions, therefore only non-MFT operators contribute to the \(s\)-channel spectral function \(\SpecNoArgs[s]\).
Direct consequence of this fact can be explicitly seen in the formulas for the anomalous dimensions presented in the following \Cref{sub:tChannelContributiontoAnomalousDimensions}.

\subsection{Contribution of \texorpdfstring{\(t\)}{t}-channel conformal blocks to anomalous dimensions}
\label{sub:tChannelContributiontoAnomalousDimensions}

Suppose we solve the theory in some kind of perturbative expansion --- we will be formulating everything in the context of large \(N\) expansion, but it is applicable more generally.
We therefore organize correlators and \CFT{} data into expansion in powers of \(1/N\), for example consider that certain part of 4-point correlator is given by
\begin{align}
    \label{eq:largeNfourPt}
    \tExchContrib = \tDisc + \frac{1}{N}\tExch + \ldots
    \eqcomma
\end{align}
where the first disconnected term is the \(t\)-channel identity contribution, and the second term is some interaction contributing at \(\bigO{1/N}\) order, for which we are able to calculate its \(t\)-channel conformal block decomposition.
Our goal is to extract the respective correction to the \CFT{} spectrum.

\paragraph{Disconnected \(t\)-channel spectral function.}
\label{par:disc_spec}

Spectral function of the \(t\)-channel identity (\(t\)-channel disconnected 4-point MFT correlator) in the \(s\)-channel was computed in its general form in~\cite{Fitzpatrick:2011dm} and later reproduced within an elegant harmonic analysis formalism~\cite{Karateev:2018oml}
\begin{equation}
    \label{eq:spec_disc_t}
    \Spec[s][\D][J]{\tDisc} =
    % \frac{\mathcolor{orange!80!black}{(-1)^{J}} 2^{J-1}}{\Snorm}
    \frac{2^{J-1}}{\Snorm}
    \frac{\G[\D-1] \G[\frac{d}{2}-\Dphi]^{2} \G[\frac{d}{2}+J] \G[\D*+J]^{\xmathstrut[0.3]{0}} }{
        \G[\Dphi]^2 \G[J+1] \G[\D - \frac{d}{2}] \G[\D+J-1]}
    \frac{\G[\frac{J+\D}{2}]^2 \G[\frac{J-\D}{2}+\Dphi] \G[\frac{\D+J-d}{2}+\Dphi]}{
        \G[\frac{J+\D*}{2}]^2 \G[\frac{2d+J-\D}{2}-\Dphi] \G[\frac{\D+J+d}{2}-\Dphi]}
    \eqcomma
\end{equation}
where \(\Snorm \equiv (-2)^{J}\!\Knorm \).

As expected, it contains poles --- which come from \(\G(\frac{J-\D}{2}+\Dphi)\) --- located at the dimensions \(\D[\MFT]_{n,J}=2\Dphi+2n+J\) corresponding to the family of MFT double-twist operators schematically given by \(\Oper[\MFT]_{n,J}=[\Ophi\dAlembertian^{n}\partial^{J}\Ophi-\text{traces}]\).
Corresponding squared OPE coefficients can be easily calculated using \eqref{eq:squaredOPEasRes}.

There are also some spurious poles coming from \(\G(\D*+J)\equiv \G(d+J-\D)\), but these are resolved as discussed in \cite[Appendix B]{Simmons-Duffin:2017nub}.

Thus, from the \(s\)-channel conformal block decomposition point of view, at the leading \(\bigO{1}\) order the \(t\)-channel identity gives rise to the aforementioned double-twist operators.
At the following \(\bigO{1/N}\) order, the connected interaction term modifies the \CFT{} data, in particular it induces anomalous dimensions \(\anomdim[]_{n,J}\) of these operators.
This is what we will focus on in the following.

\paragraph{Contribution of \(t\)-channel exchange.}%
\label{par:Contribution of t-channel exchange}

Recalling \eqref{eq:squaredOPEasRes}, the appearance of an operator \(\Ostar\) in the \(s\)-channel conformal block decomposition of the \(4\)-point correlator is reflected in the spectral function as a simple pole of the form \textcolor{darkgray}{(now given as series in \(1/N\))}
\begin{align}
    \label{eq:poleinSpec}
    \frac{-\OPEsq_{\star}\left(\frac{1}{N}\right)}{\D-\Dstar\left(\frac{1}{N}\right)}
    \in \Knorm \Spec[s][\D][J]{\tExchContrib}
    \eqend
\end{align}
Defining the leading corrections to the squared OPE coefficients and scaling dimensions as
\begin{alignat}{3}
    \label{eq:CFTdataTaylor}
    \mathllap{\mathcolor{gray}{\opesq{\Ophi\Ophi\Ostar} \equiv {}}} \OPEsq_{\star}\left(\frac{1}{N}\right) & =\OPEsq[\MFT*]_{\star} && +\frac{1}{N}\OPEsq[1]_{\star} && +\biggO{\frac{1}{N^{2}}} \eqcomma \\
    \Dstar\left(\frac{1}{N}\right)                                                                         & =\Dstar[\MFT]          && +\frac{1}{N}\anomdim_{\star}  && +\biggO{\frac{1}{N^{2}}} \eqcomma
\end{alignat}
the expansion of~\eqref{eq:poleinSpec} to the first order in \(1/N\) reads
\begin{align}
    \label{eq:poleinSpecTaylored}
    \frac{-\OPEsq_{\star}\left(\frac{1}{N}\right)}{\D-\Dstar\left(\frac{1}{N}\right)}
    = \frac{-\OPEsq[\MFT*]_{\star}}{\D-\Dstar[\MFT]}
    +\frac{1}{N}\left[
        \frac{-\OPEsq[1]_{\star}}{\D-\Dstar[\MFT]}
        +\frac{-\OPEsq[\MFT*]_{\star} \anomdim_{\star}}{\left(\D-\Dstar[\MFT]\right)^{2}}
    \right]
    +\biggO{\frac{1}{N^{2}}} \eqend
\end{align}
Anomalous dimensions are thus encoded in the coefficients of double poles in the spectral function, divided by the corresponding MFT squared OPE coefficients.
% Equivalently, we can express the anomalous dimensions as residues of simple poles by dividing the MFT (disconnected) \(t\)-channel spectral function.
% \begin{align}
%     \frac{\Spec[s][]{\FourPt[][0.8]}}{\lim\limits_{N \to \infty} \Spec[s]{\FourPt[][0.8]}}
%     = 1 + \frac{1}{N} \frac{\Spec[s][]{\tExch[][0.8]}}{\Spec[s]{\tDisc[0.8]}} + \ldots
%     = 1 + \frac{1}{N} \frac{\frac{1}{\CPWnorm} \mathcal{B}^{\Dphi}_{\D,J,\Dpr,J'}}{\Spec[s][]{\tDisc[0.8]}} + \ldots
% \end{align}
% \begin{align}
%     \frac{\Spec[s][]{\tExch[][0.8]}}{\Spec[s]{\tDisc[0.8]}} \in
%     \frac{N\Spec[s][]{\FourPt[][0.8]}}{\lim\limits_{N \to \infty} \Spec[s][]{\FourPt[][0.8]}}
% \end{align}
Since the MFT squared OPE coefficients are given by residues of the \(t\)-channel identity, we can express the \grayenclose{\(\bigO{1/N}\) part of\hspace{0.05em}} anomalous dimensions as
\begin{align}
    \label{eq:anomdim_our}
    \anomdim_{n,J} = \Res_{\D=2\Dphi+2n+J} \left( \frac{\Spec[s][\D\vspace{-0.2ex}][J]{\tExch[][0.83]}}{\Spec[s][\D\vspace{-0.2ex}][J]{\tDisc[0.83]}} \right)
    \eqend
\end{align}
If we wanted to extract the \(\bigO{1/N}\) contributions to the OPE coefficients, we would just calculate the (minus) residue without dividing by the spectral function of \(t\)-identity.

For simplicity, consider that the interaction term in \eqref{eq:anomdim_our} is composed of a single \(t\)-channel conformal block with scaling dimension \(\Dpr\) and spin \(J'\).
Utilizing \eqref{eq:tContrib_sChannel_Spec_CB}, the corresponding contribution to the anomalous dimensions is given by
\begin{align}
    \label{eq:anomdim_contrib_from_CB}
    \anomdim_{n,J} \color{darkgray} \eval_{\substack{\text{\(t\)-channel} \\\text{exchange}\\\Dpr,\.J'}} \color{black}
    = \Res_{\D=2\Dphi+2n+J} \left( \frac{\raisemath{1ex}{\CrK[\D,J][\Dpr,J']}}{\Spec[s][\D\vspace{-0.5ex}][J]{\tDisc[0.65]}} \right)
    \eqend
\end{align}
Simple poles of the expression inside residue \eqref{eq:anomdim_contrib_from_CB} --- or equivalently, double poles of \sixjsymbol{} or \(\CrK[]\) --- are not present for generic non-equal external scaling dimensions.
Nonetheless, for pairwise-equal external dimensions the crossing kernel develops double poles \cite[(3.48) and below]{Liu:2018jhs}, and hence the anomalous dimensions obtain nontrivial contributions.

\paragraph{General formulas for \(d=2\) and \(d=4\).}%
\label{par:General formulas for d2 and d4}

The \(\sixjsymbol\) and thus \(\CrK[]\) was explicitly computed in \cite[(3.36), (3.42)]{Liu:2018jhs} for \(d=2\) and \(d=4\) by methods relying on the Lorentzian inversion formula \cite{Caron-Huot:2017vep}.
Compared to our conventions specified in \eqref{eq:channel_conventions}, their \(t\)-channel and \(u\)-channel are swapped, so we include a factor of \((-1)^{J}\) when taking over their results for \sixjsymbol{}.
Moreover, they calculated also the \(t\)-channel conformal block contributions to the leading-twist \grayenclose{\(n=0\)} anomalous dimensions \cite[(3.55), (3.56)]{Liu:2018jhs}.

Here we present the general formulas applicable for arbitrary \textcolor{darkgray}{(assuming \(J>J'\))} double-twist operators \(\Oper_{n,J}\) in \(d=2\) and \(d=4\).
For the detailed calculation see the accompanying \notebook{}.
The final formulas read
\begin{equation}
    \label{eq:tchannel_anom_dim_2d}
    \begin{aligned}
        \anomdim_{n,J} \eval_{\smash[b]{\substack{\text{\(t\)-channel} \\\text{exchange}\\\Dpr,\.J'}}}
         & \eqdim{2}
        -
        \frac{2 \G[\Dphi]^{2} \G[\Dphi+J+n]^2}{\G[2(\Dphi+J+n)]}
        \frac{\sin^{2}\left(\pi\left[\Dphi-\frac{\Dpr-J'}{2}\right]\right)}{\pi^{2}}
        \frac{\G[J+n+1]}{\G[2\Dphi+J+n-1]}
        \graytimes {} \\
         & \graytimes
        \frac{1}{1+\delta_{0,J'}}
        % \left[
        %     \frac{\G[\Dpr-J']}{\G[\frac{\Dpr-J'}{2}]^2}
        %     \anomdimFfunction[d\.=\.2]{n}{\frac{\Dpr-J'}{2}}
        %     \operatorname{\Omega}_{\Dphi+J+n, \frac{\Dpr+J'}{2}, \Dphi} {}
        %     + \left( J' \exchange \widetilde{J}'\equiv -J' \right)
        % \right]
        \left[
            \frac{\G[\Dpr+\widetildesmash[-0.12ex]{J'}]}{\G[\frac{\Dpr+\widetildesmash[-0.3ex]{J'}}{2}]^2}
            \anomdimFfunction[d\.=\.2]{n}{\frac{\Dpr+\widetildesmash[-0.3ex]{J'}}{2}}
            \operatorname{\Omega}_{\Dphi+J+n, \frac{\Dpr+J'}{2}, \Dphi} {}
            + \left( J' \exchange \widetilde{J}'\equiv -J' \right)
        \right]
        \eqcomma
    \end{aligned}
\end{equation}
\begin{equation}
    \label{eq:tchannel_anom_dim_4d}
    % \begin{aligned}
    %     \anomdim_{n,J} \eval_{\smash[b]{\substack{\text{\(t\)-channel} \\\text{exchange}\\\Dpr,\.J'}}}
    %      & \eqdim{4}
    %     \frac{(-1)^{n}}{n!} \frac{2\G[\Dphi]^{2}}{\G[2-\Dphi]^{2}}
    %     \frac{\sin^{2}\left(\pi\left[\Dphi-\frac{\Dpr-J'}{2}\right]\right)}{\pi}
    %     \frac{\G[J+n+2]}{J+1} \graytimes {} \\
    %      & \qquad\graytimes
    %     \frac{\G[\Dphi+J+n]^{2}}{\G[2(\Dphi+J+n)]}
    %     \frac{\G[4-2\Dphi-n] \G[2-\Dphi-n]^{2}}{(2\Dphi+J+2n-2)\G[2\Dphi+J+n-2]}
    %     \graytimes {} \\[0.3ex]
    %      & \mspace{-110mu}\graytimes
    %     \begin{pmatrix*}[c]
    %         \displaystyle
    %         \frac{\G[\Dpr+\widetildesmash[-0.12ex]{J'}]}{\G[\frac{\Dpr+\widetildesmash[-0.3ex]{J'}}{2}]^{2}}
    %         \HypGeo*[4][3]{-n,\; -n,\; 2-\Dphi-n,\; 2-\Dphi-n}{2(2-\Dphi-n),\; 1-\frac{\Dpr+\widetilde{J}'}{2}-n,\; \frac{\Dpr+\widetilde{J}'}{2}-n}
    %         \frac{\operatorname{\Omega}_{\Dphi+J+n, \frac{\Dpr+J'}{2}, \Dphi-1}}{\sin\left(\pi\left[-n-\frac{\Dpr-J'}{2}\right]\right)} \\[3.5ex]
    %         {} - \left(J' \xleftrightarrow{\text{exchange}} \widetilde{J}'\equiv -2-J'\right)
    %     \end{pmatrix*}
    %     \eqcomma
    %     \mspace{-20mu}
    % \end{aligned}
    \begin{aligned}
        \anomdim_{n,J} \eval_{\smash[b]{\substack{\text{\(t\)-channel} \\\text{exchange}\\\Dpr,\.J'}}}
         & \eqdim{4}
        \frac{2 \G[\Dphi]^{2} \G[\Dphi+J+n]^2}{\G[2(\Dphi+J+n)]}
        \frac{\sin^{2}\left(\pi\left[\Dphi-\frac{\Dpr-J'}{2}\right]\right)}{\pi^{2}}
        \frac{\G[J+n+1]}{\G[2\Dphi+J+n-1]}
        \graytimes {} \\
         & \mspace{180mu}\graytimes
        \frac{J+n+1}{J+1}\frac{2\Dphi+J+n-2}{2\Dphi+J+2n-2}
        \graytimes {} \\[-0.4ex]
         & \mspace{20mu}\graytimes
        \left[
            \frac{\G[\Dpr+\widetildesmash[-0.12ex]{J'}]}{\G[\frac{\Dpr+\widetildesmash[-0.3ex]{J'}}{2}]^2}
            \anomdimFfunction[d\.=\.4]{n}{\frac{\Dpr+\widetildesmash[-0.3ex]{J'}}{2}}
            \operatorname{\Omega}_{\Dphi+J+n, \frac{\Dpr+J'}{2}, \Dphi-1} {}
            - \left( J' \exchange \widetilde{J}'\equiv -2-J' \right)
        \right]
        \eqend
    \end{aligned}
\end{equation}

The function \(\Omega\) appearing in the above formulas is given in~\cite[(3.38)]{Liu:2018jhs}, which after simplification for equal external dimensions takes the form
% \begin{equation}
%     \label{eq:omega_function}
%     \begin{aligned}
%         \Omega_{h,h',p} =
%         \quad\frac{\G[2h]\G[h'-p+1]^{2}\G[h-h'+p-1]}{\G[h]^{2}\G[h+h'-p+1]}
%          & \HypGeo[4][3]{h',\; h', & h'-p+1, & h'-p+1}{2h', & h+h'-p+1, & h'-h-p+2} \\
%         {} + \frac{\G[2h']\G[h+p-1]^{2}\G[h'-h-p+1]}{\G[h']^{2}\G[h'+h+p-1]}
%          & \HypGeo[4][3]{h,\; h,   & h+p-1,  & h+p-1}{2h,   & h'+h+p-1, & h-h'+p} \eqend
%     \end{aligned}
% \end{equation}
\begin{equation}
    \label{eq:omega_function_simplified}
    \begin{aligned}
        \Omega_{h,h',p}
         & = \frac{\G[2h']\G[h+p-1]^{2}\G[h'-h-p+1]}{\G[h']^{2}\G[h'+h+p-1]} \HypGeo[4][3]{h,\; h, & h+p-1, & h+p-1}{2h, & h'+h+p-1, & h-h'+p} \\
         & \mspace{130mu} {} + \bigg( h \exchange h',\ p \exchange 2-p \bigg)
        \eqend
    \end{aligned}
\end{equation}
To further simplify the expressions, we defined \textcolor{darkgray}{(for \(d=2\) and \(d=4\))}
\begin{equation} \label{eq:Function_anom_dim_n_dependence}
    \begin{aligned}
        \anomdimFfunction[d]{n}{\alpha}
        \eqdim[4.5]{d\.\in\.\{2,4\}} {}
         & \frac{(-1)^{n}}{n!}
        \frac{\G[d-2\Dphi-n]}{\G[d-2\Dphi-2n]}
        \frac{\G[\frac{d}{2}-\Dphi-n]^2}{\G[\frac{d}{2}-\Dphi]^2}
        \frac{\G[\alpha+n]}{\G[\alpha-n]}
        \graytimes {} \\
         & \mspace{80mu}\graytimes
        \HypGeo[4][3]{-n,\; -n,\; \frac{d}{2}-\Dphi-n,\; \frac{d}{2}-\Dphi-n}{d-2\Dphi-2n,\;  1-\alpha-n,\;  \alpha-n}
        \eqend
    \end{aligned}
\end{equation}
Using \(\anomdimFfunction{0}{\alpha}=1\), we can quickly check that the original leading-twist \grayenclose{\(n=0\)} results are indeed reproduced.

Note that \(d=2\) formula \cite[(3.55)]{Liu:2018jhs} considers exchange of a \(t\)-channel operator with general quantum numbers \((h',\overline{h}')\), while we directly wrote the exchange of symmetric traceless tensor \(\mathsf{STT}_{J'}\) with spin \(J'=\abs*{h'-\overline{h}'}\), which for nonzero \(J'\) reduces as \(\mathsf{STT}_{J'} = (h',\overline{h}') \oplus (\overline{h}',h')\) \cite[Section 2]{Simmons-Duffin:2017nub}.

Interestingly, both formulas \eqref{eq:tchannel_anom_dim_2d}/\eqref{eq:tchannel_anom_dim_4d} resemble each other quite well with minor modifications.
Note the appearance of terms related by the \enquote{spin shadow} affine Weyl reflection given by \(\widetilde{J}'\equiv 2-d-J'\) \cite{Kravchuk:2018htv}, which is already present at the level of conformal blocks.
% \todo{Explicitly seen for \(d=2\) and \(d=6\) \cite{Poland:2018epd}, does this hold also for higher (even) \(d\)?}
It would be interesting to see if such similar structure persists for higher dimensions as well, or perhaps whether a simple master formula for general \grayenclose{even} dimensions can be derived.

\paragraph{Summary.}%
\label{par:Summary.}

Now we come back to the case of the interaction term having a nontrivial \(t\)-channel conformal block decomposition.
The overall contribution to the anomalous dimensions of double-twist operators is then simply given by a sum of contributions for each appearing conformal block --- in \(d=2\)/\(d=4\) given by \eqref{eq:tchannel_anom_dim_2d}/\eqref{eq:tchannel_anom_dim_4d} --- weighted by the corresponding \grayenclose{squared OPE} coefficients, that is
\begin{align}
    \label{eq:anomalous_dimensions_general_formula}
    \anomdim[]_{n,J} = \!\sum_{\text{primary } \Oprstar} \OPEsq_{\Oprstar}^{\text{\color{darkgray}int}} \,
    \anomdim_{n,J} \eval_{\substack{\text{\(t\)-channel}\mspace{48mu} \\[0.1ex] \text{exchange of } \smash{\Oprstar}}}
    \eqend
\end{align}
The \(1/N\) factor associated with interaction/exchange diagram --- see \eqref{eq:largeNfourPt} --- is included in the weighting coefficients \(\OPEsq^{\text{\color{darkgray}int}}\), so the anomalous dimensions \(\anomdim[]_{n,J}\) are of order \(\bigO{1/N}\).

Note the appearance of the \(\sin(\pi[{}\cdots{}])\) factor in both of the formulas \eqref{eq:tchannel_anom_dim_2d}/\eqref{eq:tchannel_anom_dim_4d}.
They have therefore zeros at the MFT dimensions \(\Dpr=2\Dphi+2n'+J', n'\in \N_{0}\), so only non-MFT operators in the crossed channels contribute to the anomalous dimensions.
This directly reflects the property of \(\CrK[]\) mentioned at the end of \Cref{sub:CPW completeness and 6j-symbol}.

The only \emph{theory-specific} information is thus dimensions and spins of non-MFT operators together with corresponding squared OPE coefficients, or equivalently the poles and residues of \(t\)-channel spectral function of \(t\)-channel exchange.
Conformal symmetry then dictates the form of the anomalous dimensions through the structure of \(\sixjsymbol\)/\(\CrK[]\).

\paragraph{Large Spin Asymptotics.}%
\label{par:General Large Spin Asymptotics.}

Finally, we will briefly discuss the large spin asymptotics of the anomalous dimensions, and in particular their \(n\)-dependence.
More details \grayenclose{with references} can be found in \Cref{sub:Large spin asymptotics}.

One can check that large \(J\) asymptotics of the contributions \eqref{eq:tchannel_anom_dim_2d}/\eqref{eq:tchannel_anom_dim_4d} to the anomalous dimensions are given by
\begin{equation} \label{eq:regge_asymptotics}
    \anomdim_{n,J} \eval_{\smash[b]{\substack{\text{\(t\)-channel} \\\text{exchange} \\\Dpr,\.J'}}}
    \sim J^{-\tau'}
    \equiv J^{-(\Dpr-J')}
    \quad\Longrightarrow\quad
    \anomdim[]_{n,J} \simeq - \frac{c_{n}}{J^{\twistmin}} \mathcolor{gray}{+ \ldots}
    \eqcomma
\end{equation}
so the leading asymptotics of anomalous dimensions \eqref{eq:anomalous_dimensions_general_formula} are governed by the exchanged operator with the lowest twist \(\twistmin=\min \tau' \equiv \min(\Dpr-J')\).

The point we want to make here is that whatever the coefficient \(c_{0}\) for the leading-twist operators is --- for explicit formula see \eqref{eq:large_spin_c_coeff}, which is actually independent of the dimension --- inspection of \eqref{eq:tchannel_anom_dim_2d}/\eqref{eq:tchannel_anom_dim_4d} leads to
\begin{equation} \label{eq:large_spin_c_coeff_general}
    \begin{aligned}
        c_{n} & \eqdim[4.5]{d\.\in\.\{2,4\}}
        \anomdimFfunction{n}{\frac{\Dpr+\smash{J'}}{2}}
        c_{0}
        \eqcomma
    \end{aligned}
\end{equation}
where we take the operator with \((\Dpr,J')\) to be the one with the lowest twist, and \(\anomdimFfunction{n}{\alpha}\) was defined in \eqref{eq:Function_anom_dim_n_dependence}.

This follows from the fact that the \enquote{spin-shadow} terms --- with \(\operatorname{\Omega}_{\Dphi+J+n, \frac{\Dpr-J'}{2}, \Dphi}\) in \(d=2\) and \(\operatorname{\Omega}_{\Dphi+J+n, \frac{\Dpr-J'}{2}-1, \Dphi-1}\) in \(d=4\) --- are dominant for large \(J\).

Since rest of their prefactors are the same for \(n=0\), the dimension-independence of \(c_{0}\) results from their identical asymptotics.
Going now to nonzero \(n\), the modifications are:
\begin{enumerate}[(i)]
    \item in \(d=4\) we have extra rational factors which go to \(1\) for \(J\to \infty\),
    \item there is extra factor of \(\anomdimFfunction{n}{\frac{\Dpr+\smash{J'}}{2}}\) before the \enquote{spin-shadow} dominant term,
    \item in the rest of formula we replace \(J\mapsto  J+n\), which does not change the asymptotics of the form \(J^{-\tau'}\).
\end{enumerate}
The conclusion is that only (ii) changes the asymptotics, leading to \eqref{eq:large_spin_c_coeff_general}.

Just to be clear, we have defined \(\anomdimFfunction{n}{\alpha}\) in \eqref{eq:Function_anom_dim_n_dependence} only for \(d\in \{2,4\}\) by bundling similar factors appearing in both dimensions.
Whether such elegant structure for general-twist anomalous dimensions persists also for higher dimensions is an open question.
% \todo{
%     Should check if the formula matches with \cite[(2.49)]{Sleight:2018ryu}, for simplicity just \(n\leq 2\).
%     We could then claim that we provided a closed-form expressions.
%     Initial checks seem to show some inconsistencies (their formulas do not even lead to right large J asymptotics).
% }

\section{Spectrum of the \texorpdfstring{\(\OO(N)\)}{O(N)} model in \texorpdfstring{\AdS{}}{AdS}}
\label{sec:spectrum-ONinAdS}

Naturally, as the first step, in \Cref{sub:ON_irrep_decomposition} we decompose the \(4\)-point boundary correlator into irreducible representations of the global symmetry group \(\OO(N)\).
This effectively solves the \(\OO(N)\) index structure, and enables us to focus on each irrep separately.

The singlet spectrum was investigated in~\cite{Carmi:2018qzm}.
In \Cref{sec:singlet_spectrum} we summarize their main results and also extend them in some ways --- in addition to \(d=2\) we consider also \(d=4\), where a new operator can appear at strong enough coupling.
We showcase the coupling dependence of both the anomalous dimensions and OPE coefficients, and also discuss the striking pattern of the spectrum when the bulk is tuned to the criticality.
The role of singlet spectrum does not end here, since it provides crucial input for the non-singlet spectrum via methods presented in \Cref{sub:CPW completeness and 6j-symbol} and \Cref{sub:tChannelContributiontoAnomalousDimensions}.
We treat their specific implementation for non-singlet spectrum of \(\OO(N)\) model in \Cref{sub:non-singlet_spectrum}.

\subsection{Decomposition into \texorpdfstring{\(\OO(N)\)}{O(N)} irreducible representations}
\label{sub:ON_irrep_decomposition}

We want to study various operators appearing in the \(\Ophi^{\ind}\times \Ophi^{\ind}\) OPE, where \(\Ophi^{\ind}\) is the \(\CFT_{d}\) operator dual to the elementary field \(\phi^{\ind}\) in the \(\EAdS_{d+1}\) bulk.
It is well known that primaries appearing in the MFT limit (in our case \(N\to \infty\)) --- apart from the identity operator --- are the double-twist operators schematically given by \(\Oper^{\ind\ind}_{n,J}=[\Ophi^{\ind}\Ophi^{\ind}]_{n,J} \equiv [\Ophi^{\ind}\dAlembertian^{n}\partial^{J}\Ophi^{\ind}-\text{traces}]\).
They come from the crossed-channel identities, as was already mentioned in \Cref{sub:tChannelContributiontoAnomalousDimensions}.

In order to alleviate the struggle of carrying the \(\OO(N)\) indices around, we will organize the \CFT{} operators by their \(\OO(N)\) irreps, and correspondingly decompose the \(4\)-point correlator.
Without loss of generality, we choose the \(s\)-channel as the one in which we perform both the OPE and the \(\OO(N)\)-irrep decomposition.

\paragraph{Double-twist operators.}%
\label{par:Double-twist operators}

Since each \(\Ophi^{\ind}\) transforms in the vector representation \(V\) of \(\OO(N)\), the \(\Ophi^{\ind}\times \Ophi^{\ind}\) OPE decomposition under \(\OO(N)\) follows from the standard
\begin{align}
    \label{eq:ON_irrep_decomp_VxV}
    V\otimes V = \,\underbrace{\bm{1}}_{\Singlet}\, \oplus \,\underbrace{\Ext^{2}V}_{\AntiSym}\, \oplus \,\underbrace{\Odot^{2}V}_{\SymTrless}\, \eqcomma
\end{align}
where the \(\OO(N)\) irreps appearing are the singlet \(\Singlet\), the anti-symmetric \(\AntiSym\), and the symmetric traceless \(\SymTrless\) representation.
The corresponding projectors are given by
\begin{equation}
    \label{eq:projectors}
    % \PSinglet^{ijkl} = \frac{1}{N}\delta^{ij}\delta^{kl} \qquad
    % \PAntiSym^{ijkl} = \frac{1}{2}\left( \delta^{il}\delta^{jk}-\delta^{ik}\delta^{jl} \right) \qquad
    % \PSymTrless^{ijkl} = \frac{1}{2}\left( \delta^{il}\delta^{jk}+\delta^{ik}\delta^{jl}-\frac{2}{N}\delta^{ij}\delta^{kl} \right) \eqcomma
    \left(\PSinglet   \right)^{ij}_{kl} = \frac{1}{N}\delta^{ij}\delta_{kl} \eqcomma\qquad
    \left(\PAntiSym   \right)^{ij}_{kl} = \delta^{[i}_{[k}\delta^{j]}_{l]} \eqcomma\qquad
    \left(\PSymTrless \right)^{ij}_{kl} = \delta^{(i}_{(k}\delta^{j)}_{l)}-\frac{1}{N}\delta^{ij}\delta_{kl} \eqend
\end{equation}
Thus, the double-twist operators can be organized into these three \(\OO(N)\) irreps, and are schematically given as
\begin{equation}
    \label{eq:types_of_operators}
    \Oper^{\ind\ind}_{n,J} \;\cdots\mspace{-3mu}\left\{
    \begin{alignedat}{3}
        \Singlet\; \quad\  & \Oper^{\Singlet}_{n,J} &  & \equiv  {} & \PSinglet[[\Ophi^{\ind}\Ophi^{\ind}]_{n,J}] & = \frac{1}{N}\sum_{i} [\Ophi^{i}\Ophi^{i}]_{n,J} \eqcomma \xmathstrut[-1]{0.65} \\
        \AntiSym   \quad\  & \Oper^{[ij]}_{n,J}     &  & \equiv  {} & \PAntiSym[[\Ophi^{i}\Ophi^{j}]_{n,J}]       & = [\Ophi^{[i}\Ophi^{j]}]_{n,J} \eqcomma                                         \\
        \SymTrless \quad\  & \Oper^{\{ij\}}_{n,J}   &  & \equiv  {} & \PSymTrless[[\Ophi^{i}\Ophi^{j}]_{n,J}]     & = [\Ophi^{(i}\Ophi^{j)}]_{n,J}
        - \smash[b]{\frac{1}{N}\delta^{ij}\sum_{k}[\Ophi^{k}\Ophi^{k}]_{n,J}} \xmathstrut[0.75]{0.79} \eqend
    \end{alignedat}
    \right.
\end{equation}

\paragraph{Decomposition of the \texorpdfstring{\(4\)}{4}-point correlator.}%
\label{par:Decomposition 4-point correlator}

From the form of the \(\sigma\phi^{2}\) interaction vertex \eqref{eq:ON_HS_rules} that couples only two identical \(\phi^{\ind}\) fields, the correlator \eqref{eq:4pt_wd} clearly takes the form
\begin{equation}
    \label{eq:4pt_tens_str}
    \begin{aligned}
        \FourPt*
        = \delta^{ij}\delta^{kl}\left( \sDisc+\frac{1}{N}\sExch \right)
         & + \delta^{ik}\delta^{jl}\left( \tDisc+\frac{1}{N}\tExch \right) \\
         & \mspace{-100mu}+ \delta^{il}\delta^{jk}\left( \uDisc+\frac{1}{N}\uExch \right) + \biggO{\frac{1}{N^2}}
        \eqend
    \end{aligned}
\end{equation}
Diagrams on the right-hand side without indices are meant to be evaluated by taking any (but fixed) field \(\phi^{i}\) on all external legs.
We have thus decoupled the global group-theoretic index structure of the correlator from the dynamics carried by exchange of the \(\sigma\)-field.

By irreducibility (Schur's lemma or equivalently the fact that a product of non-equal projectors vanishes), only matching \(\OO(N)\) irreps on both sides of the \(s\)-channel decomposition can combine to contribute nontrivially to the \(4\)-point correlator.
Therefore, its \(\OO(N)\)-irrep decomposition reads%
\vspace*{-0.8ex}
\begin{equation}
    \label{eq:4pt_tens_irrep}
    \FourPt* = \color{black!40}\overbrace{\color{black}N\PSinglet^{ijkl}}^{\mathcolor{black!70}{\delta^{ij}\delta^{kl}}}\color{black} \underbrace{\FourPt[\Singlet*]}_{\ASinglet}
    {} + \PAntiSym^{ijkl} \underbrace{\FourPt[\AntiSym*]}_{\AAntiSym}
    {} + \PSymTrless^{ijkl} \underbrace{\FourPt[\SymTrless*]}_{\ASymTrless}
    \eqcomma
    \vspace*{-0.2ex}
\end{equation}
where indices of projectors \eqref{eq:projectors} were raised using the Kronecker delta \(\delta^{\ind\ind}\).
Together with the singlet projector \(\PSinglet\) we introduced an explicit factor of \(N\), such that together they are of order \(\bigO{1}\), same as non-singlet projectors.

Solving~\eqref{eq:4pt_tens_str} and~\eqref{eq:4pt_tens_irrep} for \(\ASinglet\), \(\AAntiSym\) and \(\ASymTrless\) yields
\begin{alignat}{2}
    \FourPt[\Singlet*]   & = \left( \sDisc \right)        + \frac{1}{N}\left( \tDisc+\uDisc+\sExch \right) && +\biggO{\frac{1}{N^{2}}} \label{eq:4-pt_singlet}   \eqcomma \\
    \FourPt[\AntiSym*]   & = \left( \tDisc-\uDisc \right) + \frac{1}{N}\left( \tExch-\uExch \right)        && +\biggO{\frac{1}{N^{2}}} \label{eq:4-pt_antisymmetric} \eqcomma \\
    \FourPt[\SymTrless*] & = \left( \tDisc+\uDisc \right) + \frac{1}{N}\left( \tExch+\uExch \right)        && +\biggO{\frac{1}{N^{2}}} \label{eq:4-pt_symmetric_traceless}   \eqcomma
\end{alignat}
which represent the three projections of the correlator onto \(\OO(N)\) irreps.
Since the whole correlator is itself of order \(\bigO{1}\), their leading order is also \(\bigO{1}\).
Next are then \(\bigO{1/N}\) corrections --- mainly coming from the interactions --- which we are about to study.
As we know, the boundary \CFT{} spectrum decomposes into the same \(\OO(N)\) irreps, and each sector can be analyzed by its associated projection of the correlator.

Authors of \cite{Carmi:2018qzm} paid thorough attention to the singlet spectrum encoded in the singlet projection \(\ASinglet = \eqref{eq:4-pt_singlet}\).
The goal of this paper is to supplement their efforts by the analysis of the remaining rank-2 non-singlet sectors governed by \(\AAntiSym = \eqref{eq:4-pt_antisymmetric}\) and \(\ASymTrless = \eqref{eq:4-pt_symmetric_traceless}\).

However, before embarking on this journey, we need to recall the results of singlet sector, as it will serve as an input for computing the anomalous dimensions for the remaining two \(\OO(N)\) irreps via crossing relations discussed in \Cref{sub:tChannelContributiontoAnomalousDimensions}.

\subsection{Singlet spectrum}
\label{sec:singlet_spectrum}

The leading \(\bigO{1}\) term in singlet sector \eqref{eq:4-pt_singlet} is just the \(s\)-channel identity contribution --- two identical scalar operators \(\Ophi^{i}\Ophi^{i}\) always contain the identity operator \(\Id\) in their OPE.

Substantially more interesting is the subleading \(\bigO{1/N}\) term, where the \(t\)-channel and \(u\)-channel disconnected diagrams meet together with the \(s\)-channel exchange diagram.
To understand this correction to the MFT picture, we need to analyze the spectral function
\begin{align}
    \label{eq:spec_sing}
    \Spec[s][]{\tDisc+\uDisc+\sExch} \eqend
\end{align}

\paragraph{Disconnected contributions.}%
\label{par:Disconnected contributions.}

The first two disconnected Witten diagrams correspond in the boundary \(\CFT\) to GFF/MFT contributions, which were already showcased in \eqref{eq:spec_disc_t}.
Slightly more generally (for even/odd combination of \(t\)-channel and \(u\)-channel) we have
\begin{equation}
    \label{eq:spec_disc_tu}
    \begin{aligned}
        \Spec[s][\D][J]{\tDisc\pm\uDisc}
         & = \left(1\pm(-1)^J\right)
        \bigg( \eqref{eq:spec_disc_t} \bigg)
        \eqcomma
    \end{aligned}
\end{equation}
with the two terms differing just by the overall sign \((-1)^J\), see the beginning of \Cref{sec:CFTgeneralities}.

In the \(\bigO{1/N}\) singlet sector spectral function \eqref{eq:spec_sing} we encounter the even combination of \eqref{eq:spec_disc_tu}, so only even \(J\) survive.
Thus, the \(\bigO{1/N}\) disconnected part of the singlet sector generates poles at the MFT dimensions \(\D[\MFT]_{n,J}=2\Dphi+2n+J\) for even spins \(J\), and the associated squared OPE coefficients are \(2/N\) times the expression one would get just from the \(t\)-channel identity \eqref{eq:spec_disc_t}.
Since the squared OPE coefficients have a factor of \(1/N\), the OPE coefficients themselves are of order \(\bigO*{1/\sqrt{N}}\).

The conformal block decomposition of the disconnected contributions to \eqref{eq:4-pt_singlet} is therefore
\begin{equation}
    \label{eq:singlet_disc_CB_decomposition}
    \frac{1}{N}\left( \tDisc+\uDisc \right) = \sum_{n=0}^{\infty}\sum_{\substack{J=0\\J\text{ even}}}^{\infty}
    \frac{2}{N}\OPEsq[\MFT*]_{n,J} \ket{\ConfBlock[2\Dphi+2n+J,\.J]} \eqcomma
\end{equation}
where \(\OPEsq[\MFT*]_{n,J}\) are MFT squared OPE coefficients coming from the \(t\)-channel identity.

\paragraph{Exchange of \(\sigma\)-field.}%
\label{par:Exchange of sigma-field.}

The next contribution in~\eqref{eq:spec_sing} comes from the \(s\)-channel exchange diagram, whose calculation was schematically outlined in \eqref{eq:exchange_diagram_computation}.
The corresponding spectral function --- including all numeric and \(\G\)-factors --- is obtained by comparing the result of \cite[(4.18)]{Carmi:2018qzm} with our convention for the integral CB decomposition \eqref{eq:CB_integral_decomposition}, yielding
\begin{align}
    \label{eq:spec_s_channel_exchange}
    \Spec[s][\D][J]{\sExch}
    = -\delta_{J,0}\.
    \frac{1}{\lambda^{-1}+2\Bubble(\D)}
    \frac{
        \G[\Dphi-\frac{\D}{2}]^2 \G[\Dphi-\frac{\D*}{2}]^2 \G[\frac{\D}{2}]^{\raisemath{0.15ex}{2}}\G[\frac{\widetildesmash[-0.2ex]{\Delta}}{2}]^{\raisemath{0.35ex}{2}}}{
        4\pi^{d}\G[\Dphi]^2 \G[1-\frac{d}{2}+\Dphi]^2 \G[\D-\frac{d}{2}] \G[\D*-\frac{d}{2}]}
    \eqend
\end{align}
Compared to \cite{Carmi:2018qzm}, we use \(\Dphi\) for external scaling dimension instead of theirs \(\D\), and we usually prefer writing formulas directly in terms of \(\D\equiv \frac{d}{2} + \I\nu\) instead of \(\nu\).
We easily see that \eqref{eq:spec_s_channel_exchange} is shadow-symmetric --- \(\D \exchange \D*\) --- provided that the bubble function \(\Bubble\) is shadow-symmetric as well, which holds generally for spectral representations of functions.

The first thing to notice is that \eqref{eq:spec_s_channel_exchange} has support only on spin \(J=0\) operators.
Already at this point we can see that \(J>0\) operators in the singlet sector have zero anomalous dimensions --- a statement valid to the \(\bigO{1}\) order actually considered, since the squared OPE coefficients already have a factor of \(1/N\), and we do not consider \(\bigO{1/N^{2}}\) corrections.
Concluding this observation, to the order \(\bigO{1/N}\), the singlet \(J>0\) operators appearing in the \(\Ophi^{\ind}\times \Ophi^{\ind}\) OPE are \(\Oper[\Singlet]_{n,J}\) with even \(J\), MFT dimensions \(\D[\Singlet]_{n,J>0} = \D[\MFT]_{n,J}\equiv 2\Dphi+2n+J\), and squared OPE coefficients as in \eqref{eq:singlet_disc_CB_decomposition}.

\paragraph{Bootstrap idea --- consistency of the spectrum.}%
\label{par:Bootstrap idea --- consistency of the spectrum.}

The case of \(J=0\) is more intricate.
The factor of \(\G^{2}(\Dphi-\frac{\D}{2})\) generates a set of double-poles at the MFT dimensions \(\D=2\Dphi+2n\), which is something one would expect from perturbative corrections to the scaling dimensions of \(\Oper[\Singlet]_{n,0}\) from the interaction.
Indeed, expanding \eqref{eq:spec_s_channel_exchange} in \(\lambda\), and comparing it to the perturbative expansion of poles in \(\SpecNoArgs[s]\) --- as in \eqref{eq:poleinSpecTaylored}, where instead of \(1/N\) we now consider series in \(\lambda\) --- one sees that \(\bigO{\lambda}\) term coming from single insertion of \(\lambda\phi^{4}\) vertex has double poles and generates \(\bigO{\lambda}\) anomalous dimensions for \(\Oper[\Singlet]_{n,0}\).

Furthermore, at \(\bigO*{\lambda^{k}}\) we would expect appearance of \((k+1)\)--degree poles at MFT dimensions.
Such a behavior can come only from higher and higher powers of bubble function \(\Bubble\) in the \(\lambda\)-expansion of \eqref{eq:spec_s_channel_exchange}.
It is thus plausible to anticipate that \(\Bubble\) has poles at MFT dimensions, such that diagrams \eqref{eq:bubble_resum} with more bubbles contribute to anomalous dimensions at corresponding orders of \(\lambda\).
At the same time, if \(\Bubble\) happened to have additional poles at non-MFT dimensions, it would indicate appearance of new operators already at the \(\bigO{\lambda^{2}}\) order.
Such operator spectrum changes are generally not expected to happen perturbatively for weak coupling, so we can conclude that the \(\Bubble\) function should have poles only at MFT dimensions.

However, in the resummed situation --- that is finite \(\lambda\) --- in addition to MFT double-poles still coming from \(\G^{2}(\Dphi-\frac{\D}{2})\) there appear poles coming from zeros of the \(\lambda^{-1}+2\Bubble\) denominator.
These poles depend continuously on the coupling \(\lambda\), so in order for this to be consistent --- without sudden appearance of a whole bunch of new operators once we turn on the coupling --- the remnants of the MFT poles must be canceled between the disconnected and exchange diagrams.
For such cancellation, the \(\G^{2}\) double-poles must be accompanied by simple zeros of the \((\lambda^{-1}+2\Bubble)^{-1}\), that is simple poles of \(\Bubble\) at MFT dimensions, as was already anticipated.
Such arguments allowed the authors of \cite{Carmi:2018qzm} to \enquote{bootstrap} the bubble function \(\Bubble\) (or more precisely its spectral representation).

\paragraph{Bubble function.}%
\label{par:Bubble function.}

This requirement of precise cancellation of the MFT poles --- together with the nice behavior at infinity for low enough dimension \(d\) --- leads to the following sum
over the \(\Delta\) poles \cite[(4.26),(4.27)]{Carmi:2018qzm}
\begin{equation}
    \label{eq:B_fn}
    \begin{aligned}
        \Bubble(\D)
         & = \frac{1}{4(4\pi)^{\frac{d}{2}}}
        \sum_{n=0}^{\infty} \frac{1}{\Dphi-\frac{\D}{2}+n}
        \frac{\G[\frac{d}{2}+n] \G[\Dphi+n]             \G[\Dphi-\frac{d}{2}+\frac{1}{2}+n] \G[2\Dphi-\frac{d}{2}+n]}
        {     \G[\frac{d}{2}]   \G[\Dphi+\frac{1}{2}+n] \G[\Dphi-\frac{d}{2}+1+n]           \G[2\Dphi-d+1+n]}
        \frac{1}{n!} + \mathrlap{\bigg( \D \exchange[\mspace{5mu}] \D* \bigg)} \\[1.2ex]
         & =\frac{\G[\Dphi]\G[\Dphi-\frac{d}{2}+\frac{1}{2}]\G[2\Dphi-\frac{d}{2}]}{4(4\pi)^{\frac{d}{2}}} \graytimes {} \\[-0.6ex]
         & \quad\ \graytimes
        \begin{pmatrix*}[c]
            \G[\Dphi-\frac{\D}{2}]
            \HypGeo*[5][4]{
                &\Dphi-\frac{\D}{2},   &\frac{d}{2},\mspace{15mu} \Dphi,\; &\Dphi-\frac{d}{2}+\frac{1}{2}, &2\Dphi-\frac{d}{2}}{
                &\Dphi-\frac{\D}{2}+1, & \Dphi+\frac{1}{2},  &\Dphi-\frac{d}{2}+1,           &2\Dphi-d+1} \\[3ex]
            {} + \left(\D \xleftrightarrow{\text{exchange}} \D*\right)
        \end{pmatrix*} \eqcomma
    \end{aligned}
\end{equation}
where after extracting some \(\G\)-factors in front we recognized the \emph{regularized} generalized hypergeometric function \(\HypGeoNoArgs*[5][4]\).
The bubble function \(\Bubble(\D)\) --- even for unequal masses --- has been previously computed also by different methods in~\cite{Bros:2011vh} (see also~\cite{Cacciatori:2024zbe}).

The summands in \eqref{eq:B_fn} behave asymptotically as \(n^{d-4}\) for \(n\to \infty\), so the sum becomes divergent for \(d+1\ge 4\).
Same as in flat space, this corresponds to the UV divergence of the bubble diagram, and one must regularize it somehow.
By subtracting sufficient number of terms (up to and including degree \(d-3\)) in the Taylor expansion of the summands around \(\nu=0 \Leftrightarrow \D=\frac{d}{2}\), the series becomes convergent and can be summed up in principle.
Note that only even powers of \(\nu \equiv -\I(\D-\frac{d}{2})\) are present, since \eqref{eq:B_fn} is shadow-symmetric.
At the end, to account for the subtraction, an even polynomial in \(\nu\) of the corresponding degree with arbitrary coefficients (in principle depending on \(\Dphi\)) must be added back to the regularized sum.
This was also discussed for the spin \(1\) bubble function in \cite[Section 3.3]{Ankur:2023lum}.

The sum simplifies in even dimensions (see \notebook{} for details).
For example, taking \(d=2\), the bubble function does not need any regularization, and it evaluates to
\begin{equation}
    \label{eq:B2d}
    \Bubble(\D) \eqdim{2}
    \frac{\I}{4 \pi} \frac{\digamma(\Dphi - \frac{1+\I\nu}{2}) - \digamma(\Dphi - \frac{1-\I\nu}{2})}{2\nu}
    \equiv \frac{1}{4 \pi} \frac{\digamma[\Dphi - \frac{\D}{2}] - \digamma[\Dphi - \frac{\D*}{2}]}{\D*\xmathstrut[-1]{0.3} - \D} \eqcomma
\end{equation}
where \(\digamma[z]\equiv \digamma(z)\equiv \dv{z}\ln{\G(z)}\) is the digamma function.

In the case of \(d=4\), the bubble function is UV divergent and requires a regularization as we described above.
Due to the shadow-symmetry, it is enough to subtract the constant term in the expansion of the summands around \(\D=\frac{d}{2}\), and the result is
\begin{equation}
    \label{eq:B4d}
    \Bubble(\D) \eqdim{4}
    \frac{
        \nu\left[
            2(2\Dphi - 5)\nu
            - \I\Big(4(\Dphi-2)^{2}+\nu^{2}\Big) \left(
                \digamma[\Dphi-\frac{\D}{2}]-\digamma[\Dphi-\frac{\D*}{2}\xmathstrut[0.4]{-1}]
            \right)
        \right]
    }{128\pi^{2}\left(1+\nu^{2}\right)}
    + a_{0}(\Dphi)
    % +\left(-\frac{1}{16 \pi^{2}}\right)
    \eqend
    %%% following has specific \D=d/2<->\nu=0 piece (dependent on \Dphi), such that \lambda=\infty is bulk-conformal; with \digamma
    % \frac{
    %     4 + 6 \left( \frac{\nu}{2} \right)^{2} - \Dphi \left(2 + \nu^{2}\right)
    %     - \I\nu\left((\Dphi-2)^{2}+\left( \frac{\nu}{2} \right)^{2}\right) \left(
    %         \digamma[\Dphi-\frac{\D}{2}]-\digamma[\Dphi-\frac{\D*}{2}]
    %     \right)}{
    %     32\pi^{2}\left(1+\nu^{2}\right)} \eqend
    %%% following is subtracted such that \D=d/2<->\nu=0 piece is 0
    % \frac{\D*-\D}{64 \pi^{2}} \frac{(\D-\D*)(2\Dphi-5) + 4\left( 1-\Dphi+\frac{\D}{2} \right)\left( 1-\Dphi+\frac{\D*}{2} \right) \left( \digamma(\Dphi-\frac{\D}{2}) - \digamma(\Dphi-\frac{\D*}{2})  \right) }{4-\left( \D-\D* \right)^{2}}
\end{equation}

There are no clear physical constraints which would enable us to fix the subtraction ambiguity in the calculation of the regularized bubble function, in the case of \(d=4\) being just the undetermined constant \(a_{0}\).
This can be seen from the fact that \(\lambda\) itself is not renormalization scheme independent, and only the combination \(\lambda^{-1}+2\Bubble\) in \eqref{eq:spec_s_channel_exchange} has invariant meaning directly connected with the shape of the physical spectrum.

In the following we will set \(a_{0}\equiv -\frac{1}{16\pi^{2}}\).
In \Cref{sub:Criticality in the bulk} we will discuss the rationale behind this choice.
Choosing different \(a_{0}\) can always be reabsorbed in the coupling \(\lambda\).

One question still remains --- what is the range of physically admissible values of \(\lambda\)?
Without a more detailed analysis of the phase structure in \(d=4\) we are unable to provide a definitive answer, and will assume that any non-negative value of \(\lambda\) is allowed.

% \paragraph{Generalization to \(1/N\) correction.}
% \todo{
% Can we somehow comment on \cite[Subsection 4.1.2]{Carmi:2018qzm} with the same title?
% We actually computed (although not entirely analytically) the \(s\)-channel CB decomposition of the \(t\)-channel exchange diagram.
% It again yields a collection of double-twist operators (unshifted MFT values of dimensions, different OPE coefficients).
% Physical assumption is however that they are not present in the spectrum, since presumably \(\bigO{1/N^{2}}\) \(s\)-channel diagram contains additional operators, and thus should cancel the MFT ones.
% But I don't see how this is enough to fix it also here without performing any diagrammatic computation.
% If this was possible, it would seem there is only one possible theory which at \(\bigO{1/N}\) matches the \(\OO(N)\) model, but I suspect there is more freedom (can add in Lagrangian anything at the appropriate order?).
% }

\subsection{Analysis of the singlet sector}%
\label{sub:Analysis of singlet sector}

Now that we presented all relevant formulas for the singlet sector, we can analyze it in more detail.
Since the \(J>0\) operators are unaffected by the \(s\)-channel exchange diagram, in the following we will focus solely on the \(J=0\) operators.

In particular, we can extract the scaling dimensions of the singlet operators contributing in the subleading \(\bigO{1/N}\) order from the poles of the spectral function~\eqref{eq:spec_sing}, and also the corresponding squared OPE coefficients from the residues of these poles.

\paragraph{Scaling dimensions in the singlet sector.}%
\label{par:Scaling dimensions in the singlet sector.}

As we already discussed, due to the interplay of the disconnected diagrams and the \(s\)-channel exchange, the \(J=0\) MFT poles \(\D[\MFT]_{n,0}\equiv 2\Dphi+2n\) are canceled in the complete spectral function \eqref{eq:spec_sing}.
Instead, they are replaced by their finite-shifted counterparts
\(\D[\Singlet]_{\dind,0}\), which are roots of
\begin{align}
    \label{eq:singlet_poles-bubble_roots}
    \lambda^{-1}+2\Bubble(\D[\Singlet]_{\dind,0})=0 \eqcomma
\end{align}
such that \eqref{eq:spec_s_channel_exchange} has a pole at the corresponding location.
We use \(\dind\) to indicate the place for indexing this new \grayenclose{infinite} family of singlet \(J=0\) non-MFT primary operators denoted suggestively as \(\Osigma_{\dind}\), since they are induced by the exchange of the \(\sigma\)-field.
As we will see soon, not all of them need to be continuously connected to the MFT operators.

For generic \(\lambda\) and \(\Dphi\), the key equation \eqref{eq:singlet_poles-bubble_roots} is transcendental, and its roots have to be found numerically.
Nevertheless, this can be easily done to a very high precision using standard numerical methods, for example with the help of \wmathematica{}.

\begin{figure}[!ht]
    \centering
    \begin{adjustbox}{center}
        \includegraphics[width=1.05\textwidth]{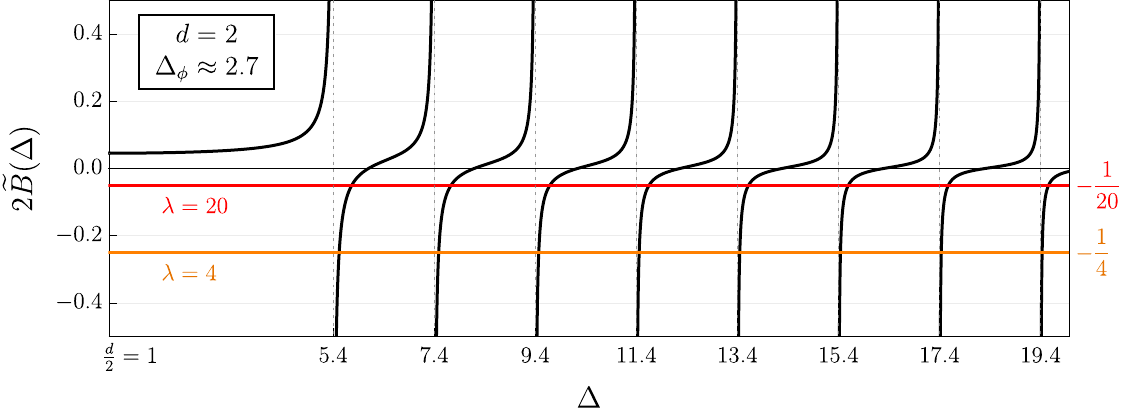}
    \end{adjustbox}
    \vspace*{-0.2ex}

    \begin{adjustbox}{center}
        \includegraphics[width=1.05\textwidth]{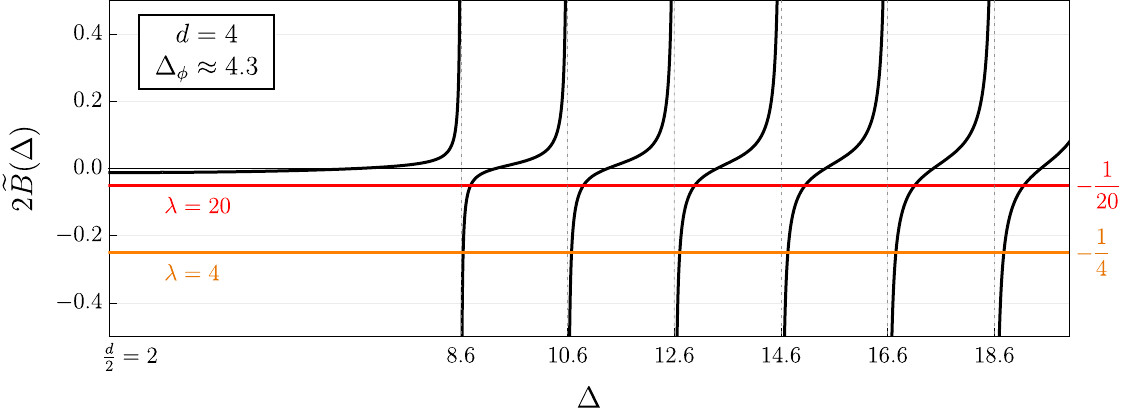}
    \end{adjustbox}
    \caption{
        Graphical representation of scalar singlet spectrum equation~\eqref{eq:singlet_poles-bubble_roots} in \(d=2\) and \(d=4\).
        The graph of (twice) the bubble function \(\Bubble\)~\eqref{eq:B2d}/\eqref{eq:B4d} is drawn by solid black lines.
        The term \(-\lambda^{-1}\) for \(\lambda=4\) and \(\lambda=20\) is represented by the orange and red lines, respectively.
        Their intersection points with black lines correspond to the \(J=0\) operators \(\Osigma_{\dind}\) in the singlet sector.
        Most of these operators (all in \(d=2\)) are associated with the MFT spectrum as can be seen by their asymptotic convergence to MFT values (gray dashed lines) for \(\lambda\to 0\).
        In the case of \(d=4\), there is however a potential emergent operator at strong enough coupling in the region \(\D\lesssim 2\Dphi\), where the graph of \(\Bubble\) is reaching slightly below the horizontal axis.
        The plot was evaluated at indicated external scaling dimensions, corresponding to a massive \(\phi\)-field with Dirichlet boundary conditions.
    }
    \label{fig:Bubble2dand4d}
\end{figure}

Examples of the bubble functions \(\Bubble\) in \(d=2\) and \(d=4\), together with some particular choices of \(\lambda\), are displayed in \Cref{fig:Bubble2dand4d}.
Intersection points in the plots correspond to singlet scalar operators \(\Osigma_{\dind}\).
At small \(\lambda\), we can clearly identify them as finite deformations of the MFT operators \(\Oper[\Singlet]_{n,0}\simeq [ \Ophi^{\ind}\dAlembertian^{n}\Ophi^{\ind} ]^{\Singlet} \).
With increasing \(\lambda\), the anomalous dimensions grow, and eventually become of order \(\bigO{1}\) in both \(\lambda\) and \(1/N\).

There is a subtlety in \(d=4\), where for a strong enough coupling \(\lambda\) a new operator possibly appears that is not continuously connected with the MFT spectrum.
Such an operator would be then associated to a bound state in \(\AdS\).
As we will see, its emergence is crucial for the bulk theory to be critical, which will be discussed in \Cref{sub:Criticality in the bulk}.

At large conformal dimensions \(\D\gg 1\), which concerns the operators \(\Oper[\Singlet]_{n,0}\) with \(n\gg 1\), the bubble functions have asymptotics \textcolor{darkgray}{(dots include subleading \(1/\D\) terms)}
\begin{equation} \label{eq:Bubble_large_dim_asymptotics}
    \Bubble(\D) \eqdim{2} \frac{\cot(\pi(\Dphi-\frac{\D}{2}))}{8\D} \mathcolor{darkgray}{+ \dots} \eqcomma \quad
    \Bubble(\D) \eqdim{4} \frac{\D\cot(\pi(\Dphi-\frac{\D}{2}))}{128\pi} \mathcolor{darkgray}{+ \dots} \eqend
\end{equation}
For finite \(\lambda\), the inspection of \eqref{eq:singlet_poles-bubble_roots} gives us following asymptotic values of the anomalous dimensions (governed by the infinities or zeros of \(\cot(\cdots)\) in \(d=2\) or \(d=4\), respectively)
\begin{equation}
    \label{eq:singlet_anom_dim_asymptotics}
    \mathcolor{darkgray}{(0<\lambda<\infty)} \quad
    \lim_{n \to \infty} \anomdim[\Singlet*]_{n,0} =
    \begin{cases}
        0 & \text{for } d=2 \eqcomma \\
        1 & \text{for } d=4 \eqend
    \end{cases}
\end{equation}
Of course, for \(\lambda=0\) all anomalous dimensions vanish, and for \(\lambda=\infty\) the different \(\D\) scaling in \eqref{eq:Bubble_large_dim_asymptotics} does not play a role, and we have \(\lim_{n \to \infty}\anomdim[\Singlet*]_{n,0} = 1\) for both \(d=2\) and \(d=4\).
% \Note{
%     Is there any possible (general) explanation for this different large conformal dimension behavior?
%     Perhaps something to do with relevancy/irrelevancy of \(\lambda\) coupling?
% }

\begin{figure}[!ht]
    \centering
    \begin{adjustbox}{center}
        \includegraphics[width=0.5\linewidth]{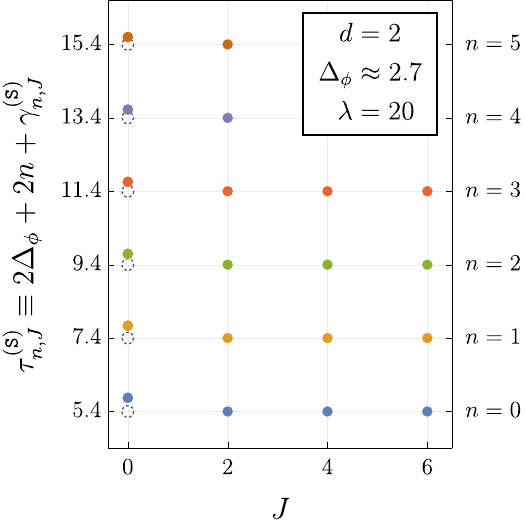}
        \hspace{0.03\linewidth}
        \includegraphics[width=0.5\linewidth]{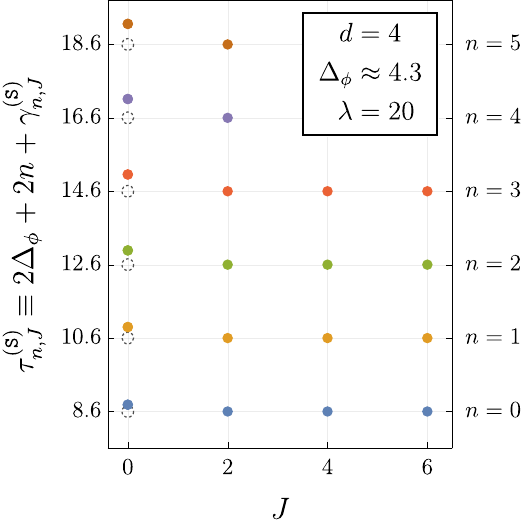}
    \end{adjustbox}
    \caption{
        Twist--spin plots of the singlet spectrum for \(d=2\) and \(d=4\) given by the poles of the complete spectral function~\eqref{eq:spec_sing}.
        Only spin \(J=0\) operators get \(\bigO{1}\) anomalous dimensions in large \(N\) expansion.
        The plots correspond to~\Cref{fig:Bubble2dand4d}, so the non-MFT scaling dimensions of scalar operators are precisely given by the intersection points of black and red lines in that figure.
    }
    \label{fig:singlet_sector_twists}
\end{figure}

In summary, the complete singlet spectrum given by the poles of the spectral function~\eqref{eq:spec_sing} consists of non-MFT scalar \((J=0)\) operators supplemented by MFT operators supported at even spins \(J\geq 2\).
The corresponding singlet twist--spin plot is displayed in~\Cref{fig:singlet_sector_twists}, with the twist being defined as \(\tau^{\Singlet}_{n,J}\equiv \D[\Singlet]_{n,J}-J\equiv 2\Dphi+2n+\anomdim[\Singlet*]_{n,J}\).

\begin{figure}[!ht]
    \centering
    \begin{adjustbox}{center}
        \includegraphics[width=1.0\linewidth]{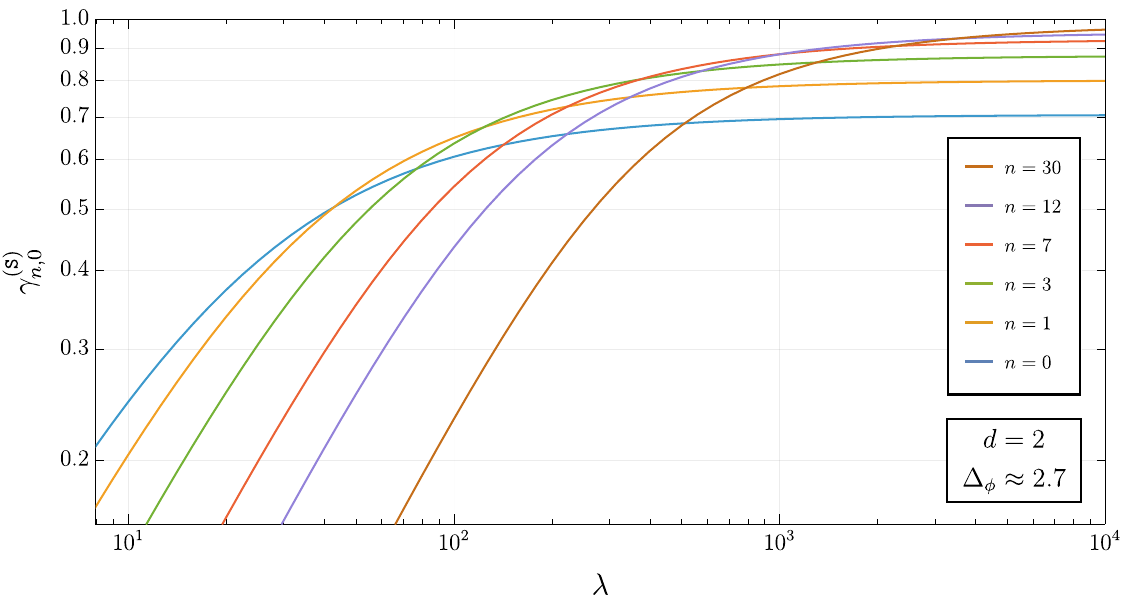}
    \end{adjustbox}
    \vspace*{0.8ex}

    \begin{adjustbox}{center}
        \includegraphics[width=1.0\linewidth]{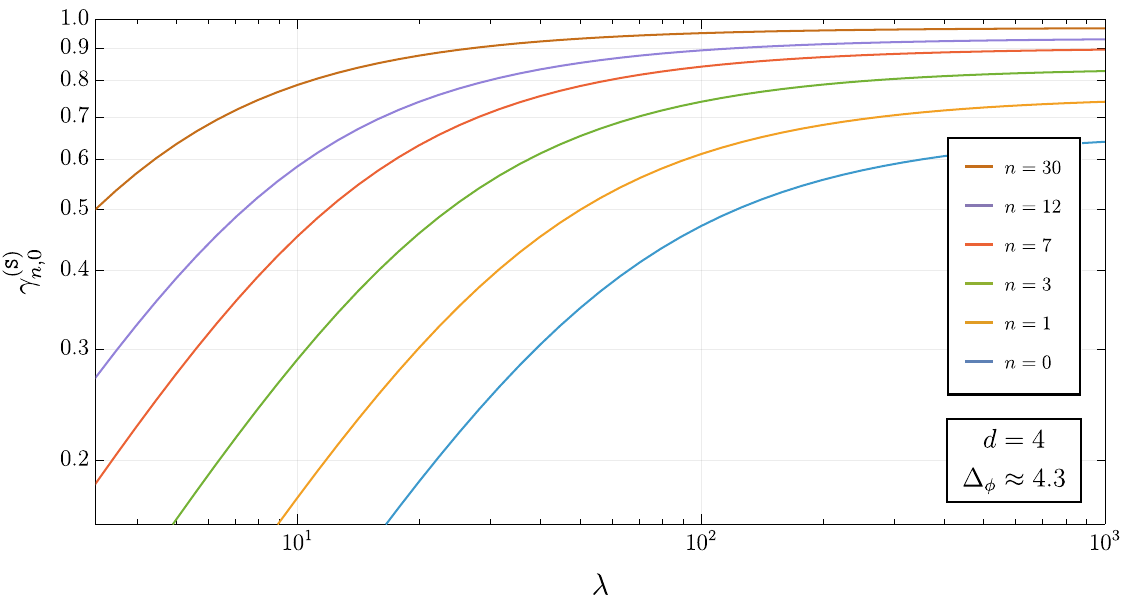}
    \end{adjustbox}
    \vspace*{-0.5ex}
    \caption{
        Coupling dependence of the singlet scalar anomalous dimensions at indicated fixed external scaling dimensions coinciding with previous plots.
        In the free limit \(\lambda=0\), they correspond to primary MFT operators of the schematic form \([\Ophi^{\ind}\dAlembertian^{n}\Ophi^{\ind}]^{\Singlet}\) (we do not include here the possible new operator in \(d=4\)).
        For any fixed \(n\), the anomalous dimensions in the singlet sector are all positive and approach a constant value (bounded above by \(1\)) at sufficiently strong coupling.
        Positiveness of anomalous dimensions indicates that the interaction in bulk \AdS{} has repulsive character in the singlet sector.
        Different behavior when increasing \(n\) for constant \(\lambda\) in \(d=2\) and \(d=4\) follows from the large conformal dimension asymptotics discussed in~\eqref{eq:singlet_anom_dim_asymptotics}.
    }
    \label{fig:singlet_anom_dim_on_coupling}
    \vspace*{0.2ex}
\end{figure}

The precise dependence of the singlet scalar anomalous dimensions on the coupling is plotted in \Cref{fig:singlet_anom_dim_on_coupling}.
The plot mainly focuses on the strong coupling region, and shows in what fashion the finite constant values of the anomalous dimensions are approached at infinite coupling.
The weak coupling regime is not entirely captured in these plots, but the asymptotics in that region are simple.
Anomalous dimensions are approximately linear in the coupling there, which follows from standard perturbation theory, namely the \(\phi^{4}\) contact Witten diagram contributes to the \(J=0\) anomalous dimensions at order \(\bigO{\lambda}\).

\paragraph{Squared OPE coefficients in the singlet sector.}%
\label{par:Squared OPE coefficients in the singlet sector.}

Having identified the locations of physical poles \(\D[\Singlet]_{\dind,0}\) in the spectral function~\eqref{eq:spec_s_channel_exchange}, we can now compute the squared OPE coefficients of the corresponding singlet operators \(\Oper[\Singlet]_{\dind,0}\).

Excluding the isolated case where \(\lambda\) is tuned precisely to the critical value when a new operator emerges at \(\D=\frac{d}{2}\), all physical poles at solutions of~\eqref{eq:singlet_poles-bubble_roots} are simple.
Since the rest of the remaining factors in~\eqref{eq:spec_s_channel_exchange} together with \(\Knorm[\D*,0]\) are holomorphic at these poles, the corresponding squared OPE coefficients are given by \eqref{eq:squaredOPEasRes} as
\begin{equation}
    \label{eq:ope_sq}
    \begin{alignedat}{2}
        \opesq{\PSinglet[\Ophi^{\ind}\Ophi^{\ind}] \Oper[\Singlet]_{\dind,0}}
         & = -\Res_{\D=\D[\Singlet]_{\dind,0}}\left(\Knorm[\D*\mathcolor{gray}{,0}]\frac{1}{N}\Spec[s]{\sExch}\right)
         &                                                                                                            & +\biggO{\frac{1}{N^{2}}} \\
         & =
        \left.
        \frac{1}{N} \frac{1}{2\Bubble*(\D)}
        \frac{
            \G[\Dphi-\frac{\D}{2}]^2 \G[\Dphi-\frac{\D*}{2}]^2 \G[\frac{\D}{2}]^4}{
            4\pi^{\frac{d}{2}}\G[\Dphi]^2 \G[1-\frac{d}{2}+\Dphi]^2 \G[\D-\frac{d}{2}] \G[\D]}
        \right\vert_{\mathrlap{\D=\D[\Singlet]_{\dind,0}}}
         &                                                                                                            & +\biggO{\frac{1}{N^{2}}}
        \eqcomma
    \end{alignedat}
    \setlength\belowdisplayskip{4pt}%
\end{equation}
where \(\Bubble* \equiv \dv{\Bubble}{\D}\) denotes the derivative of the bubble function.
Note that we included the factor of \(1/N\) with which the exchange diagram enters into the correlator.

\begin{figure}[!ht]
    \centering
    \begin{adjustbox}{center}
        \includegraphics[width=1.0\linewidth]{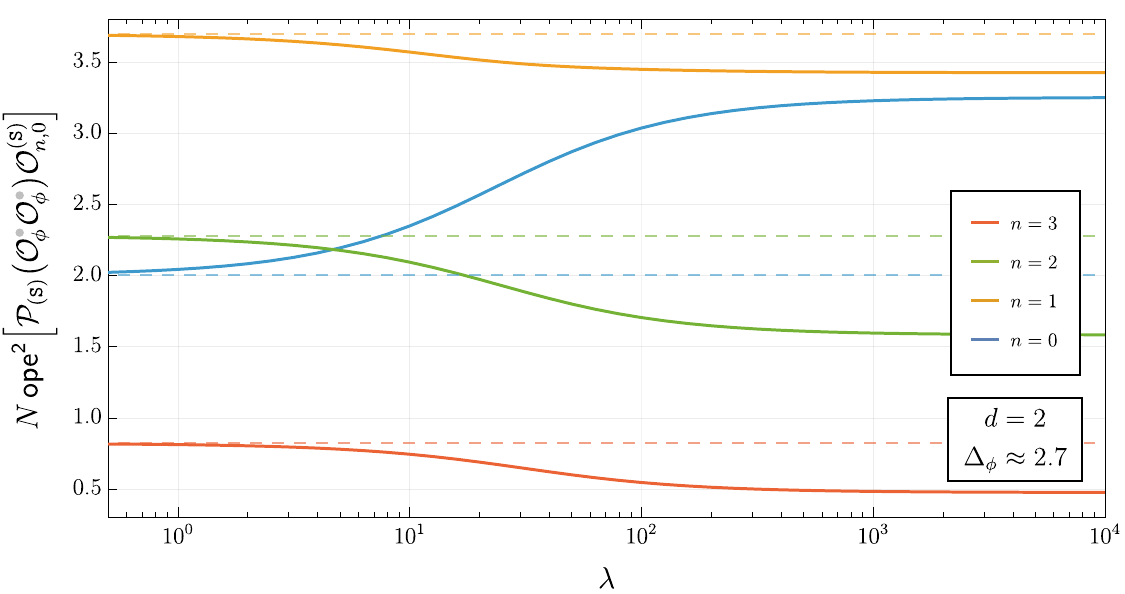}
    \end{adjustbox}
    \vspace*{-0.8ex}

    \begin{adjustbox}{center}
        \includegraphics[width=1.0\linewidth]{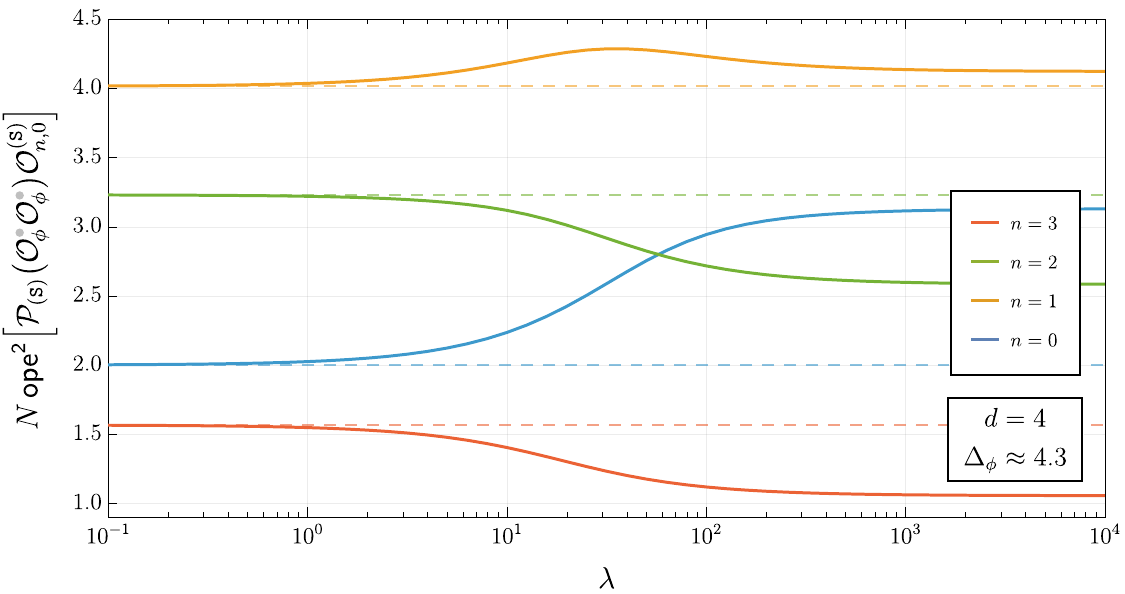}
    \end{adjustbox}
    \caption{
        Coupling dependence of squared OPE coefficients of singlet scalar operators appearing in the \(\Ophi^{\ind}\times\Ophi^{\ind}\) OPE.
        They are of order \(1/N\) as is clear from~\eqref{eq:ope_sq}, so we plot the expression multiplied by this factor.
        In the free limit \(\lambda\to 0\), they reduce to MFT values (dashed lines) given by minus residues of~\eqref{eq:spec_disc_t}, as they should.
        All OPE coefficients approach constant values at strong coupling, with \(\bigO{1}\) corrections to the corresponding MFT values.
        No regularities can be inferred from these plots, neither in \(d=2\) nor in \(d=4\).
        Corrections of both signs occur, their strength is not ordered by scaling dimensions of the singlet scalar operators, moreover their magnitudes can cross.
    }
    \label{fig:ope2}
\end{figure}

\pagebreak
Plots for the coupling dependence of squared OPE coefficients~\eqref{eq:ope_sq} for some of the \(J=0\) singlet operators \(\Oper[\Singlet]_{n,0}\) associated to MFT ones in the \(\lambda\to 0\) limit --- we do not show possible new operator in \(d=4\) --- is displayed in \Cref{fig:ope2}.

\subsection{Criticality in the bulk}
\label{sub:Criticality in the bulk}

By appropriately tuning the couplings of the theory, a \emph{critical point} in the bulk of \(\AdS\) can be reached.
Its existence and the evidence for the bulk conformal symmetry was first discussed in \cite[Section 5]{Carmi:2018qzm}.
Since the critical theory in \(\EAdS_{d+1}\) describes an interacting \(\CFT[B]_{d+1}\) by performing a Weyl transformation to a flat half-space \(\R^{d}\times \R_{\ge}\), we start by recalling the results for the large \(N\) critical \(\OO(N)\) model obtained there \cite{McAvity:1995zd}.

As already mentioned in \Cref{sub:Generalities of ON model}, an appropriately tuned \(\OO(N)\) model in \(\R^{d+1}\) with dimension ranging in \(2<d+1<4\) flows in the IR to an interacting \(\CFT_{\text{IR}}\) describing the second-order phase transition separating the broken and the unbroken phase.
Considering now the theory on a flat half-space, we can obtain different \CFT[B]s in the IR by imposing different conformal boundary conditions --- for the definition of the \emph{ordinary}, \emph{special}, and \emph{extraordinary} transitions of the \(\OO(N)\) model on half-space see~\cite{Metlitski:2020cqy,Padayasi:2021sik,Giombi:2020rmc}.
Our choice of boundary conditions in \(\AdS\) corresponds to the Dirichlet boundary conditions leading to the ordinary transition.

As explained for example in~\cite{Giombi:2020rmc}, the ordinary transition can be reached by setting \(\Dphi=d-1\) and sending the coupling \(\lambda\to\infty\).
In the critical point, the scaling dimensions of boundary operators \(\Osigma_{n}\) induced by the bulk \(\sigma\)-field are given by \(\Delta_{\Osigma_{n}}=d+1+2n\).
The leading boundary operator \(\Osigma_{0}\) corresponds to the displacement operator generally present in any \CFT[B] with a protected scaling dimension \(\Delta_{\Osigma_{0}}=d+1\) --- see \cite[Section 3.2]{Liendo:2012hy}.

As was shown already in~\cite{Carmi:2018qzm} for \(d=2\) (\(\AdS_{3}\)), the singlet spectrum --- or equivalently the spectral representation of the \(\sigma\)-propagator \eqref{eq:sigma_exact_propagator} or the bubble function \eqref{eq:B2d} --- simplifies greatly in the critical point (\(\Dphi=1,\lambda\to\infty\)).
Recalling the equation governing the singlet spectrum \eqref{eq:singlet_poles-bubble_roots}, we just need to find the zeros of the simplified bubble function
\begin{equation} \label{eq:B2d_conformal}
    \Bubble(\D)\eval_{\Dphi=1} \eqdim{2} -\frac{\cot({\frac{\pi}{2}}\D)}{8(\D-1)} \eqend
\end{equation}
Thus, the induced boundary singlet scalar operators \(\Osigma_{n}\) indeed turn out to have the expected scaling dimensions \(\Delta_{\Osigma_{n}}=3+2n\).

Although the preceding discussion assumed \(2<d+1<4\), and in particular \(d=2\), we found a similar striking pattern exhibited also for other even \(d\), suggesting the continuation of the ordinary transition above its upper critical dimension \(d+1=4\).
Here we however need to discuss the subtraction ambiguity in the definition of bubble function~\eqref{eq:B_fn}, which for example in \(d=4\) is just the constant \(a_{0}\) in~\eqref{eq:B4d}.

This turns out to be tied with the appearance of a new operator in \(d=4\).
Taking \(\Dphi=d-1=3\), all singlet scalar operators continuously connected with the MFT spectrum have scaling dimensions \(\D>2\Dphi=6\), so the only candidate for the displacement operator with \(\Delta_{\Osigma_{0}}=5\) is the one coming from the branch reaching slightly below the horizontal axis for small \(\D\), see~\Cref{fig:Bubble2dand4d}.

If we want to obtain the critical behavior at \(\lambda\to \infty\), the requirement of \(\Delta_{\Osigma_{0}}=5\) fixes the subtraction constant as \(a_{0}=-\frac{1}{16\pi^{2}}\) --- we choose it independent of \(\Dphi\), such that the critical value of the coupling \(\lambda_{\star}\equiv 8\pi^{2}\) when the new operator emerges is fixed.
It just so happens that the rest of operators \(\Osigma_{n}\) with \(n\geq 1\) have scaling dimensions \(\Delta_{\Osigma_{n}}=5+2n\), so they complete the family of boundary operators induced by the bulk \(\sigma\)-field.

\begin{figure}[!ht]
    \centering
    \begin{adjustbox}{center}
        \hspace*{-0.015\linewidth}
        \adjincludegraphics[width=0.51\textwidth, Clip={0\width} {0.15\height} {0\height} {0\height}]{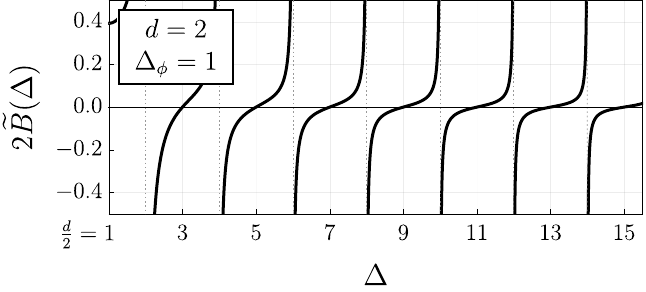}
        \hspace{0.01\linewidth}
        \adjincludegraphics[width=0.51\textwidth, Clip={0.09\width} {0.15\height} {0\height} {0\height}]{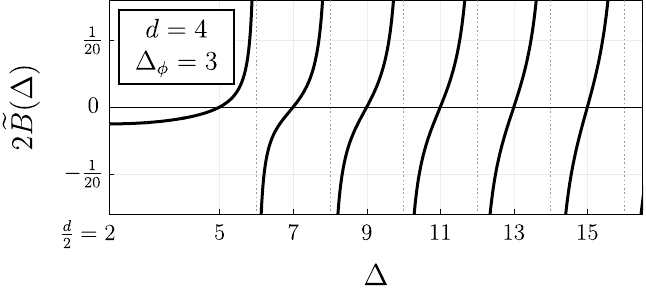}
    \end{adjustbox}
    \vspace*{-1.2ex}

    \begin{adjustbox}{center}
        \hspace*{-0.015\linewidth}
        \adjincludegraphics[width=0.51\textwidth]{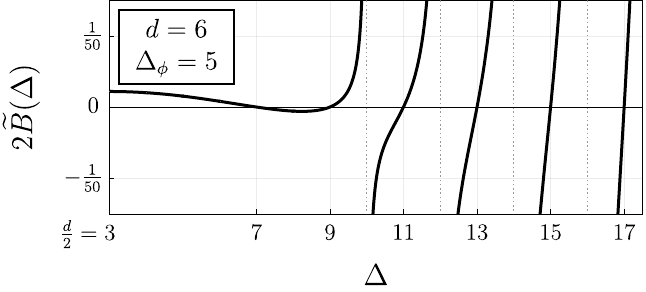}
        \hspace{0.01\linewidth}
        \adjincludegraphics[width=0.51\textwidth, Clip={0.09\width} {0\height} {0\height} {0\height}]{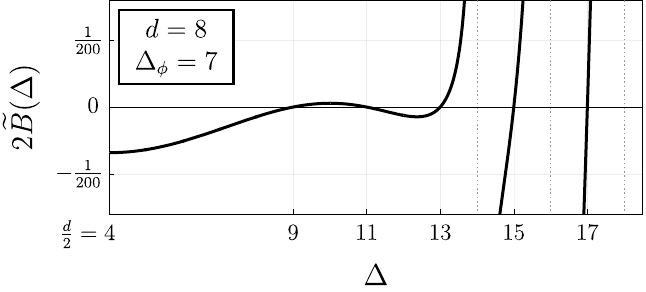}
    \end{adjustbox}
    \caption{
        Bubble functions for \(\Dphi=d-1\) in even dimensions up to \(d=8\).
        The critical point is reached by sending the coupling \(\lambda\to\infty\), that is by looking at the zeros of the bubble function.
        The scale of \(y\)-axis is adapted in each dimension to make the nuanced appearance of \(\frac{d}{2}-1\) new operators more visible.
        All zeros are at dimensions \(\D=d+1+2n\), the first corresponding to the displacement operator, and possibly some other emergent operators below the MFT dimension \(2\Dphi=2(d-1)\).
        The rest are finite deformations of the MFT operators.
    }
    \label{fig:BubbleFunctionCriticalPoint}
\end{figure}

A similar story is true also for higher dimensions.
Let us first present the observed formula for simplified bubble functions at critical point in even dimensions, and comment on the fixing of the subtraction ambiguity after.
We found
\begin{equation} \label{eq:Bubble_conformal_general_dimension}
    \Bubble(\D)\eval_{\Dphi=d-1} \mathrel{\overset{\mathcolor{black!70}{d\text{ even}}}{\scalebox{3.5}[1]{\(=\)}}}
    -\frac{\cot({\frac{\pi}{2}}\D)}{2^{2d-1} \pi^{\frac{d}{2}-1} \G(\frac{d}{2})}
    % \frac{\displaystyle\prod\limits_{\substack{i=4-d\\i\text{ even}}}^{2(d-2)} (\D-i)}
    % {\displaystyle\prod\limits_{\substack{i=1\\i \text{ odd}}}^{d-1} (\D-i)}
    \left(
        \,
        \prod_{\substack{a\.=\.4-d\\a\text{ even}}}^{2(d-2)} (\D-a)
        \middle/
        \prod_{\substack{b\.=\.1\\b \text{ odd}}}^{d-1} (\D-b)
    \right)
    \eqcomma
\end{equation}
so for example we have explicitly
\begin{equation}
    \label{eq:Bubble_conformal_d246}
    \Bubble(\D)\eval_{\Dphi=d-1} =\mspace{-11mu}
    \begin{dcases}
        \mathcolor{darkgray}{\eqdim{2}}
        -\frac{\cot({\frac{\pi}{2}}\D)}{8}
        \frac{1}{(\D-1)} \eqcomma                                           \\
        \mathcolor{darkgray}{\eqdim{4}}
        -\frac{\cot({\frac{\pi}{2}}\D)}{128\pi}
        \frac{(\D \mathcolor{gray}{-0})(\D-2)(\D-4)}{(\D-1)(\D-3)} \eqcomma \\
        \mathcolor{darkgray}{\eqdim{6}}
        -\frac{\cot({\frac{\pi}{2}}\D)}{4096\pi^{2}}
        \frac{(\D+2)(\D\mathcolor{gray}{-0})(\D-2)(\D-4)(\D-6)(\D-8)}{(\D-1)(\D-3)(\D-5)} \eqend
    \end{dcases}
\end{equation}
Plots of the bubble functions at the critical point can be seen in \Cref{fig:BubbleFunctionCriticalPoint}.

Expressions for \(d=2\) and \(d=4\) in \eqref{eq:Bubble_conformal_d246} can be directly obtained from general formulas \eqref{eq:B2d} and \eqref{eq:B4d} (with the appropriate choice of \(a_{0}\) already discussed) by substituting \(\Dphi=d-1\).
In the case of \(d=6\), the subtraction ambiguity is a polynomial of the form \(a_{0} + a_{1}\nu^{2}\).
There are now two emergent operators, and requiring their dimensions to be \(\Delta_{\Osigma_{0}}=7\) and \(\Delta_{\Osigma_{1}}=9\) fixes the subtraction constants \(a_{0}\) and \(a_{1}\).
As before, continuous deformations of MFT operators then complete the family \(\Osigma_{n}\) with scaling dimensions \(\Delta_{\Osigma_{n}}=7+2n\).

Similar procedure was applied to all even dimensions up to \(d=12\).
Although we were not able to perform an explicit resummation of the critical bubble function for dimensions \(d \ge 8\) at general \(\D\) to explicitly compare with \eqref{eq:Bubble_conformal_general_dimension}, direct evaluation at any chosen \(\D\) confirms the expected structure.

The product in the numerator of \eqref{eq:Bubble_conformal_general_dimension} generates shadow-symmetric zeros at even integers below the MFT dimension \(\D=2\Dphi=2(d-1)\), canceling the \enquote{spurious} poles of \(\cot(\frac{\pi}{2}\D)\).
The product in the denominator generates shadow-symmetric poles at odd integers below the displacement operator dimension \(\D=d+1\), canceling the \enquote{spurious} zeros of \(\cot(\frac{\pi}{2}\D)\).
We are thus left with poles at MFT dimensions and zeros at \(\Delta_{\Osigma_{n}}=d+1+2n\).

\subsection{Non-singlet spectrum}
\label{sub:non-singlet_spectrum}

The non-singlet part of the spectrum is described by the spectral decomposition of~\eqref{eq:4-pt_antisymmetric} and~\eqref{eq:4-pt_symmetric_traceless}.
Both \SymTrless{} and \AntiSym{} cases can be treated simultaneously, as we have
\begin{align}
    \label{eq:spec_nonsinglet}
    \Spec[s][]{\FourPtNonSinglet}
    =\left(1\pm(-1)^{J}\right)\left(\Spec[s][]{\tDisc}+\frac{1}{N}\Spec[s][]{\tExch}+\biggO{\frac{1}{N^{2}}}\right)
    \eqcomma
\end{align}
where similarly to \eqref{eq:spec_disc_tu} we used that \(u\)-channel diagrams are related to the \(t\)-channel ones by a factor of \((-1)^{J}\).
Now we will apply methods of \Cref{sub:CPW completeness and 6j-symbol} and \Cref{sub:tChannelContributiontoAnomalousDimensions} just focusing on \(t\)-channel diagrams of \eqref{eq:spec_nonsinglet}, and the subsequent addition of \(u\)-channel diagrams merely multiplies all of the squared OPE coefficients by \(\left(1\pm (-1)^{J}\right)\).
We instantly see, that non-singlet sectors \SymTrless{}/\AntiSym{} contain only even/odd spins, respectively.

Decomposition of the \(t\)-channel exchange diagram into \(t\)-channel conformal blocks is clearly the same as the \(s\)-channel conformal block decomposition of the \(s\)-channel exchange diagram.
Alternatively, the \(t\)-channel spectral function \(\SpecNoArgs[t]\) of \(t\)-channel exchange diagram is given by the right-hand side of \eqref{eq:spec_s_channel_exchange}.

Even though \eqref{eq:spec_s_channel_exchange} has two types of poles/operators --- MFT ones originating from the \(\G^{2}\)-factor and the non-MFT ones located at solutions of~\eqref{eq:singlet_poles-bubble_roots} --- only the non-MFT poles/operators contribute to the \(s\)-channel spectral function \(\SpecNoArgs[s]\) of \(t\)-channel exchange diagram.
This was discussed after equation \eqref{eq:tContrib_sChannel_Spec_CB}, which we now use to write the \(s\)-channel spectral function of the \(t\)-channel exchange diagram (including the \(1/N\) prefactor) as
\begin{align}
    \label{eq:sChannel_Spec_tExchange}
    \begin{aligned}
        \frac{1}{N} \Spec[s][\D][J]{\tExch}
         & = \int_{\frac{d}{2}+\I\R}\frac{\dd{\Dpr}}{2\pi\I}
        \CrK[\D,J][\Dpr\mathcolor{gray}{,0}] \Knorm[\Dpr*\mathcolor{gray}{,0}] \frac{1}{N} \Spec[t][\Dpr]{\tExch} \\
         & = \sum_{\Oper[\Singlet]_{\dind,0}}
        \opesq{\Ophi\Ophi\Oper[\Singlet]_{\dind,0}} \CrK[\D,J][\D[\raisemath{0.3ex}{\Singlet}]_{\smash{\dind,0}}\mathcolor{gray}{,0}]
        \eqcomma
    \end{aligned}
\end{align}
where the sum runs only over the non-MFT operators exchanged in the \(t\)-channel, all of which are scalar singlets \(\Oper[\Singlet]_{\dind,0}\) with dimensions given by roots of \eqref{eq:singlet_poles-bubble_roots} and corresponding squared OPE coefficients given by \eqref{eq:ope_sq}.

\paragraph{Anomalous dimensions in the non-singlet sector.}%
\label{par:Anomalous dimensions in the non-singlet sector.}

We are precisely in the setting of \Cref{sub:tChannelContributiontoAnomalousDimensions}, and the formula for the anomalous dimensions of double-twist operators from the \(t\)-channel exchange \eqref{eq:anomalous_dimensions_general_formula} applied to the non-singlet sector of \(\OO(N)\) model reads
\begin{align}
    \label{eq:anomalous_dimensions_non-singlet}
    \anomdim[]^{\NonSinglet}_{n,J} =
    \sum_{\Oper[\Singlet]_{\dind,0}}
    \opesq{\Ophi\Ophi\Oper[\Singlet]_{\dind,0}}
    \, \anomdim_{n,J} \eval_{\substack{\text{\(t\)-channel}\mspace{62mu} \\[0.1ex]\text{exchange of }\smash{\Oper[\Singlet]_{\dind,0}}}}
    \eqcomma
\end{align}
where again only the non-MFT operators contribute.
Definition of \(\anomdim_{n,J}\) was given in~\eqref{eq:anomdim_contrib_from_CB} and explicit expressions in \(d=2\) and \(d=4\) were presented in~\eqref{eq:tchannel_anom_dim_2d} and~\eqref{eq:tchannel_anom_dim_4d}.
Remember that the squared OPE coefficients \eqref{eq:ope_sq} of the exchanged non-MFT operators are of order \(\bigO{1/N}\), and thus are also the anomalous dimensions \(\anomdim[]^{\NonSinglet}_{n,J}\).
Even though the same formula \eqref{eq:anomalous_dimensions_non-singlet} is applicable for both \NonSinglet{} sectors, only operators with even/odd spins \(J\) actually appear in the \NonSinglet{} sector, respectively.

The whole next section is dedicated to the analysis of the \(\bigO{1/N}\) non-singlet spectrum corrections governed by~\eqref{eq:anomalous_dimensions_non-singlet}.

\section{Analysis of the non-singlet sector}
\label{sec:analysis_of_the_non-singlet_sector}

Methods established in previous sections allow us to compute the \emph{complete non-singlet spectrum} occurring in the \(\Ophi^{\ind}\times \Ophi^{\ind}\) OPE to the order \(1/N\) considered.
Almost all ingredients in the master formula~\eqref{eq:anomalous_dimensions_non-singlet} determining the non-singlet spectrum contributions are known analytically, exception being the precise values of singlet scalar scaling dimensions \(\D[\Singlet]_{\dind,0}\), which need to be found numerically.

Therefore, in~\Cref{sub:Numerical caluclation of anomalous dimensions in the non-singlet sector} we first describe how we chose to truncate the sum initially going over infinitely many singlet operators.
See the accompanying \notebook{} for the code implementing all of the calculations.

We then analyze the obtained non-singlet spectrum in a series of plots.
The principal one --- a twist--spin plot --- is discussed in~\Cref{sub:Twists in the non-singlet sector} and verifies a known Regge trajectory structure of the spectrum.
Asymptotic behaviors of two important sections of this plot are investigated further.
First, \Cref{sub:Dependence of anomalous dimension on n} studies the asymptotics at large twist for a fixed spin.
Second, the large spin asymptotics for a fixed twist family is investigated in~\Cref{sub:Large spin asymptotics}, where a theorem determining the asymptotic behavior is verified as a consistency cross-check of the obtained spectral data.

Finally, dependence on the external scaling dimension and the coupling is examined in~\Cref{sub:Dependence of anomalous dimensions on the coupling}.
In particular, strong and weak coupling asymptotics are explored, from which it is recognized that scaling dimensions of \SymTrless{} scalar operators were not computed correctly due to a known limitation of the Lorentzian inversion formula entering the derivation of~\eqref{eq:anomalous_dimensions_non-singlet}.

\subsection{Numerical calculation of anomalous dimensions in the non-singlet sector}%
\label{sub:Numerical caluclation of anomalous dimensions in the non-singlet sector}

As anticipated, the practical problem with the sum~\eqref{eq:anomalous_dimensions_non-singlet} is its infinite support on numerical solutions to~\eqref{eq:singlet_poles-bubble_roots} labeled by non-negative integers.
The exchanged \(t\)-channel singlet \(J'=0\) operators will be labeled by \(k\) as \(\Oper[\Singlet]_{k,0}\), and we reserve \(n\) to label the non-singlet twist family whose anomalous dimensions we are computing.
For simplicity, we exclude the possible contribution of the emergent operator in \(d=4\).

For numerical evaluation, we chose to simply truncate the sum over exchanged singlet operators at some maximal value \(k_{\text{max}}\).
This corresponds to a pole of the \(t\)-channel spectral function of some maximal scaling dimension \(\D[\Singlet]_{k_{\text{max}}}\).
To be completely explicit, let us reproduce here the practical truncated formula for the non-singlet anomalous dimensions
\begin{align}
    \label{eq:anomlous_dimensions_non-singlet_truncated}
    \anomdim[]^{\NonSinglet}_{n,J} =
    \sum_{k=0}^{k_{\text{max}}}
    \opesq{\Ophi\Ophi\Oper[\Singlet]_{k,0}}
    \, \anomdim_{n,J} \eval_{\substack{\text{\(t\)-channel}\mspace{62mu} \\[0.1ex]\text{exchange of }\smash{\Oper[\Singlet]_{k,0}}}}
    \eqend
\end{align}
Before discussing its behavior on the choice of \(k_{\text{max}}\), or equivalently the convergence of original sum \eqref{eq:anomalous_dimensions_non-singlet}, let us make a remark about \(\anomdim_{n,J}\) provided in~\eqref{eq:tchannel_anom_dim_2d}/\eqref{eq:tchannel_anom_dim_4d}.
Its evaluation sometimes involves performing a difference of two terms that are numerically huge but almost equal.
This forced us to keep a high \texttt{WorkingPrecision} in the \notebook{}, namely we set it to a value of \(100\), sufficient for our range of computations.

We are still left with a legitimate question --- how large should \(k_{\text{max}}\) be, such that the anomalous dimensions~\eqref{eq:anomlous_dimensions_non-singlet_truncated} of non-singlet operators are calculated with sufficient precision?

\paragraph{Convergence.}%
\label{par:Convergence}

\begin{figure}[!ht]
    \centering
    \includegraphics[width=\linewidth]{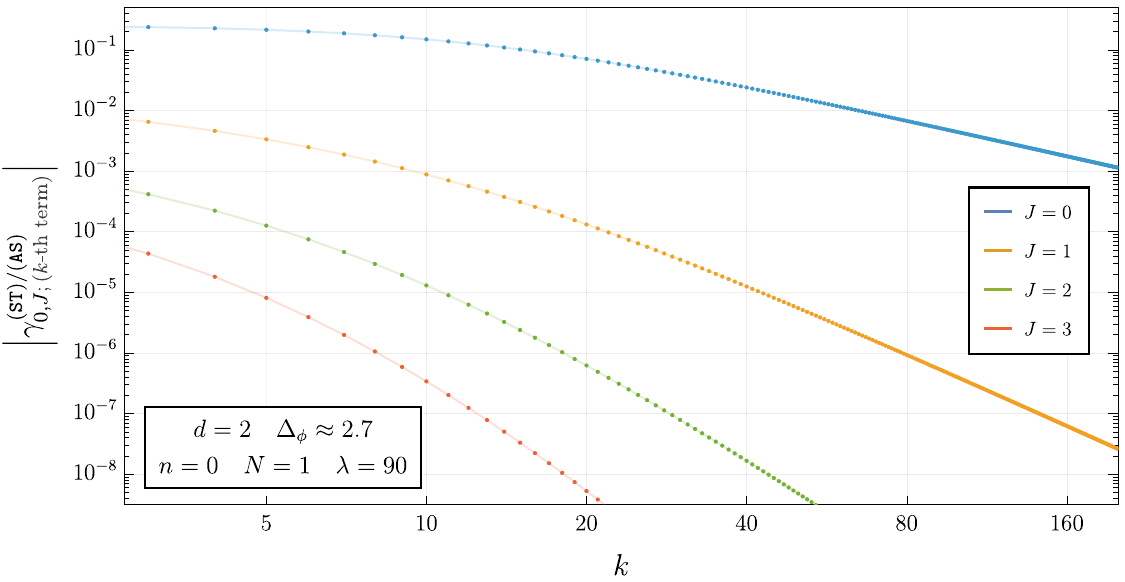}
    \caption{Log--log plot of magnitudes of individual terms in~\eqref{eq:anomlous_dimensions_non-singlet_truncated} helping to assess the convergence of the sum calculating the non-singlet anomalous dimensions.
        Fixed external parameters are summarized in the boxes.
        For large \(k\), the summands falloff as \(\,\sim k^{-2(1+J)}\).
        % Similar behavior is seen in \(d=4\).
    }
    \label{fig:anomalous_dim_convergence}
\end{figure}

Obtaining the asymptotic behavior of the summands in \eqref{eq:anomalous_dimensions_non-singlet} is not straightforward --- in particular for general \(n\) --- so we chose a more pedestrian approach.
To assess the convergence, we plotted individual terms in~\eqref{eq:anomlous_dimensions_non-singlet_truncated} and analyzed their behavior for large \(k\).
See for example the \(n=0\) case shown in~\Cref{fig:anomalous_dim_convergence}.

By creating more examples of such log--log scale plots, where a power law falloff can be fitted conveniently, we observed an asymptotic behavior \(\.\sim k^{-2(1+J)}\) for large \(k\).
Hence, the convergence improves drastically with increasing spin, and not many \(t\)-channel terms are required to obtain sufficiently precise results.
Based on this analysis, we chose the value \(k_{\text{max}}=50\) for spinning \((J\geq 1)\) non-singlet operators and \(k_{\text{max}}=200\) for scalar \((J=0)\) operators, where the convergence is much slower.

As will be explained later, \(\SymTrless\) operators with \(J=0\) are problematic for yet another reason --- a limitation present already in the derivation of~\eqref{eq:anomalous_dimensions_non-singlet} --- so we did not attempt to employ resummation techniques that would take care of the neglected tail of the sum.

\subsection{Twists in the non-singlet sector}%
\label{sub:Twists in the non-singlet sector}

\begin{figure}[!ht]
    \centering
    \begin{adjustbox}{center}
        \includegraphics[width=1.0\textwidth]{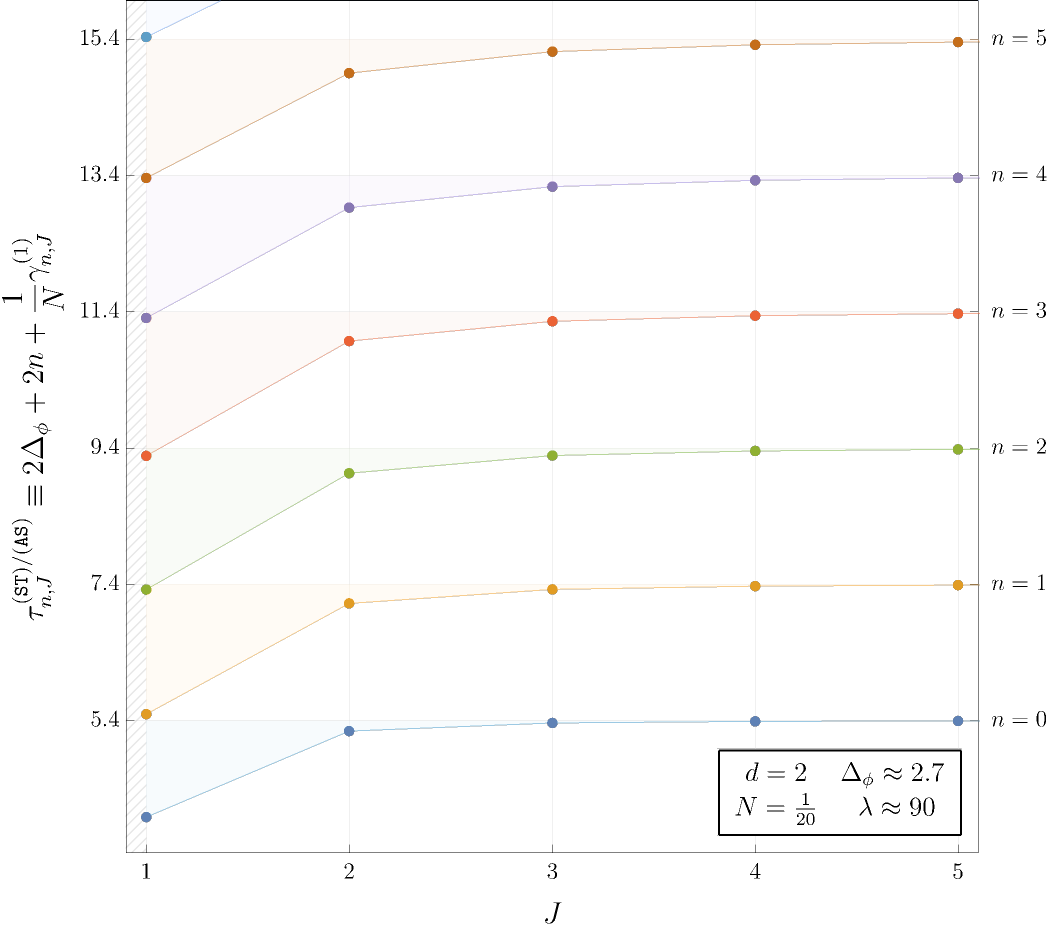}
    \end{adjustbox}
    \caption{
        The twist--spin plot of non-singlet spectrum in \(d=2\) at finite coupling \(\lambda\approx 90\).
        The chosen dimension \(\Dphi\) corresponds to a massive \(\phi\)-field in \(\AdS_{3}\) with Dirichlet boundary conditions.
        The plot shows operators in both non-singlet irreps of \(\OO(N)\) --- \SymTrless{} operators are supported on even spins, while \AntiSym{} operators on odd spins.
        Anomalous dimensions for \(J=0\) are comparatively huge compared to those for \(J\ge1\), so we excluded the scalar operators from this plot (furthermore, they are not reliably computed by~\eqref{eq:anomalous_dimensions_non-singlet} due to a limitation in its derivation).
        The essential feature of the spectrum --- its organization into Regge trajectories concave in spin --- is discussed in the main text.
    }
    \label{fig:NonSingletTwists2d}
\end{figure}

\begin{figure}[!ht]
    \centering
    \begin{adjustbox}{center}
        \includegraphics[width=1.0\textwidth]{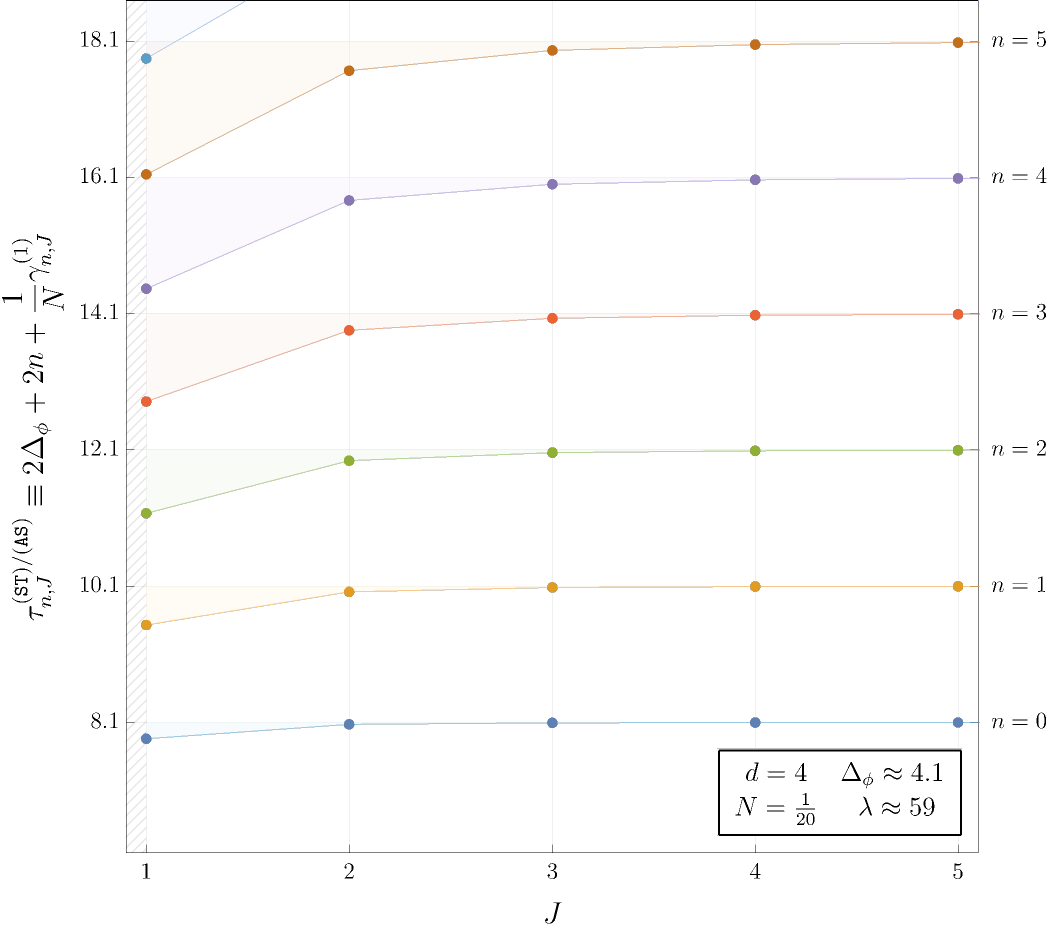}
    \end{adjustbox}
    \caption{
        The twist--spin plot of non-singlet spectrum in \(d=4\) at finite coupling \(\lambda\approx 59\).
        The chosen dimension \(\Dphi\) corresponds to a massive \(\phi\)-field in \(\AdS_{5}\) with Dirichlet boundary conditions.
        The plot shows operators in both non-singlet irreps of \(\OO(N)\) --- \SymTrless{} operators are supported on even spins, while \AntiSym{} operators on odd spins.
    }
    \label{fig:NonSingletTwists4d}
\end{figure}

Roughly since~\cite{Costa:2012cb,Komargodski:2012ek,Fitzpatrick:2012yx}, it was known that the proper organizing principle for a \CFT{} spectrum is in terms of twist/Regge trajectories labeled by \(n\in\mathbb{N}_{0}\) that are analytic in spin.
This idea was definitely confirmed by~\cite{Caron-Huot:2017vep}.
Therefore, we present the main plots of the non-singlet spectrum in the twist--spin plane, for a fixed value of the coupling and external scaling dimension.
The plot for \(d=2\) (\(\AdS{}_{3}\)) is shown in~\Cref{fig:NonSingletTwists2d}, while the one for \(d=4\) (\(\AdS{}_{5}\)) is in~\Cref{fig:NonSingletTwists4d}.

To prepare the plots we had to choose also a particular value for \(N\).
It was fixed to a rather small unphysical value \(N=\tfrac{1}{20}\), otherwise corrections of twists by anomalous dimensions (multiplied by a factor \(1/N\)) from their MFT values would be barely visible.
So the functional dependence of the anomalous dimensions on \((n,J)\) is faithfully displayed, but their absolute value is exaggerated by extrapolating them far beyond the validity of the large \(N\) expansion.

Both plots clearly exhibit the discussed organization of the spectrum into twist trajectories labeled by \(n\in\N_{0}\) that are concave as a function of the spin \(J\) and asymptote to the MFT spectrum for \(J\to\infty\).
This asymptotic shape is valid above some minimal spin \(J_{\text{crit}}\) that depends on a concrete theory (in our case \(J_{\text{crit}}=1\)).

These claims are by now a theorem valid in any CFT, which was established in a series of papers~\cite{Komargodski:2012ek,Fitzpatrick:2012yx,Alday:2016njk,Pal:2022vqc,vanRees:2024xkb}.
Applied to our setting, it states that if \(\Ophi^{i}\) is in the spectrum, then so must be (at least for \(J\geq 1\)) the double-twist families of schematic form \([\Ophi^{i}\dAlembertian^{n}\partial^{J}\Ophi^{j}]\).

For the MFT theory at \(N\to\infty\), they all have twists \(\tau_{n,J}\equiv \D_{n,J}-J=2\Dphi+2n\).
Deforming from the strict large \(N\) limit, they receive anomalous dimensions and the large spin asymptotics of their twist trajectories was shown in~\cite[(1.7)]{Komargodski:2012ek} and~\cite[(1)]{Fitzpatrick:2012yx} to be of the form
\begin{align}
    \label{eq:large_spin_asympt}
    \tau_{n,J} \simeq 2\Dphi+2n-\frac{c_{n}}{J^{\twistmin}} + \cdots \eqend
\end{align}
% \todo{
%     Is this also valid for singlet sector?
%     It is even exactly for \(J>0\).
%     But can't we also calculate \(c\) for singlet sector, and it would not be zero, right?
%     But it would be \(\bigO{1/N}\) from squared OPE coefficients, so it seems compatible.
%     Does allow us to compute large \(J\) asymptotic of \(1/N\) anomalous dimensions for \(J>0\) singlet operators?
% }
This theorem serves us as a check of consistency done in \Cref{sub:Large spin asymptotics}, where we also discuss which \enquote{minimal twist} operator governs the asymptotics in our case.

\subsection{Dependence of anomalous dimensions on the coupling}%
\label{sub:Dependence of anomalous dimensions on the coupling}

The main advantage of the large \(N\) approach is that it outputs observables as exact functions of the coupling \(\lambda\).
We will exploit the known functional dependence in this section.
However, we should emphasize that the coupling dependence is complicated --- entering implicitly via numeric solutions to~\eqref{eq:singlet_poles-bubble_roots} --- thus the best we can do is sample over a finite set of coupling values.

Of particular interest are the limiting cases.
At which rate do the anomalous dimensions approach weak/strong coupling?
Such question is best answered by a log--log plot, where a power law approach is captured by the slope of the graph.

\begin{figure}[!ht]
    \centering
    \begin{adjustbox}{center}
        \includegraphics[width=1.05\textwidth]{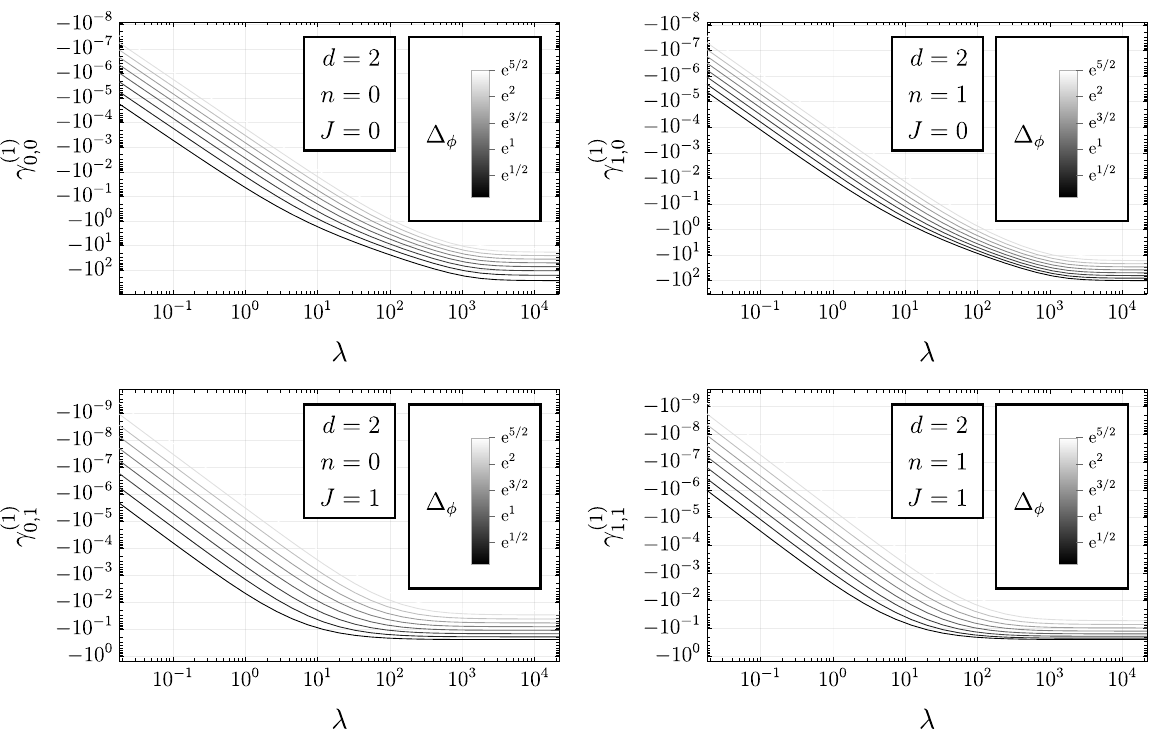}
    \end{adjustbox}
    \caption{
        Dependence of anomalous dimensions \(\anomdim_{n,J}\) on the coupling in \(d=2\).
        The top row shows the \(J=0\) operators in the \SymTrless{} irrep for the first two Regge trajectories --- \(n=0\) and \(n=1\).
        The bottom row displays the same for the \(J=1\) operators in the \AntiSym{} irrep.
        Note, that the plots are in a log--log scale and cover a range of indicated external scaling dimensions starting at the Breitenlohner--Freedman bound, above which Dirichlet boundary conditions are applicable.
    }
    \label{fig:AnomDimOnCouplingGrid2d}
\end{figure}

We explore this question for \(d=2\) in \Cref{fig:AnomDimOnCouplingGrid2d}.
It shows the dependence of anomalous dimensions on the coupling for \NonSinglet{} operators with the lowest spins \(J=0\)/\(J=1\) belonging to the first two leading Regge trajectories.
As is clear from the plot, non-singlet anomalous dimensions approach a constant value at very strong coupling.
% \Note{We should study the \(\lambda\to\infty\) asymptotics, it should be similar to the large spin asymptotics.
%     The dependence is probably \(\gamma(\lambda)\underset{\lambda\to\infty}{\sim}c+a\lambda^{-b}\).
%     Then we subtract the constant \(c\), take a logarithm \(\left[\log(\gamma(\lambda)-c)\sim \log a-b\log \lambda\right]\) and fit \(a,b\) to get the asymptotics.
%     In fact, the main advantage of large \(N\) is that we can confidently approach \(\lambda\to\infty\), so we should take advantage of that and figure out the asymptotic behavior for both singlet and non-singlet anomalous dimensions.}
On the other hand, the slope at very weak coupling is two, thus non-singlet anomalous dimensions start to decrease at a quadratic rate from their vanishing values in the free theory.
This is expected for the \(J=1\) \AntiSym{} operators, as the \(\bigO{\lambda}\) contact Witten diagram does not contribute to anomalous dimensions of \(J\geq 1\) operators.
However, we should see \(\bigO{\lambda}\) contribution to \(J=0\) operators, an issue to which we turn next.

\paragraph{Contribution of contact Witten diagram to \SymTrless{} spin zero operators.}
\label{par:contactWD}

As we will review, the contact diagram contributes to anomalous dimensions of \(J=0\) \SymTrless{} operators, thus their weak coupling asymptotics must be linear.
It implies that the method on which we based our computations missed at least this contribution for this particular class of operators.

To locate the error, let us briefly comment on the Lorentzian inversion formula (LIF) that was used in~\cite{Liu:2018jhs} to derive the key formulas on which we build upon in~\Cref{sec:CFTgeneralities}.
Its purpose is to output the \(s\)-channel spectral density (encoding the anomalous dimensions we are after) from the \(t\)-channel scalar exchange Witten diagram.

% Let's imagine expanding the \(t\)-channel scalar exchange in perturbation series and applying LIF at each order in the coupling.
% Leading order corresponds to the contact diagram.

The first step of LIF extracts from a given diagram (\(4\)-point correlator) its double-discontinuity, an operation that has a property of projecting out all contributions of MFT operators.
However, since the conformal block decomposition of the contact diagram --- the only \(\bigO{\lambda}\) contribution --- contains solely MFT operators, it is completely annihilated by LIF.
Below, in~\eqref{eq:anomdim_contact_symtrless}, we will provide its contribution to anomalous dimensions, which would otherwise be missed by LIF.

Higher order diagrams have a nonvanishing double-discontinuity, and are therefore taken into account by LIF.
Still, there might be a subtlety connected with convergence.
LIF integrates the double-discontinuity against a certain kernel and this integral might not converge.
In fact, inverting \(J^{\prime}=0\) operators in the \(t\)-channel, it is guaranteed to converge only for spin \(J\geq 1\) operators in the \(s\)-channel.
Yet, when extracting the anomalous dimensions from the spectral function computed by LIF, a brief comment was made in~\cite{Liu:2018jhs} that such combined operation should yield convergent integrals for all spins.
If this is true, the contact diagram would be the only omission, and its inclusion in~\eqref{eq:full-ST-spin0-anomdim} would give correct anomalous dimensions also for \(J=0\) \SymTrless{} operators.

Having motivated the necessity of including the contact diagram contribution by hand, we proceed by recalling its contributions to anomalous dimensions.
As is well known, it affects only scaling dimensions of spin zero operators, both \Singlet{} and \SymTrless{}.
Its conformal block decomposition determining also anomalous dimensions was studied in detail, and an explicit formula for \Singlet{} operators can be found for instance in~\cite[(4.81),\,(4.85--4.87)]{Giombi:2020rmc}.
Adapted to our conventions, the final expression reads \textcolor{darkgray}{(ignoring subleading \(1/N\) corrections)}
\begin{equation} \label{eq:anomdim_contact_singlet}
    \anomdim[]^{\Singlet\.\text{contact}}_{n,\mathcolor{gray}{J=}0} =
    \lambda \frac{1}{(4\pi)^{\frac{d}{2}} n!}
    \frac{\G[\frac{d}{2}+n] \G[\Dphi+n] \G[\Dphi-\frac{d}{2}+\frac{1}{2}+n] \G[2\Dphi-\frac{d}{2}+n]}
         {\G[\frac{d}{2}] \G[\Dphi+\frac{1}{2}+n] \G[\Dphi-\frac{d}{2}+1+n] \G[2\Dphi-d+1+n]}
\end{equation}
where we used the dictionary \(\lambda_{\text{\cite{Giombi:2020rmc}}}=\tfrac{2}{N}\lambda_{\text{\textcolor{gray}{us}}}\) and \(d_{\text{\cite{Giombi:2020rmc}}}=d_{\text{\textcolor{gray}{us}}}+1 \).

Alternatively, we could obtain \eqref{eq:anomdim_contact_singlet} without the explicit form of the contact diagram CB decomposition by looking at \eqref{eq:singlet_poles-bubble_roots} in the regime of weak coupling, where to the first order only a single pole of bubble function \eqref{eq:B_fn} at \(\D = 2\Dphi + 2n\) plays a role.
Ignoring other poles in \(\Bubble\) we immediately land at the same expression --- easily checked by the identical appearance of \(\Gamma\)-factors --- providing a successful cross-check.

By comparing the \SymTrless/\Singlet{} decompositions \eqref{eq:4-pt_symmetric_traceless}/\eqref{eq:4-pt_singlet} and focusing only on terms up to \(\bigO{\lambda}\), we readily obtain the contact contribution for \(J=0\) \SymTrless{} operators as
\begin{equation} \label{eq:anomdim_contact_symtrless}
    \anomdim[]^{\SymTrless\.\text{contact}}_{n,\mathcolor{gray}{J=}0} =
    \frac{2}{N} \anomdim[]^{\Singlet\.\text{contact}}_{n,\mathcolor{gray}{J=}0}
    \eqend
\end{equation}
If the contact Witten diagram contribution is the only needed correction to LIF --- which seems plausible based on comments in~\cite{Liu:2018jhs} --- then the complete anomalous dimensions of \(J=0\) \SymTrless{} operators are given by
\begin{align}
    \label{eq:full-ST-spin0-anomdim}
    \anomdim[]^{\SymTrless}_{n,\mathcolor{gray}{J=}0} =
    \anomdim[]^{\SymTrless\.\text{LIF}}_{n,\mathcolor{gray}{J=}0} +
    \anomdim[]^{\SymTrless\.\text{contact}}_{n,\mathcolor{gray}{J=}0}
    =
    \frac{1}{N}\left(
        \anomdim[]^{(1)\,\text{LIF}}_{n,\mathcolor{gray}{J=}0} +
        2\anomdim[]^{\Singlet\.\text{contact}}_{n,\mathcolor{gray}{J=}0}
    \right)
    \mathcolor{gray}{+ \biggO{\frac{1}{N^{2}}}}
    \eqcomma
\end{align}
where the first piece is computed by LIF, see \eqref{eq:anomalous_dimensions_non-singlet}.

\subsection{Dependence of anomalous dimensions on Regge trajectory label \texorpdfstring{\(n\)}{n}}%
\label{sub:Dependence of anomalous dimension on n}

Inspecting a cut through \Cref{fig:NonSingletTwists2d} at fixed spin \(J\geq 1\), it appears that the anomalous dimensions are at first growing (in absolute value) as a function of the twist trajectory label \(n\).
To establish their fate at large \(n\), it is interesting to study the anomalous dimensions as a function of the twist trajectory label at fixed spin.

\begin{figure}[!ht]
    \centering
    \begin{adjustbox}{center}
        \includegraphics[width=1.05\textwidth]{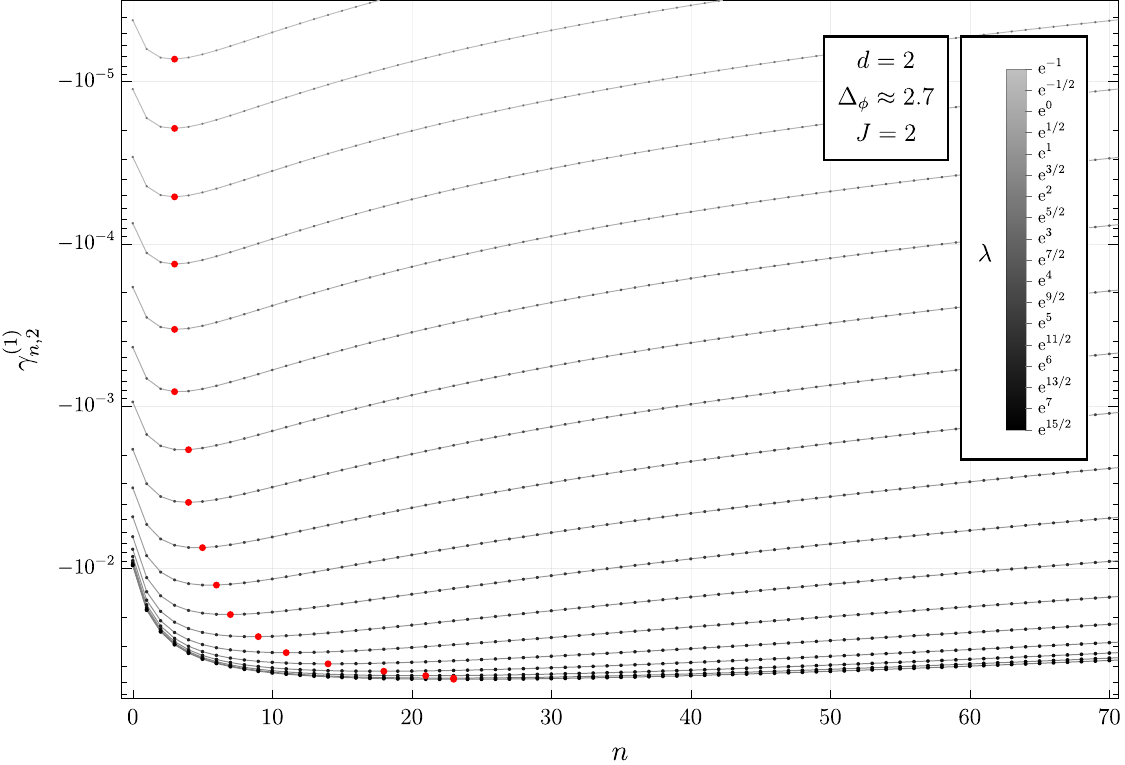}
    \end{adjustbox}
    \caption{
        Dependence of anomalous dimensions on the twist trajectory label \(n\) in \(d=2\) for varying values of \(\lambda\).
        A fixed spin \(J=2\) was chosen --- operators are transforming in the \SymTrless{} irrep of \(\OO(N)\) --- and \(\Delta_{\phi}\) corresponds to the value in \Cref{fig:NonSingletTwists2d}.
        For the chosen parameters, the weak-coupling shape valid until \(\lambda\approx\E^{2}\) has the minimum at \(n_{*}=4\).
        For higher \(\lambda\) we see a finite coupling deformation to take place, eventually (at \(\lambda\approx \E^{7}\)) reaching a \(\lambda\to \infty\) shape with the minimum at \(n_{*}=23\).
    }
    \label{fig:Non-Singlet_AnomDim_nDep_2d}
\end{figure}

A representative plot of this dependence is shown in \Cref{fig:Non-Singlet_AnomDim_nDep_2d} for the case \(d=2\) and spin \(J=2\) corresponding to the \Cref{fig:NonSingletTwists2d} (for various values of coupling \(\lambda\)), but conclusions are generic.
The anomalous dimensions (in the absolute value) increase at first until reaching a maximum for a critical twist family \(n_{*}\) --- the precise value depends on \(\lambda\), for example in the case of~\Cref{fig:NonSingletTwists2d} it is slightly higher than shown --- and then start monotonically decreasing.
This asymptotic fall off for very subleading Regge trajectories (\(n\to\infty\)) is in fact what one would expect for the asymptotically free \(\OO(N)\) model in \(\AdS{}_3\).

Recalling~\eqref{eq:anomalous_dimensions_non-singlet}, it is clear that dependence on \(n\) resides only in the second piece inside the sum, given by~\eqref{eq:tchannel_anom_dim_2d} or \eqref{eq:tchannel_anom_dim_4d}.
Those are known analytic functions whose large \(n\) asymptotics can in principle be determined exactly.

\paragraph{Qualitative features of the \(n\)-dependence.}
\label{par:n-dep-qualitative}
Dependence of anomalous dimensions on the Regge trajectory label \(n\), in particular their aymptotic behavior for large \(n\), was studied in detail in~\cite {Fitzpatrick:2010zm}.
It should be emphasized that their analysis was at leading order at weak coupling, based on tree level unitarity in \AdS{}.
In fact, one of the most prominent suggestions for future directions, was extension to loops.
In our case we work with a partially resummed perturbation series at large \(N\), hence the details of their computations cannot be imported without modifications.
Nevertheless, two main qualitative conclusions of their analysis should remain valid:
\begin{itemize}
    \item The fall-off/growth of anomalous dimensions at large \(n\) should be connected with relevance/irrelevance of interactions of the \AdS{} theory  understood as an EFT.
          They worked around free theory (Gaussian fixed point of the renormalization group) in \AdS{}.
          The prime example in their paper was the \(4\)-point function of light scalars \(\phi\) with a tree level exchange of a heavy scalar \(\chi\) governed by the interaction \(g\.\phi^2\chi\).
          For instance in \(\AdS{}_3\), canonical dimension of the relevant coupling \(g\) is \([g]=\tfrac{3}{2}\) and the exchange diagram (only \(s\)-channel analyzed) has two powers of the coupling, thus the anomalous dimensions of the dual exchanged operators fall-off at large \(n\) as \(n^{-2[g]}=n^{-3}\).
          This simple dimensional analysis has to be modified in our case, as we do not work around a free \AdS{} theory, but still the asymptotic large \(n\) behavior of the anomalous dimensions should be connected with renormalized dimensions of operators/couplings in the \AdS{} interaction Lagrangian.
    \item It was convincingly demonstrated on a number of examples that the large \(n\) asymptotics (when the external dimensions are also taken large at the same time) of anomalous dimensions encodes the \(S\)-matrix of Minkowski spacetime when the flat space limit \(\AdS{}_{d+1}\to\mathbb{R}^{1,d}\) is taken.
          To be precise, the asymptotic behavior of anomalous dimensions captures (up to precisely defined kinematic factors) partial waves associated with the flat space scattering amplitudes.
          Exchange of a particle --- for example we can think about the heavy scalar \(\chi\) --- in \AdS{} then manifests itself as a sharp peak in the \(n\)-dependence of anomalous dimensions.
\end{itemize}

At the moment, we can offer only a very preliminary study of the features mentioned above (without showing all the relevant plots):
\begin{enumerate}
    \item \(\AdS{}_3\;(d=2)\) --- We observe peaks in the \(n\)-dependence of anomalous dimensions both for \NonSinglet{} (all couplings) and \Singlet{} (for sufficiently strong coupling), followed by a fall-off at large \(n\).
          They are not as sharp as for the heavy scalar \(\chi\) in~\cite{Fitzpatrick:2010zm}, and based on the discussion above they should probably be interpreted as exchanges of \(2\)-particle bound states associated with the large \(N\) Hubbard-Stratonovich field \(\sigma\sim\phi^2\).
    \item \(\AdS{}_5\;(d=4)\) --- For \Singlet{}-sector anomalous dimensions we observe a monotonic growth that asymptotes to a constant value \(1\) at large \(n\), no peaks seem to be present at any strength of the coupling.
          The case of \NonSinglet{} anomalous dimensions reaches a very shallow minimum/plateau followed by what seems to be a really slow fall-off of their absolute values at large \(n\).
\end{enumerate}

The above points should not be taken as conclusive results, but hopefully could help to guide potential future investigations.
It would be interesting to make the connection between the large \(n\) asymptotic behavior of the anomalous dimensions and (the partial wave expansion of) the flat space \(S\)-matrix at large \(N\) more precise.

\subsection{Large spin asymptotics}%
\label{sub:Large spin asymptotics}

Similarly to the previous subsection, also the large spin asymptotics --- taking \(n\) fixed --- is fully specified by the second piece in~\eqref{eq:anomalous_dimensions_non-singlet}, given by analytic formulas \eqref{eq:tchannel_anom_dim_2d} and \eqref{eq:tchannel_anom_dim_4d} for \(d=2\) and \(d=4\), respectively.

The large spin asymptotics were already derived in~\cite{Komargodski:2012ek,Fitzpatrick:2012yx}, as presented in~\eqref{eq:large_spin_asympt}.
The authors even computed the coefficient \(c_{0}\) determining the leading-twist deviation from MFT at large spin \cite[(1.8)]{Komargodski:2012ek}.
It is given in terms of OPE data associated with the operator of minimal twist \(\twistmin \equiv \Dmin - \Jmin\) and spin \(\Jmin\).
Since it turns out to be important for determining the correct \enquote{minimal twist} operator \(\Oper[\Repre]_{\Dmin,\Jmin}\) governing the large spin asymptotics, we reproduce it here in the form
\begin{align}
    \label{eq:large_spin_c_coeff}
    c_{0} =
    \frac{2 \G[\twistmin+2\Jmin]\G[\Dphi]^{2}}
    {\G[\frac{\twistmin}{2}+\Jmin]^{2} \G[\Dphi-\frac{\twistmin}{2}]^{2}}
    \opesq{\PRepre[\Ophi^{\ind}\Ophi^{\ind}]\, \Oper[\Repre]_{\Dmin,\Jmin}}
    \eqend
\end{align}
We take all operators involved to have 2-point functions normalized to unity, and compared to the cited formula we left out a factor of \(2^{-\Jmin}\) due to different normalization of conformal blocks.

Using \eqref{eq:tchannel_anom_dim_2d}/\eqref{eq:tchannel_anom_dim_4d} we were able to derive \(c_{n}\) even for non-leading \grayenclose{\(n>0\)} twist families, see \eqref{eq:large_spin_c_coeff_general} for the expression in terms of \(c_{0}\).

Prefactors in \eqref{eq:large_spin_c_coeff} can be traced back to \eqref{eq:tchannel_anom_dim_2d}/\eqref{eq:tchannel_anom_dim_4d}, in particular \(\xmathstrut[0.1]{0}\smash[b]{\G[\Dphi-\frac{\twistmin}{2}]^{-2}}\) is related via \(\G\)-function identities to \(\sin^{2}\left(\pi\left[\Dphi-\frac{\Dpr-J^{\prime}}{2}\right]\right)\) after identification \(\twistmin \leftrightarrow \Dpr-J'\).
These factors are responsible for producing double zeros at values of MFT operators, which will be important in the following.

Let us now recall possible candidates for the minimal twist operators, and discuss their respective orders at which they contribute in~\eqref{eq:large_spin_c_coeff}.
The candidates are clearly the three minimal twist families \([\Ophi^{i}\partial^{J}\Ophi^{j}]^{\Repre}\), \(\Repre \in \{\Singlet,\AntiSym,\SymTrless\}\) (up to a possible emergent operator in \(d=4\) that would take over the role of a minimal twist operator at sufficiently strong coupling):
\begin{itemize}
    \item[\Singlet]
          The discussion in singlet sector splits into scalar and spinning operators (see~\Cref{fig:singlet_sector_twists}).
          \begin{itemize}
              \item Scalar operators get \(\bigO{1}\) deformations from MFT scaling dimensions, thus their coefficients \(c_{n}\) are \(\bigO{1/N}\) due to the squared OPE coefficients in the singlet sector \eqref{eq:ope_sq}.
              \item Spinning operators have MFT scaling dimensions \grayenclose{at \(\bigO{1}\) order}, therefore their \(c_{n}\) coefficients are strongly suppressed by the double zeros emphasized above.
                    Given they receive some \(\bigO{1/N}\) anomalous dimensions by extending the computation of the correlator to order \(\bigO{1/N^{2}}\), their contributions to coefficient \(c_{n}\) turn out to be \(\bigO{1/N^{3}}\) --- a \(\bigO{1/N^{2}}\) suppression coming from the double-pole factor, and an additional \(\bigO{1/N}\) from their squared OPE coefficients.
          \end{itemize}
    \item[\(\smash{\mathllap{\substack{\textstyle\SymTrless\\\textstyle\AntiSym}}}\)]
          Both types of non-singlet operators have \(\bigO{1/N}\) anomalous dimensions, therefore their contributions to \(c_{n}\) coefficients are of order \(\bigO{1/N^{2}}\), since their leading OPE coefficients are \(\bigO{1}\).
\end{itemize}
The important corollary of this analysis is that although there are the \NonSinglet{} leading-twist families with \(\tau^{\NonSinglet}<2\Dphi\) needed for unitarity of the boundary \CFT{}, their contribution to large spin asymptotics~\eqref{eq:large_spin_asympt} is strongly suppressed as \(\bigO{1/N^{2}}\).
Hence, the large spin asymptotics are governed by the singlet scalar operator \(\Osigma_{0}\mathcolor{darkgray}{\sim [\Ophi^{\ind}\Ophi^{\ind}]^{\Singlet}}\) of \enquote{minimal} non-MFT twist \(\tau_{\min}=2\Dphi+\smash{\anomdim[\Singlet*]_{0,0}}\).

By evaluating (\ref{eq:large_spin_c_coeff},\,\ref{eq:large_spin_c_coeff_general}) for \(\bigl(\twistmin=2\Dphi+\smash{\anomdim[\Singlet*]_{0,0}},\Jmin=0\bigr)\), the formula \eqref{eq:large_spin_asympt} provides a complete theoretical prediction for the large spin asymptotics of the \NonSinglet{} spectrum, that can be checked against data generated by our code.

\begin{figure}[!ht]
    \centering
    \begin{adjustbox}{center}
        \hspace*{-0.015\linewidth}
        \includegraphics[width=0.51\textwidth]{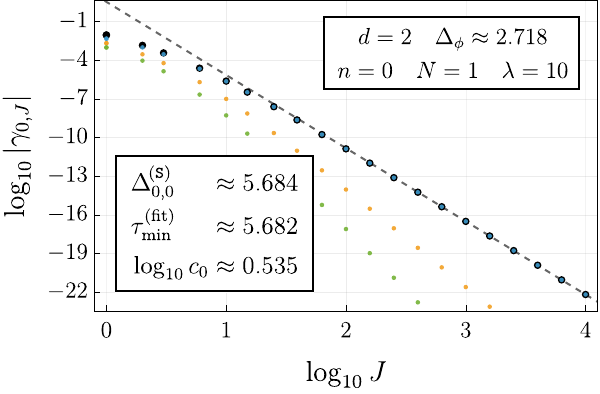}
        \hspace{0.01\linewidth}
        \includegraphics[width=0.51\textwidth]{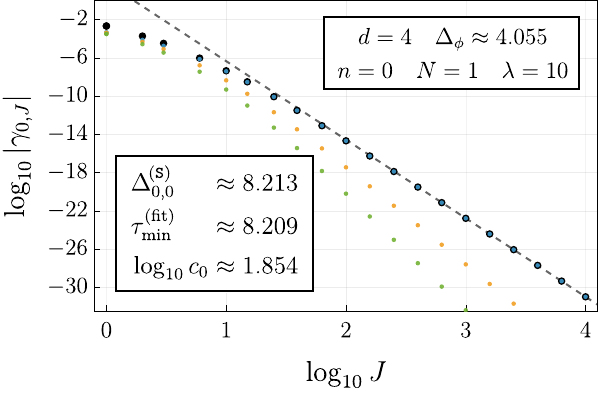}
    \end{adjustbox}
    \vspace*{-1.2ex}

    \begin{adjustbox}{center}
        \hspace*{-0.015\linewidth}
        \includegraphics[width=0.51\textwidth]{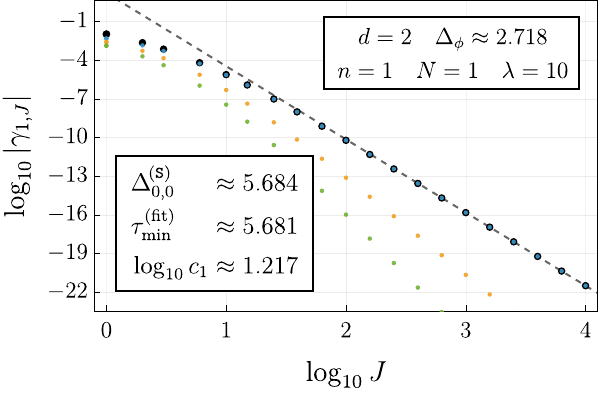}
        \hspace{0.01\linewidth}
        \includegraphics[width=0.51\textwidth]{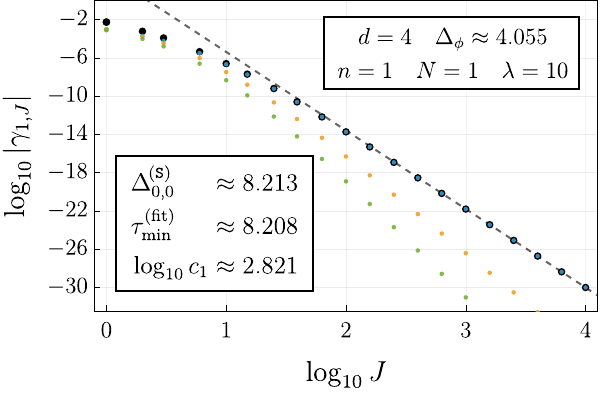}
    \end{adjustbox}
    \caption{
        Log--log plots displaying the large spin asymptotics of non-singlet anomalous dimensions for first two twist trajectories (\(n=0\) and \(n=1\)) in \(d=2\) and \(d=4\).
        Choices of the parameters used in each plot are shown in the upper right boxes.
        The last few data (black) points were fitted by a linear function~\eqref{eq:fitted_large_spin_dependence} (dashed line) with parameters given in the bottom left boxes.
        The slopes of dashed lines represent the calculated minimal twists, and are in good agreement with the theory value, thus verifying the large spin asymptotics theorem.
    }
    \label{fig:LargeSpinAsympt}
\end{figure}

We turn to describing the setup of these tests and presenting their results in the rest of this section.
Starting from~\eqref{eq:large_spin_asympt}, taking its logarithm and plugging in the relation \(\tau_{n,J} \simeq 2\Dphi+2n+\anomdim[]_{n,J}\), we considered the following functional dependence for the asymptotics
\begin{align}
    \label{eq:fitted_large_spin_dependence}
    \log_{10}\abs{\anomdim[]_{n,J}} \sim \log_{10}c_{n}-\twistmin\log_{10}J
    \eqend
\end{align}
The left-hand side was evaluated for \(\log_{10}J\in\left[0,4\right]\) with a step \(\tfrac{1}{5}\), and last few data points were then fitted.
The results for first two twist families are shown in~\Cref{fig:LargeSpinAsympt}, but we performed the same analysis also for higher twist families with similar results.

The leading asymptotics \(J^{-\twistmin}\) given by the slope of the linear fit in these log--log plots is the same for all Regge trajectories.
As explained, the value \(\twistmin^{(\text{fit})}\) determined by the fits should be compared against the scaling dimension \(\D[\Singlet]_{0,0}\) of the non-MFT \enquote{minimal} twist operator \(\Osigma_{0}\).
We confirmed almost a perfect match, which only improves with higher \(J\).
Similarly, the coefficients \(c_{n}\) show a good agreement with theoretical values \eqref{eq:large_spin_c_coeff}, \eqref{eq:large_spin_c_coeff_general}.

In spirit, this verifies that general formulas \eqref{eq:tchannel_anom_dim_2d}/\eqref{eq:tchannel_anom_dim_4d} have an appropriate form reproducing the large spin asymptotics theorem.
The specific results for the \(\OO(N)\) model then necessarily follow.

It is worthwhile to highlight another feature implemented in~\Cref{fig:LargeSpinAsympt}.
The black points represent \enquote{complete} anomalous dimensions \eqref{eq:anomlous_dimensions_non-singlet_truncated} summed up to \(k_{\text{max}}=20\), while the colored (blue, yellow, green) points correspond in order to contributions of the first few members \(\{\Osigma_{0},\,\Osigma_{1},\,\Osigma_{2}\}\) of the scalar singlet family.
It is satisfying to visually see how the \enquote{minimal} twist operator \(\Osigma_{0}\) completely saturates the asymptotics, and how the subdominant contributions are suppressed at large \(J\).
Even if one resummed the whole infinite tower of remaining operators \(\Osigma_{n\geq 1}\), their asymptotics would still remain subleading compared to the one associated with \(\Osigma_{0}\).

Finally, we should comment that the theorem is applicable for a generic \(\CFT{}_{d}\) only in \(d>2\).
The problem with \(d=2\) is that the stress tensor is in the same conformal Virasoro multiplet as the identity operator (it is its first nontrivial descendant at level two) and thus has the same twist \(\tau=0\).
Without a twist gap between the identity and an operator of minimal twist, the theorem does not apply.
Yet, we are dealing with a holographic theory without dynamical gravity in the bulk, and therefore with a non-local dual \(\CFT_{d}\) on the boundary without a stress tensor.
This implies that the whole Virasoro multiplet of the identity operator reduces just to the identity operator, and explains why we are able to use the asymptotic behavior~\eqref{eq:large_spin_asympt} also in \(d=2\), which~\Cref{fig:LargeSpinAsympt} confirms.

\section{Summary and outlook}%
\label{sec:summary}

This project arose from the desire to examine associativity of the OPE (crossing) in a \CFT{} that is at finite distance (in the sense of its spectral and OPE data) from an MFT.
Holographic theories are a fruitful playground, since especially in \(d>2\) there are not many other interacting examples that can be handled analytically, without resorting to the numerical conformal bootstrap.

We chose the \(\OO(N)\) model at finite coupling in the bulk, as it is a reasonably simple theory and its input data for crossing, in the form of the singlet spectrum, were already available thanks to~\cite{Carmi:2018qzm}.
The set of techniques employed is applicable to any other theory as long as one has partial knowledge of its \CFT{} data.

In the course of this research we obtained two classes of results which we summarize in the following.
We remind the interested reader that various detailed computations and the implementation of main formulas can be found in the accompanying \notebook{}.

\paragraph{Model-independent --- \(t\)-channel conformal block inversion for higher twists.}

Imagine a situation when one has certain control of the \CFT{} data in one channel, say the \(t\)-channel.
To complete the analysis, one should try to deduce from it \CFT{} data in the nontrivial crossed \(s\)-channel (as \(u\)-channel is related in a simple way).
In particular, for analyzing the \CFT{} spectrum it is important to know how a single \(t\)-channel conformal block contributes to \(s\)-channel anomalous dimensions.
This contribution needs to be further weighted by \(t\)-channel (squared) OPE coefficients to obtain the final result for \(s\)-channel anomalous dimensions, hence both scaling dimensions and OPE coefficients are needed as input in one of the crossing channels.

This computation requires the knowledge of conformal blocks, whose analytic expressions are known in \(d=2\) and \(d=4\) dimensions.
For these spacetime dimensions,~\cite{Liu:2018jhs} provided a closed-form formula for anomalous dimensions of leading-twist (\(n=0\)) double-twist operators \([\Ophi^{\ind}\Ophi^{\ind}]_{n,J}\), once \CFT{} data associated with the exchange of an operator in the crossed channel are specified.
In this paper we present the generalization of these formulas to arbitrary twists \(n\in\N_{0}\), namely see~\eqref{eq:tchannel_anom_dim_2d} and~\eqref{eq:tchannel_anom_dim_4d}.
Moreover, we found an elegant formula \eqref{eq:large_spin_c_coeff_general} for the coefficients of the corresponding large spin asymptotics.

Quite generally, once the direct channel spectrum is resolved, contributions from the crossed channel can be incorporated by the techniques presented in \Cref{sec:CFTgeneralities}.
We have demonstrated them in the context of the \(\OO(N)\) model, but they are applicable also for other \CFT{}s, perhaps after some generalization.
The closest candidates are \CFT{}s on the boundary of \(\AdS\) in the large \(N\) expansions of the Gross--Neveu model or the scalar QED.

\paragraph{Model-specific --- non-singlet spectrum of the \(\OO(N)\) model at finite coupling.}

The main output of this paper is the structure of the OPE of two boundary operators \(\Ophi^{\ind}\) associated with the fundamental fields \(\phi^{\ind}\) in the bulk, considering the phase with unbroken \(\OO(N)\) global symmetry.
It was obtained at a finite coupling \(\lambda\) and up to the first nontrivial order of \(1/N\) in the large \(N\) expansion.

It can be viewed as a deformation of the two limiting cases where the quartic interaction in the action \eqref{eq:Action_ON} is turned off --- either by setting \(N=\infty\) or \(\lambda=0\).
In both cases, all correlators are just products of two-point functions, and the corresponding MFT OPE can be schematically decomposed into \(\OO(N)\) irreps \grayenclose{denoted by the superscripts} as
\begin{equation}
    \label{eq:ope_MFT}
    \begin{aligned}
        \Ophi^{i}\times\Ophi^{j}
        \, & \overset{N=\infty}{\sim}\, \Id^{\Singlet}
        \oplus \left[\Ophi^{ [i}\dAlembertian^{n}\partial^{J}_{\text{\color{gray}\tiny odd}} \Ophi^{j]}\right]^{\AntiSym}_{\MFT*}
        \oplus \left[\Ophi^{\{i}\dAlembertian^{n}\partial^{J}_{\text{\color{gray}\tiny even}}\Ophi^{j\}}\right]^{\SymTrless}_{\MFT*}
        \eqcomma \\[0.6ex]
        \Ophi^{i}\times\Ophi^{j}
        \, & \mspace{5.5mu} \overset{\lambda=0}{\sim} \mspace{6mu} \, \Id^{\Singlet}
        \oplus \frac{1}{\sqrt{N}}\left[\Ophi^{\ind}\dAlembertian^{n}\partial^{J}_{\text{\color{gray}\tiny even}}\Ophi^{\ind}\right]^{\Singlet}_{\MFT*} \mathcolor{gray}{\oplus} {} \\
           & \mspace{80mu}
        \oplus \left[\Ophi^{ [i}\dAlembertian^{n}\partial^{J}_{\text{\color{gray}\tiny odd}} \Ophi^{j]}\right]^{\AntiSym}_{\MFT*}
        \oplus \left[\Ophi^{\{i}\dAlembertian^{n}\partial^{J}_{\text{\color{gray}\tiny even}}\Ophi^{j\}}\right]^{\SymTrless}_{\MFT*}
        \eqend
    \end{aligned}
\end{equation}
Both of them are obtained from the disconnected parts of the 4-point function decomposition in (\ref{eq:4-pt_singlet},\,\ref{eq:4-pt_antisymmetric},\,\ref{eq:4-pt_symmetric_traceless}).
As the singlet sector double-twist operators have a factor of \(1/N\) in their squared OPE coefficients, OPE coefficients are of order \(\bigO*{1/\sqrt{N}}\).
For \(N=\infty\) they vanish, and the singlet double-twist operators are transferred fully to the \SymTrless{} part, see also \eqref{eq:types_of_operators}.

At finite \(\lambda\) and large (but finite) \(N\), this picture is modified --- only leading corrections in large \(N\) are shown --- to the following schematic form
\begin{equation}
    \label{eq:ope_result}
    \begin{aligned}
        \Ophi^{i}\times\Ophi^{j}
        \,\sim\, \Id^{\Singlet}
         & \oplus \frac{1}{\sqrt{N}}\left[\Osigma_{\dind} \mathcolor{darkgray}{\sim \text{\enquote{\(\Ophi^{\ind}\dAlembertian^{n}\Ophi^{\ind}\)}}}\right]^{\Singlet}_{\text{fin}}
        {}\oplus \frac{1}{\sqrt{N}}\left[\Ophi^{\ind}\dAlembertian^{n}\partial^{J>0}_{\text{\color{gray}\tiny even}}\Ophi^{\ind}\right]^{\Singlet}_{\MFT*} \mathcolor{gray}{\oplus} {} \\
         & \oplus \left[\Ophi^{ [i}\dAlembertian^{n}\partial^{J}_{\text{\color{gray}\tiny odd}} \Ophi^{j]}\right]^{\AntiSym}_{1/N}
        \oplus    \left[\Ophi^{\{i}\dAlembertian^{n}\partial^{J}_{\text{\color{gray}\tiny even}}\Ophi^{j\}}\right]^{\SymTrless}_{1/N}
        \eqend
    \end{aligned}
\end{equation}
Discussion of singlet operators in the first line of~\eqref{eq:ope_result} splits according to the spin.
Scalar boundary operators \(\Osigma_{\dind}\) are induced by the composite interacting \(\sigma\)-field in the bulk --- they are given by the poles of the spectral function associated with the exact \(\sigma\)-propagator.
As indicated, their scaling dimensions get finite shifts from MFT, and their OPE coefficients are \(\bigO*{1/\sqrt{N}}\) to the order considered.
Spinning singlet operators do not receive such \(\bigO{1}\) corrections to their MFT scaling dimensions.
The formula for scaling dimensions of the \(\OO(N)\) singlet sector was already given in~\cite{Carmi:2018qzm}.

The non-singlet operators in the second line of~\eqref{eq:ope_result} get \(1/N\) shifts to their scaling dimensions.
Leading OPE coefficients in this sector are of order \(\bigO{1}\), and acquire \(\bigO*{1/\sqrt{N}}\) deformations.
Both of these corrections correspond to a \(1/N\) modification of the correlator.
In this work we did not compute the corrections to the OPE coefficients, even though the setup is ready for it.
It is just a matter of computing residues of fairly complicated functions at known poles, and we might come back to it in the future.

The main new contribution of this work are the scaling dimensions for the non-singlet double-twist families transforming in the symmetric traceless \SymTrless{} and anti-symmetric \AntiSym{} irreps of the \(\OO(N)\) group.
They were obtained for \(d=2\) (\(\AdS_{3}\)) and \(d=4\) (\(\AdS_{5}\)), as a function of the twist/Regge trajectory label \(n\), the spin \(J\), the external scaling dimension \(\Delta_{\phi}\), and the coupling \(\lambda\).

It is important to realize that they are given as a sum of partly factorized expressions --- the model-dependent squared OPE coefficients, times a model-independent contribution of a crossed channel block, which however still needs to be evaluated at model-dependent values.
The relevant formula is \eqref{eq:anomalous_dimensions_non-singlet}, which we reproduce here with a particular emphasis on the functional dependence
\begin{align}
    \label{eq:anomlous_dimensions_non-singlet_res}
    \anomdim[]^{\NonSinglet}_{n,J}(\Dphi,\lambda) = \!
    \sum_{\Oper[\Singlet]_{\dind,0}(\Dphi,\lambda)} \!
    \opesq{\Ophi\Ophi\Oper[\Singlet]_{\dind,0}}(\Dphi,\lambda)
    \, \anomdim_{n,J}(\Dphi) \color{darkgray}
    \eval_{\substack{\text{\(t\)-channel}\mspace{115mu} \\[0.1ex]\text{exchange of }\smash{\Oper[\Singlet]_{\dind,0}(\Dphi,\lambda)}}} \color{black}
    \eqend
\end{align}
The dependence on the external scaling dimension \(\Dphi\) and the coupling \(\lambda\) is complicated, in particular due to an implicit dependence in \(\anomdim\) via solutions of a transcendental equation~\eqref{eq:singlet_poles-bubble_roots}.
However, the dependence on the Regge trajectory label \(n\) and the spin \(J\) is completely isolated in the model-independent piece.
Especially, asymptotic behavior for large \(J\) or \(n\) can be in principle analytically determined from~\eqref{eq:tchannel_anom_dim_2d} or~\eqref{eq:tchannel_anom_dim_4d}, and is displayed in~\Cref{fig:LargeSpinAsympt} and \Cref{fig:Non-Singlet_AnomDim_nDep_2d}.
The large spin asymptotics are verified by a theorem independently predicting it.

The most significant projection of these data --- after choosing some fixed coupling and desired external scaling dimension --- is a plot in the twist--spin plane.
A combined plot of the anti-symmetric and symmetric traceless double-twist families is shown in \Cref{fig:NonSingletTwists2d} for \(d=2\), and in \Cref{fig:NonSingletTwists4d} for \(d=4\).
Both plots confirm organization of the non-singlet double-twist spectrum into Regge trajectories concave in spin.

Unlike in the singlet sector where anomalous dimensions are strictly positive for \(J=0\) operators --- indicating a repulsive interaction in the bulk --- and zero otherwise, anti-symmetric and symmetric traceless double-twist families have negative anomalous dimensions --- indicating an attractive interaction in the bulk.

In the process of computing non-singlet anomalous dimensions, the (squared) OPE coefficients \eqref{eq:ope_sq} in the scalar singlet sector were needed.
We plotted them in \Cref{fig:ope2} as a function of the coupling.
In the limit \(\lambda\to 0\) they approach MFT values as they must, which provides another successful test of implementation.
Compared to anomalous dimensions displaying fixed sign and convexity properties, OPE coefficients show a rather unorganized behavior.
Different \(\Osigma_{\dind}\) operators get corrected by different signs and their relative strength changes with the coupling.

Based on the self-consistency scrutiny of the spectrum summarized in the previous paragraphs and also additional checks done in~\cite{Carmi:2018qzm} we believe that the results are reliable and ready to be used in applications (at least in \(d=2\) which has a well understood phase structure and no UV divergence).

\paragraph{Future directions.}
\label{par:Future directions}
There are still various unresolved questions and possible future directions.
One concerns the anomalous dimensions of \(J=0\) \SymTrless{} operators, for which we have indications that they were not properly captured by our general method of computation.
While it is possible to manually take into account the missed contribution of the contact Witten diagram, see~\eqref{eq:full-ST-spin0-anomdim}, it is still unclear whether \sixjsymbol{} (employing under the hood the Lorentzian inversion formula) is not missing some other contributions.

Another challenge (of a technical nature) is the computation of the OPE coefficients for the non-singlet operators.
Since in principle we can calculate anomalous dimensions in the non-singlet sector for any \(n\) and \(J\), this opens a possibility to perform a more detailed analysis of the spectrum and its properties.

Following \cite{DiPietro:2023inn}, it would be interesting to calculate the non-singlet spectrum of the \(\OO(N)\) model in \emph{de Sitter} (\dS{}) background.

A further valuable extension of our work concerns the critical theory in the bulk.
Its careful investigation particularly attracts our attention for a future project --- either in the \(\CFT[B]{}_{d+1}\) setting~\cite{Giombi:2020rmc} linked directly to \(\AdS{}_{d+1}\) or in the \(\CFT[D]{}_{n+m}\) setting~\cite{Cuomo:2021kfm} corresponding to a theory on \(\AdS_{n}\times \Sphere^{m}\) (and requiring thus the Kaluza--Klein reduction on the sphere).
We hope to report on some additional observables in these theories that could be inferred from the results obtained in this work.

\acknowledgments{We thank Jozef Csipes for pointing out a few minor typographical errors.
    This work was supported by the Ministry of Education, Youth and Sports of the Czech Republic under the grant \texttt{FORTE CZ.02.01.01/00/22\_008/0004632} and by the Grant Agency of Czech Republic under the grant \texttt{GA-CR 24-11722S}.}

\appendix
\crefalias{section}{appendix}

%% Bibliography/References
\bibliography{ref}
\bibliographystyle{JHEP}

\end{document}